\documentclass{article}
\usepackage{amssymb}
\usepackage{amsmath}

\input{tcilatex}
\begin{document}

\begin{center}
{\LARGE The unitary dynamical state-locking process, the HSSS quantum search
process, and the quantum-computing speedup theory}%
\begin{equation*}
\end{equation*}

{\large Xijia Miao\footnote{%
Email: miaoxijia@yahoo.com}}

{\large Somerville, Massachusetts}

{\large Date: September, 2013\bigskip }

{\Large An extended abstract}
\end{center}

A unitary dynamical state-locking (UNIDYSLOCK) process is a unitary process
that transforms simultaneously two or more orthogonal quantum states to
their corresponding non-orthogonal states whose differences may be
arbitrarily small. Its inverse process, i.e., the quantum-state-difference
amplification (QUANSDAM) process could be used to realize an exponential
quantum-computing speedup in solving an unstructured search problem in the
quantum-computing speedup theory (X. Miao, arXiv: quant-ph/1105.3573
(2011)). In this paper the principle and mechanism of how a UNIDYSLOCK
process and a QUANSDAM process work in a quantum system are described in
detail on the basis of the quantum-computing speedup theory. A UNIDYSLOCK
process is a unitary quantum dynamical process and also obeys the
mathematic-logical principle of the search problem. In pure quantum
mechanics ($QM$) a UNIDYSLOCK\ process does not exist at all, because any $%
QM $ unitary operator is not able to change the quantum-state difference
between a pair of quantum states, as shown in the paper. It must be
comprehended from the viewpoint of the mathematical-logical principle. This
unique property of a UNIDYSLOCK process reflects the fundamental importance
of the interaction between the quantum physical laws (i.e., the unitary
quantum dynamics and the Hilbert-space symmetric structure and property) and
the mathematical-logical principle that a computational problem obeys in the
quantum-computing speedup mechanism (or theory). The reversible functional
operation of the search problem is originally responsible for the
quantum-state-difference change in a UNIDYSLOCK\ (or QUANSDAM) process. But
this interaction leads to that though a $QM$ unitary operator itself cannot
change the quantum-state difference, it could promote a UNIDYSLOCK\ (or
QUANSDAM) process to change greatly the quantum-state difference. It is also
shown that any quantum-state effects can not cause the
quantum-state-difference change during a UNIDYSLOCK\ (or QUANSDAM) process
in a quantum system, here the quantum-state effects include the
quantum-state superposition, coherence interference, entanglement and
nonlocal effect, correlation and so on. This conclusion is universal in the
frame of unitary quantum dynamics for any quantum computational process that
employs a UNIDYSLOCK\ (or QUANSDAM) process to realize its quantum-computing
speedup. A UNIDYSLOCK\ (or QUANSDAM) process is the characteristic quantum
computational process (or sub-process) of the quantum-computing speedup
theory. Even if a conventional quantum computation that is reversible or
unitary contained a UNIDYSLOCK (or QUANSDAM) process, the contribution of
the UNIDYSLOCK (or QUANSDAM) process to the quantum-computing speedup of the
conventional quantum computation would be secondary or negligible. The
reason for this is that if the contribution of the UNIDYSLOCK (or QUANSDAM)
process were dominating, then the contribution of the quantum-state effects
to the quantum-computing speedup would be secondary or negligible. Generally
the quantum-state effects are considered to be essentially responsible for
the quantum-computing speedup in the conventional quantum computational
theory based on the quantum parallel principle. Therefore, whether or not a
UNIDYSLOCK (or QUANSDAM) process can make an essential contribution to a
quantum-computing speedup can distinguish the quantum-computing speedup
theory from any conventional quantum computational theory. A QUANSDAM\
process is the second step of the two consecutive steps of the $HSSS$
quantum search process, while the first step is the search-space dynamical
reduction. The $HSSS$ quantum search process is used to solve an
unstructured search problem, which is essentially different from a
conventional quantum search algorithm. Its quantum-computing speedup
mechanism is studied in the paper. Its quantum-computing speedup is original
from the interaction mentioned above and hence the interaction is also
responsible for the search-space dynamical reduction. In particular, the
fundamental quantum-computing resource, i.e., the symmetric structure of the
Hilbert space of a composite quantum system is necessary to realize the
search-space dynamical reduction. The $HSSS$ quantum search process works in
both the Hilbert space of the quantum system and the math Hilbert space of
the search problem. In contrast, a conventional quantum search algorithm
works only in the Hilbert space of the quantum system. The math Hilbert
space of a computational problem is a fundamental concept in the
quantum-computing speedup theory. It is mathematical and does not have any
physical meaning. No concept of the math Hilbert space exists in
conventional quantum computation. Due to the math Hilbert space the manner
to realize a reversible functional operation in the quantum-computing
speedup theory is essentially different from the one in conventional quantum
computation. A mathematical-parallel functional operation is the
characteristic manner in the quantum-computing speedup theory. In contrast,
a quantum-parallel functional operation is the characteristic manner in the
conventional quantum computational theory. Due to the interaction between
the math Hilbert space and the Hilbert space a reversible functional
operation is able to change the quantum-state difference. This leads to that
there is a UNIDYSLOCK (or QUANSDAM) process in the quantum-computing speedup
theory. \pagebreak \newline
{\Large 1. Introduction}

According to the quantum-computing speedup theory [1] the unitary quantum
dynamics is considered as the universal quantum driving force to speed up a
quantum computation, while the symmetric structure of the Hilbert space of a
composite quantum system is the fundamental quantum-computing resource.%
\footnote{%
Tensor product of the Hilbert space of a composite quantum system could lead
to that the energy used by a quantum computation in the quantum system may
not be exponentially large, here the energy is governed by the
tensor-product Hamiltonian of the quantum system. This property of tensor
product could be important for a quantum computation, but it may not be
related to the fundamental quantum-computing resource to speedup essentially
a quantum computation.} Here the Hilbert-space symmetric structure is not
owned by classical computation and hence it is responsible for an
exponential quantum-computing speedup. Both the unitary quantum dynamics and
the fundamental quantum-computing resource have been thought of as the two
pillars to build an efficient quantum search process (See the early author's
works [5, 12]). The interaction between the quantum-physical laws (i.e., the
unitary quantum dynamics and the Hilbert-space symmetric structure and
property) and the mathematical-logical principle obeyed by a computational
problem to be solved [5, 6, 1, 2] is really fundamentally important to
realize an exponential quantum-computing speedup in solving the (hard)
computational problem and understand the mechanism of the exponential
quantum-computing speedup in quantum computation. The theory proposes a
two-step quantum-computing process to solve efficiently an unstructured
search problem, which may be called the unstructured quantum search process
based on the unitary quantum dynamics and the $\mathit{H}$ilbert-$\mathit{s}$%
pace $\mathit{s}$ymmetric $\mathit{s}$tructure or briefly the $HSSS$ quantum
search process. The first step of the $HSSS$ quantum search process is an
efficient search-space dynamical reduction that the exponentially large
unstructured search space of the search problem is reduced dynamically to a
polynomially small subspace. In this step the unstructured search space is
eliminated and at the same time a polynomially small subspace is generated
that carries the information of the component state of the solution state of
the search problem \footnote{%
The definition of the solution state of a search problem in this paper is
consistent with those in a conventional quantum search algorithm [3, 13].
Given a Boolean function $f:\{0,1,...,N-1\}$ $\rightarrow \{0,1\}$ there
exists a unique element $x_{0}\in \{0,1,...,N-1\}$ such that $f(x)=1\ $if $%
x=x_{0}$ and $f(x)=0$ if $x\neq x_{0}.$ Then the solution state is defined
as the quantum state corresponding to the element $x_{0}.$ Sometimes the
solution state also is called the marked state, etc..}. This step is
performed in the frame of unitary quantum dynamics and with the help of the
Hilbert-space symmetric structure of the quantum system. Its detailed
mechanism has been described in Refs. [1, 2] and its efficient realization
in an $n-$qubit quantum system is described in Ref. [2]. The second step of
the process is involved in the $\mathit{uni}$tary $\mathit{dy}$namical $%
\mathit{s}$tate-$\mathit{lock}$ing ($UNIDYSLOCK$) process and its inverse
process, i.e., the $\mathit{quan}$tum-$\mathit{s}$tate-$\mathit{d}$ifference 
$\mathit{am}$plification ($QUANSDAM$) process. It mainly employs a QUANSDAM
process to extract efficiently the information of all the component states
of the solution state. The solution to the search problem then is obtained
completely from this solution information. The present work is for this
second step. It is devoted to investigating the principle and mechanism of
how a UNIDYSLOCK process and a QUANSDAM process work in a quantum system. In
the future paper [36] the author will report in detail an exponential
QUANSDAM process in a single-atom system.

The $HSSS$ quantum search process is essentially different from the
conventional quantum search algorithm (i.e., the Grover's quantum search
algorithm [3]). Though they obey the same mathematical-logical principle
that the unstructured search problem obeys, in algorithm they essentially
differ from one another mainly in the two aspects as follows. The first
aspect is that what the conventional quantum search algorithm searches for
is the solution state of the search problem as a whole, while what the $HSSS$
quantum search process searches for is any single component state of the
solution state. This difference really reflects that there must be the
search-space dynamical reduction for the $HSSS$ quantum search process. The
second aspect is that as far as their basic building blocks, i.e., the
black-box functional operations are concerned, they obey essentially
different mathematical-logical principles of black-box functional operation,
respectively. The oracle operation (i.e., the black-box functional
operation) of a conventional quantum search algorithm [3, 13] is a quantum
parallel functional operation in the Hilbert space of the quantum system. In
contrast, the black-box functional operation of the $HSSS$ quantum search
process works not only in the Hilbert space of the quantum system but also
in the mathematical Hilbert space (i.e., the math Hilbert space) of the
search problem. In the Hilbert space of the quantum system it is considered
as a single black-box functional operation and hence does not have the
mathematical-logical meaning of quantum parallel operation and in the math
Hilbert space it may be thought of as a mathematical-parallel black-box
functional operation that is essentially different from a conventional
quantum parallel functional operation [11, 10, 9]. This difference indicates
that there is the quantum-state-difference amplification for the $HSSS$
quantum search process, which is essentially different from the amplitude
amplification of a conventional quantum search algorithm.

The $HSSS$ quantum search process does not employ the quantum parallel
principle of the conventional quantum computation theory [11a] (the
essential aspect of the principle is the quantum-state entanglement and
nonlocal effect) to achieve its quantum-searching speedup. Instead, its
quantum-searching speedup is original from the interaction between the
quantum physical laws (i.e., the unitary quantum dynamics and the
Hilbert-space symmetric structure and property) and the mathematical-logical
principle that the unstructured search problem obeys [1]. Because the $HSSS$
quantum search process has its own quantum-computing speedup mechanism
essentially different from the one of a conventional quantum search
algorithm, in which it eliminates the unstructured search space and obeys
the mathematical-logical principle of black-box functional operation that is
essentially different from the one obeyed by a\ conventional quantum search
algorithm (See the preceding paragraph), the well-known square speedup limit
[4a] on a conventional quantum search algorithm could not be applied to it.
Therefore, it could have a super-square or even an exponential
quantum-searching speedup.

It is well known that a conventional quantum search algorithm to solve an
unstructured search problem can achieve a square search speedup [3], while
its classical counterpart can achieve at best a linear search speedup; and
moreover, this square speedup is the maximum one in the worst case [4]
(i.e., the square speedup limit). It has been believed extensively that
hardness to solve an unstructured search problem is closely related to the
unstructured symmetric property of search space of the search problem both
in classical computation and in quantum computation. Intuitively one may say
that due to its unstructured property the unstructured search space could
prohibit any conventional quantum and classical search algorithms from
achieving an exponential searching speedup. A reasonable explanation for
this phenomenon is simply given below.

A quantum search process (or algorithm) obeys not only the unitary quantum
dynamics but also the mathematical-logical principle such as the symmetric
structure of unstructured search space of the search problem [5, 6, 1].
There is no exception for a conventional quantum search algorithm. When the
conventional quantum search algorithm is performed in a quantum system, the
unstructured symmetric property of search space may make a severe constraint
on the time evolution process of the quantum system. As a consequence of
this constraint, the quantum system could be forced to run into the regime
of\ classical physics, here or there and early or later, during the quantum
search process, no matter that it is a single quantum system such as a
single-atom system or a composite quantum system such as an $n-$qubit
quantum system, and no matter which quantum state it takes at the initial
time (its quantum state at any later time is determined completely by the
initial state according to the unitary quantum dynamics). This consequence
implies that a conventional quantum search algorithm is semiclassical [1].
Ultimately it results in that the quantum-searching speedup of the algorithm
is greatly suppressed and the algorithm becomes inefficient in the worst
case.

If the unstructured search space is not eliminated and the classical effect
hidden behind the unstructured search space is not purged, any unstructured
quantum search algorithm is not able to achieve an exponential speedup.
Therefore, the very first task for any possible efficient unstructured
quantum search process (or algorithm) is naturally to eliminate this
unstructured search space.

The search-space dynamical reduction, i.e., the first step of the $HSSS$
quantum search process is just used to eliminate this unstructured search
space. Obviously, it is not a usual computational complexity reduction. The
latter means that one computational problem may be solved by reducing it to
another which may be solved conveniently on quantum computer. In concept it
has nothing to do with the fundamental quantum-computing resource.

It is not free to eliminate an unstructured search space. First of all, the
fundamental quantum-computing resource is necessary to realize a
search-space dynamical reduction. Then a search-space dynamical reduction
even makes it harder to extract the information of the component states of
the solution state of a search problem from the amplitude-amplification
point of view of a conventional quantum search algorithm. It is known [3]
that the initial amplitude (i.e., the minimum amplitude) of the solution
state is proportional to $1/\sqrt{N}$ in a conventional quantum search
algorithm with search-space dimension $N=2^{n}$. However, in the $HSSS$
quantum search process, after a search-space dynamical reduction (here the
unitary state-transformation description is used) the quantum state of the
polynomially-small reduction subspace (with dimension two) that carries the
solution information of the component state has an amplitude proportional to 
$1/N$ [2]. This amplitude is exponentially small. Moreover, it is even
exponentially smaller than its counterpart ($1/\sqrt{N}$) in a conventional
quantum search algorithm. It seems that a search-space dynamical reduction
makes thing worse. This, of course, concludes from the
amplitude-amplification viewpoint of a conventional quantum search algorithm.

Here an exponentially-small amplitude for the final quantum state in a
search-space dynamical reduction that carries the solution information means
an exponentially-small quantum-state difference between the possible quantum
states created at the end of the search-space dynamical reduction. The
latter, i.e., the quantum-state difference, is closely related to the
characteristic physical quantity, i.e., the quantum-state-difference varying
rate, of a UNIDYSLOCK process and a QUANSDAM process in the $HSSS$ quantum
search process [1].

After the search-space dynamical reduction, the $HSSS$ quantum search
process performs the inverse of a UNIDYSLOCK process, i.e., the QUANSDAM
process to amplify the quantum-state difference [1], so that the solution
information can be obtained for the search problem. This is just the second
step of the $HSSS$ quantum search process. Since the unstructured search
space is eliminated, there could not exist such a severe constraint as the
square speedup limit [4] on the $\mathit{q}$uantum-$\mathit{s}$tate-$\mathit{%
d}$ifference ($QSD$) amplification ability (i.e., the $QSD$ varying speed)
of the QUANSDAM process in the $HSSS$ quantum search process. Because the $%
QSD$ amplification ability is original from the interaction between the
quantum physical laws (i.e., the unitary quantum dynamics and the
Hilbert-space symmetric structure and property) and the mathematical-logical
principle of the search problem, there could be a QUANSDAM\ process that has
a super-square or even an exponential $QSD$\ amplification ability. If the $%
QSD$ amplification ability is large enough, then it can compensate not only
the amplitude loss due to the search-space dynamical reduction (See the
paragraphs above) but also it enables the $HSSS$ quantum search process as a
whole to achieve a super-square or even an exponential quantum-searching
speedup. Therefore, the $QSD$ amplification ability of a QUANSDAM process is
a key factor to measure whether the $HSSS$ quantum search process is
successful or not.

Originally the concept of unitary dynamical state-locking (UNIDYSLOCK)
process is connected with the search-space dynamical reduction in the $HSSS$
quantum search process [5, 6]. It was made an attempt in Ref. [6] to use the
reversible and unitary halting protocol to realize the search-space
dynamical reduction,\footnote{%
So far there still lacks a rigorous mathematical proof to show that a
reversible and unitary halting protocol can realize a search-space dynamical
reduction. Obviously, any conventional halting operation (or protocol) that
is involved in quantum measurement is not suited to realize a search-space
dynamical reduction.} while it was proposed to use a UNIDYSLOCK process [6,
7] to realize the reversible and unitary halting protocol.

Initially a UNIDYSLOCK process was constructed in Ref. [7] with the help of
a single-atom system. It was used to convert unitarily the solution state
(or its component state) of the search problem into the final state that is
sufficiently close to the desired (known) state with the help of the
reversible and unitary halting protocol. It was thought at that time that
the final state might be prepared more easily. Then the inverse of the
UNIDYSLOCK process could change unitarily the final state back to the
solution state (or its component state) with the help of the inverse of the
reversible and unitary halting protocol. Therefore, a UNIDYSLOCK process and
its inverse process, i.e., the QUANSDAM process could be used to solve the
unstructured search problem [6, 7]. It is known that the solution state can
be any state of the unstructured search space of the search problem. This
implies that a UNIDYSLOCK process could convert unitarily many different (or
orthogonal) quantum states into their corresponding non-orthogonal final
states at the same time, respectively, which are sufficiently close to each
other and to the desired state. Unfortunately, the research into a
UNIDYSLOCK process in past years found that in pure quantum mechanics a
UNIDYSLOCK process does not make sense and cannot have an ability to convert
unitarily many orthogonal quantum states into their corresponding
non-orthogonal states at the same time.

Actually the research also finds that in classical reversible computation
[8] and any conventional quantum computation [9, 10, 11] any computational
process which is reversible (or unitary) can not transform many orthogonal
quantum states to their corresponding non-orthogonal states at the same
time. This raises question as to whether a UNIDYSLOCK process does exist
really in quantum computation.

In nature only a non-equilibrium irreversible process is able to change many
orthogonal quantum states to their corresponding non-orthogonal states at
the same time. It is well known that in reality there is a huge conflict
between a non-equilibrium irreversible process and a reversible process or a
quantum-mechanical unitary dynamical process. The latter two processes are
not able to change the quantum-state difference. The largest puzzle for a
UNIDYSLOCK process then is that it is a unitary quantum dynamical process,
but it wants to change simultaneously many orthogonal quantum states to
their corresponding non-orthogonal states.

A UNIDYSLOCK\ process [6, 7] is not only a unitary quantum dynamical process
but it also obeys the mathematical-logical principle of the computational
problem to be solved (i.e., an unstructured search problem here). Therefore,
a complete description for a UNIDYSLOCK process must consider both the
aspect of the unitary quantum dynamics and the aspect of the
mathematical-logical principle [1]. This is quite like that a microscopic
particle is described by the wave-particle dualism in quantum mechanics,
that is, a microscopic particle can be described completely only when one
considers both the particle aspect and the wave aspect of the same particle.
Both the aspects of the unitary quantum dynamics and the
mathematical-logical principle together and especially their interaction for
a UNIDYSLOCK\ process are really key to understanding completely what a
UNIDYSLOCK process means exactly and how it works and to answering all its
puzzles.

On the basis of the quantum-computing speedup theory [1] this paper is
devoted to revealing the principle and mechanism of how a UNIDYSLOCK process
and a QUANSDAM process\ work in a quantum system. A UNIDYSLOCK (or QUANSDAM)
process is the only nontrivial and unitary quantum-physical process that can
change the quantum-state difference between a pair of quantum states. It is
essentially different from a usual non-equilibrium irreversible process
which could contain the usual quantum measurement. It is the characteristic
quantum computational process (or sub-process) of the quantum-computing
speedup theory. Whether or not its contribution to a quantum-computing
speedup is dominating can distinguish the quantum-computing speedup theory
from any conventional quantum computational theory. In the paper the
research is also carried out on the quantum-computing speedup mechanism of
the $HSSS$ quantum search process. The $HSSS$ quantum search process works
in both the Hilbert space of a quantum system and the math Hilbert space of
the search problem. In contrast, a conventional quantum search algorithm [3,
13] works only in the Hilbert space of the quantum system. Here the math
Hilbert space of a computational problem is a fundamental concept of the
quantum-computing speedup theory. The concept does not exist in conventional
quantum computation. Due to the math Hilbert space a mathematical-parallel
operation is the characteristic manner to realize a reversible functional
operation in the quantum-computing speedup theory. In contrast, a quantum
parallel functional operation is the characteristic manner in the
conventional quantum computational theory based on the quantum parallel
principle [11a]. The fundamental interaction between the quantum physical
laws (i.e., the unitary quantum dynamics and the Hilbert-space symmetric
structure and property) and the mathematical-logical principle that a
computational problem obeys is essentially responsible for a
quantum-computing speedup in the quantum-computing speedup theory. This is a
fundamental principle of the theory. It is available not only for a
UNIDYSLOCK (or QUANSDAM) process but also for the whole $HSSS$ quantum
search process.

There are six sections in the paper. The section 2 describes the relation
between the search-space dynamical reduction and the information-carrying
unitary operator. The section 3 is devoted to investigating the principle
and mechanism of how a UNIDYSLOCK process and a QUANSDAM process\ work in a
quantum system. The section 4 describes how to prepare the
information-carrying unitary propagators of a quantum system and especially
a single-atom system. The section 5 describes the mechanism of the $HSSS$
quantum search process and the essential difference between the $HSSS$
quantum search process and the conventional quantum search algorithm. The
final section is devoted to discussing the essential difference between the
quantum-computing speedup theory and any conventional quantum computational
theory.\newline
\newline
{\Large 2. The efficient search-space dynamical reduction and the
information-carrying unitary operators}

The search-space dynamical reduction was proposed in Ref. [5a] for the first
time. It was realized efficiently in Ref. [2]. Its purpose is to realize
that the exponentially large unstructured search space of a search problem
is reduced dynamically to a polynomially small subspace. The unstructured
search space of a search problem is mathematical. It is different from the
usual Hilbert space of a quantum system used to solve the search problem.
From the mathematical-logical viewpoint a direct reduction of the
unstructured search space is not allowed [1]. However, the quantum-computing
speedup theory [1] does allow one to reduce indirectly and dynamically the
unstructured search space through the unitary oracle selective diagonal
operator in the frame of unitary quantum dynamics. This reduction needs to
use the fundamental quantum-computing resource and is shaped by the
Hilbert-space symmetric structure of the quantum system. A search-space
dynamical reduction shows that in the quantum-computing speedup mechanism
[1] one needs to consider explicitly not only the quantum-physical laws
(i.e., the unitary quantum dynamics and the Hilbert-space symmetric
structure and property) that governs the quantum system used to solve the
search problem but also the mathematical-logical principle (such as the
unstructured symmetric property of the search space) obeyed by the search
problem and especially the interaction between the quantum-physical laws and
the mathematical-logical principle. This is essentially different from the
conventional quantum computational theory [11].

There are two different Hilbert spaces that need to be considered explicitly
in the quantum-computing speedup theory [1]. One of which is the Hilbert
space of a quantum system that is used to solve a computational problem. It
is quantum physical. Its symmetric structure is the fundamental
quantum-computing resource. Another is of the computational problem to be
solved in the quantum system. It is a linear complex vector space that
accommodates the mathematically operational space of function of the
computational problem. It is mathematical and does not have any physical
meaning. For convenience, hereafter this linear complex vector space is
called the mathematical (or math) Hilbert space of the computational
problem. The math Hilbert space of a computational problem is a fundamental
concept in the quantum-computing speedup theory. There does not exist
independently the concept of the math Hilbert space in any conventional
quantum computation. The very reason why the quantum-computing speedup
theory needs to consider explicitly both the Hilbert spaces is that in the
theory an essential quantum-computing speedup is original from the
interaction between the quantum-physical laws (i.e., the unitary quantum
dynamics and the Hilbert-space symmetric structure and property) obeyed by a
quantum system and the mathematical-logical principle obeyed by a
computational problem to be solved in the quantum system.

The quantum-computing speedup theory [1] then pointed out that the math
Hilbert space (e.g., the unstructured search space) may be considered as the
Hilbert space of the quantum system in conventional quantum computation.
This is the special case. Equivalently it may be considered that the Hilbert
space of a quantum system also acts as the math Hilbert space in
conventional quantum computation, and then both the Hilbert space and the
math Hilbert space are treated together as one space. Clearly there is not a
separate math Hilbert space in conventional quantum computation. There is
only the Hilbert space of the quantum system. There is no consideration of
the interaction between the two Hilbert spaces. These results show that
there is no exploitation of the Hilbert-space symmetrical structure of a
quantum system to speed up a quantum computation in conventional quantum
computation [9, 10, 11].

In contrast, in the quantum-computing speedup theory one has to treat
explicitly and separately the Hilbert space of a quantum system and the math
Hilbert space of a computational problem, because the $HSSS$ quantum search
process needs first to perform the search-space dynamical reduction [5, 6,
1, 2] for the unstructured search space which is just the math Hilbert space
of an unstructured search problem. Moreover, one has to set up suitably the
interaction between the two Hilbert spaces so that the Hilbert-space
symmetric structure of the quantum system (i.e., the fundamental
quantum-computing resource) can be harnessed to speed up a quantum
computation.

The difference for how to treat these two different Hilbert spaces mentioned
above between the quantum-computing speedup theory [1] and the conventional
quantum computational theory based on the quantum parallel principle [11a]
really reflects two different understandings for the origin of
quantum-computing speedup. For the quantum-computing speedup theory an
essential quantum-computing speedup is original from\ the interaction
between the quantum-physical laws (i.e., the unitary quantum dynamics and
the Hilbert-space symmetric structure and property) and the
mathematical-logical principle that a computational problem obeys, while for
the conventional quantum computational theory\ an essential
quantum-computing speedup is original from\ the quantum-state effects which
may include the quantum-state superposition, coherence interference,
entanglement and non-local effect, correlation and so on. (Among these
quantum-state effects the essential one is the quantum entanglement and
non-local effect.) The difficulties and limitations for the conventional
quantum computational theory [11] have been exposed in detail in Ref. [1].

A typical example is given below to help understand both the Hilbert space
of a quantum system and the math Hilbert space of a computational problem.
Suppose that the search space of an unstructured search problem is given by $%
\left\{ 0,1,...,N-1\right\} $ in the (mathematical) number representation or
by $\left\{ |0),|1),...,|N-1)\right\} $ in the vector (or state)
representation with the usual computational basis vectors $\{|k)\}$ ($0\leq
k<N$). The latter may be called the math Hilbert space of the search
problem. It is a linear complex vector space with dimensional size $N$. It
is mathematical and unstructured. It does not have any physical meaning,
although here one uses the vector (or state) set $\left\{
|0),|1),...,|N-1)\right\} $ to stand for it. It is different from the usual
Hilbert space of a quantum system. The latter is quantum physical. As usual,
here the Hilbert space of a quantum system with dimensional size $N$ may be
given by $\left\{ |0\rangle ,|1\rangle ,...,|N-1\rangle \right\} ,$ where $%
\{|k\rangle \}$ ($0\leq k<N$) are the usual computational basis states of
the quantum system.

As shown in the section 5 below, both the Hilbert space of a quantum system
and the math Hilbert space of an unstructured search problem are treated in
different manners by a conventional quantum search algorithm [3, 13] and the 
$HSSS$ quantum search process, respectively. In a conventional quantum
search algorithm an $n-$qubit quantum system is usually employed to solve an
unstructured search problem with $2^{n}-$dimensional search space [3] and
its Hilbert space $\left\{ |0\rangle ,|1\rangle ,...,|N-1\rangle \right\} $
with $N=2^{n}$ just acts as the unstructured search space $\left\{
|0),|1),...,|N-1)\right\} $ of the search problem \footnote{%
Obviously in addition to this pivotal $n-$qubit quantum system and the
unstructured search space there could be other auxiliary quantum systems and
mathematically operational spaces of function in the quantum search
algorithm. They are mainly used to help realize the quantum search algorithm.%
}. Therefore, both the Hilbert space of the quantum system and the math
Hilbert space of the search problem are the same one and coincident
completely with one another. They are treated together as one space [3, 13]
and any one of the Hilbert space and the math Hilbert space is no longer
considered separately in the algorithm. Therefore, one may say that a
conventional quantum search algorithm works in the Hilbert space of a
quantum system only. However, in the $HSSS$ quantum search process both the
Hilbert spaces have to be treated separately [1, 6, 5], although they are
the same one. This means that the $HSSS$ quantum search process works in two
different Hilbert spaces, one being of the quantum system and another of the
search problem.

The interaction between the two Hilbert spaces may be best understood
through the unitary oracle selective diagonal operator which is the basic
building block of the $HSSS$ quantum search process. Below this research
topic is described simply. A more detailed description for the research
topic should refer Refs. [5, 6, 1, 2] and the section 5 below. The unitary
oracle selective diagonal (or phase-shift) operator $C_{S}\left( \theta
\right) $ may be written as [5, 2]%
\begin{equation}
C_{S}\left( \theta \right) =\exp \left( -i\theta D_{S}\right) .  \tag{2.1}
\end{equation}%
Here $D_{S}$ is called the oracle (or black-box) selective diagonal
operator. It is defined by (2.3) below or equivalently by $D_{S}=|S\rangle
\langle S|$ with the usual computational basis state $|S\rangle $ given by
(2.4) below. It is required here that the state $|S\rangle $ take any
possible (or candidate) solution state of the unstructured search space of
the search\ problem. The reason why the state $|S\rangle $ represents any
candidate solution state is given in the section 5. Sometimes $C_{S}\left(
\theta \right) $ is called the unitary oracle selective phase-shift operator
[5a]. The unitary oracle selective diagonal operator $C_{S}\left( \theta
\right) $ in the $HSSS$ quantum search process is functionally equivalent to
the usual oracle operation $U_{o}\left( \theta \right) $ (i.e., the basic
building block) in a conventional quantum search algorithm. The latter is
defined by [3, 13]%
\begin{equation}
U_{o}\left( \theta \right) :|x\rangle \rightarrow \left\{ 
\begin{array}{ccc}
\exp \left( -i\theta \right) |x\rangle & if & f\left( x\right) =1 \\ 
|x\rangle & if & f\left( x\right) =0%
\end{array}%
\right.  \tag{2.2}
\end{equation}%
where the function $f\left( x\right) $ is defined by $f\left( x\right) =1$
if $x=x_{0}$ and $f\left( x\right) =0$ if $x\neq x_{0}$ and $|x_{0}\rangle $
denotes the unique solution state of the search problem (See also Footnote 3
in this paper). Therefore, here $C_{S}\left( \theta \right) $ and $%
U_{o}\left( \theta \right) $ also may be called the black-box functional
operations of the $HSSS$ quantum search process and a conventional quantum
search algorithm, respectively. Note that the original oracle operation $%
U_{o}\left( \pi \right) $ [3] is basic,\ (two $U_{o}\left( \pi \right) $
plus one auxiliary qubit are sufficient to generate any $U_{o}\left( \theta
\right) $), and the mathematical-logical meaning of $C_{S}\left( \theta
\right) $ is determined completely by the Boolean functional operation (or
equivalently by the mathematical-logical principle of the search problem),
as shown in the section 5. The mathematical-logical meaning for $C_{S}\left(
\theta \right) $ and $U_{o}\left( \theta \right) $ then has nothing to do
with the pure parameter $\theta $.

As shown in the section 5, both $C_{S}\left( \theta \right) $ and $%
U_{o}\left( \theta \right) $ can represent faithfully the Boolean
function-operational sequence $BFSEQ$ in $(5.3)$ or $(5.4)$, indicating that
they obey the same mathematical-logical principle of the search problem.
Therefore, each one of $C_{S}\left( \theta \right) $ and $U_{o}\left( \theta
\right) $ can be used as the basic building block or the black-box
functional operation to construct a black-box quantum search algorithm to
solve the search problem. However, in quantum computation the unitary oracle
selective diagonal operator $C_{S}\left( \theta \right) $ is essentially
different from the usual oracle operation $U_{o}\left( \theta \right) .$ It
is well known that the usual oracle operation $U_{o}\left( \theta \right) $
[3, 13] is a quantum parallel operation that works in the Hilbert space of a
quantum system and it is irreducible in the mathematical-logical meaning of
unstructured search. In contrast, as shown in the section 5, the unitary
oracle selective diagonal operator $C_{S}\left( \theta \right) $ works not
only in the Hilbert space of a quantum system but also in the math Hilbert
space of the search problem. It is a single black-box functional operation
in the Hilbert space of the quantum system and does not have the
mathematical-logical meaning of quantum parallel operation, and it may be
considered as a mathematical-parallel black-box functional operation in the
math Hilbert space. Moreover, it is allowed to be reducible in the
mathematical-logical meaning of unstructured search. Note that a
mathematical-parallel functional operation is essentially different from a
quantum-parallel functional operation (See the section 5). These substantial
mathematical-logical differences between $C_{S}\left( \theta \right) $ and $%
U_{o}\left( \theta \right) $ result in that the $HSSS$ quantum search
process is essentially different from a conventional quantum search
algorithm, although they solve the same unstructured search problem.

According to the quantum-computing speedup theory there are $N$ candidate
solution states if the unstructured search space of the search problem is $%
N- $dimensional. Here one may employ the $N$ usual computational basis
vectors $\{|S)\}$ ($0\leq S<N)$ of the unstructured search space $\left\{
|0),|1),...,|N-1)\right\} $ to represent these $N$ candidate solution
states, respectively. One also may use the $N$ usual computational basis
states $\{|S\rangle \}$ ($0\leq S<N)$ of the Hilbert space $\left\{
|0\rangle ,|1\rangle ,...,|N-1\rangle \right\} $ of a quantum system to
represent these $N$ candidate solution states, respectively. In either case
all these $N$ candidate solution states are clearly orthogonal to each
other. The theory demands to set up suitably the interaction between the
math Hilbert space (i.e., the unstructured search space) and the Hilbert
space of the quantum system used to solve the search problem. It will be
shown below that the interaction may be set up through the oracle selective
diagonal operator $D_{S}$ or the unitary oracle selective diagonal operator $%
C_{S}\left( \theta \right) .$ First consider the conventional case that the
unstructured search space is $2^{n}-$dimensional, i.e., $N=2^{n}.$ Then in
usual quantum computation the search problem usually may be solved in an $n-$%
qubit quantum system [3, 13]. On the one hand, the $n-$qubit quantum system
provides the fundamental quantum-computing resource. For convenience, here
it is set to an $n-$qubit spin (or pseudospin) system, so that the oracle
selective diagonal operator $D_{S}$ may be characterized by the $z-$%
component spin$-1/2$ (or pseudospin$-1/2$) operators $\{I_{1z},$ $I_{2z},$ $%
...,$ $I_{nz}\}$ of the $n-$qubit spin system. On the other hand, every one
of these $N$ candidate solution states $\{|S)\}$ ($0\leq S<N)$ can be
represented uniquely by a unit-number vector (or the double-value logical
number vector) $\left\{ a_{1}^{s},\text{ }a_{2}^{s},...,\text{ }%
a_{n}^{s}\right\} $ with the unit number (or the double-value logical
number) $a_{k}^{s}=\pm 1$ ($1\leq k\leq n$). Note that these $N$ candidate
solution states, which are the $N$ usual computational basis vectors, form
an orthonormal basis set of the unstructured search space. Therefore, the
unit-number vector $\left\{ a_{1}^{s},\text{ }a_{2}^{s},...,\text{ }%
a_{n}^{s}\right\} $ can represent faithfully the unstructured search space $%
\left\{ |0),|1),...,|N-1)\right\} $. Now in the unit-number representation
it can prove that the oracle selective diagonal operator $D_{S},$ which also
is given by $D_{S}=|S\rangle \langle S|,$ may be expressed as [5, 2] 
\begin{equation}
D_{S}=(\frac{1}{2}E_{1}+a_{1}^{s}I_{1z})\tbigotimes (\frac{1}{2}%
E_{2}+a_{2}^{s}I_{2z})\tbigotimes ......\tbigotimes (\frac{1}{2}%
E_{n}+a_{n}^{s}I_{nz}).  \tag{2.3}
\end{equation}%
Here the unit-number vector $\left\{ a_{1}^{s},\text{ }a_{2}^{s},...,\text{ }%
a_{n}^{s}\right\} $ also may be used to stand for any candidate solution
state $|S\rangle =|s_{1}\rangle |s_{2}\rangle ...|s_{n}\rangle $ in the
Hilbert space of the $n-$qubit spin system, and the latter may be expressed
as%
\begin{equation}
|S\rangle =(\frac{1}{2}|T_{1}\rangle +a_{1}^{s}|S_{1}\rangle )\tbigotimes (%
\frac{1}{2}|T_{2}\rangle +a_{2}^{s}|S_{2}\rangle )\tbigotimes
......\tbigotimes (\frac{1}{2}|T_{n}\rangle +a_{n}^{s}|S_{n}\rangle ) 
\tag{2.4}
\end{equation}%
where $|T_{k}\rangle =|0_{k}\rangle +|1_{k}\rangle $ and $|S_{k}\rangle =%
\frac{1}{2}(|0_{k}\rangle -|1_{k}\rangle ),$ and $|s_{k}\rangle
=|0_{k}\rangle \ $or $|1_{k}\rangle $ is the $k-$th component state of the
candidate solution state; and the operator $E_{k}$ and $I_{kz}$ are the unit
operator and the $z-$component spin operator of the $k-$th spin$-1/2$ of the 
$n-$qubit spin system, respectively. The eigenvalue equation for the spin
operator $I_{kz}$ \footnote{%
The spin operator $I_{\mu }$ of a single spin$-1/2$ is related to the Pauli
operator $\sigma _{\mu }$ by $I_{\mu }=\frac{1}{2}\sigma _{\mu }$ for $\mu
=x,y,z$} is $I_{kz}|0_{k}\rangle =\frac{1}{2}|0_{k}\rangle $ and $%
I_{kz}|1_{k}\rangle =-\frac{1}{2}|1_{k}\rangle .$ It is easy to find that
the number (or index) $S$ that corresponds to the candidate solution state $%
|S\rangle $ is given by 
\begin{equation}
S=\frac{1}{2}(1-a_{1}^{s})\times 2^{n-1}+\frac{1}{2}(1-a_{2}^{s})\times
2^{n-2}+...+\frac{1}{2}(1-a_{n}^{s})\times 2^{0}.  \tag{2.5}
\end{equation}%
These formulae (2.1), (2.3), (2.4), and (2.5) set up the one-to-one
corresponding relations between the unitary oracle selective diagonal
operator $C_{S}\left( \theta \right) ,$ the oracle selective diagonal
operator $D_{S},$ the candidate solution state $|S\rangle ,$ and the (index)
number $S$. Here the unit-number vector $\left\{ a_{1}^{s},\text{ }%
a_{2}^{s},...,\text{ }a_{n}^{s}\right\} $ plays the central role in setting
up these corresponding relations. Similarly, one also may set up these
one-to-one corresponding relations in the math Hilbert space and here one
may consider that the `$n-$qubit spin system' associated with the math
Hilbert space is completely coincident with the $n-$qubit spin system.
Therefore, the expression of (2.4) also may be used to represent the
candidate solution state $|S)$ of the math Hilbert space as long as the `$n-$%
qubit spin system' is considered as the $n-$qubit spin system.

A complete understanding for the unit number $a_{m}^{s}$ ($1\leq m\leq n$)
must be from the two aspects; One aspect is the logical number $\left(
a_{m}^{s}=\pm 1\right) $ and another the information of the component state $%
\left( |s_{m}\rangle \right) $ of the solution state. Lacking either aspect
will result in a non-complete understanding. In particular, one must not
consider that the unit number $a_{m}^{s}$ is merely a pure parameter.
Otherwise there is no mathematical-logical meaning for $C_{S}\left( \theta
\right) .$

In quantum computation this type of unitary selective diagonal operators $%
\{C_{k}\left( \theta \right) \}$ [5, 25b, 12] could have an extensive
application in future to solving many hard problems (See, for example, Refs.
[3, 13, 25, 26]). Such an application was reported in Ref. [5a] for the
first time, where the unstructured search problem is a research topic, and
the search-space dynamical reduction is realized theoretically in the
unit-number representation $\left\{ a_{1}^{s},\text{ }a_{2}^{s},...,\text{ }%
a_{n}^{s}\right\} \ $in which $C_{k}\left( \theta \right) $ ($k=S$) is the
so-called unitary oracle selective diagonal operator $C_{S}\left( \theta
\right) $ mentioned above. The mathematical-logical meaning for the unitary
oracle selective diagonal operator $C_{S}\left( \theta \right) $ in the $%
HSSS $ quantum search process is described in the section 5. The key point
in the application is that the (mathematical) number $k$, the computational
basis state $|k\rangle ,$ and the unitary selective diagonal operator $%
C_{k}\left( \theta \right) =\exp \left( -i\theta D_{k}\right) $ or the
selective diagonal operator $D_{k}=|k\rangle \langle k|$ must be one-to-one
correspondent to each other; and moreover, the basis state $|k\rangle $ and
the selective diagonal operator $D_{k}=|k\rangle \langle k|$ or $C_{k}\left(
\theta \right) $ must reflect the symmetric structure of the tensor-product
Hilbert space of the quantum system used to solve a computational problem.

The unitary oracle selective diagonal operator acts as a bridge to connect
the mathematical-logical principle obeyed by the search problem and the
quantum physical laws (i.e., the unitary quantum dynamics and the
Hilbert-space symmetric structure and property) obeyed by the quantum
system. Now the oracle selective diagonal operator of (2.3) can tell ones
that there exists the interaction between the Hilbert space of the $n-$qubit
spin system and the math Hilbert space of the search problem. Note that in
the quantum-computing speedup theory the math Hilbert space is not
coincident with the Hilbert space completely. On the one hand, as shown
above, the unit-number vector $\left\{ a_{1}^{s},\text{ }a_{2}^{s},...,\text{
}a_{n}^{s}\right\} $ in the oracle selective diagonal operator $D_{S}$ is
able to represent the unstructured search space of the search problem. On
the other hand, all these $n$ $z-$component spin operators $\left\{
I_{kz}\right\} $ in the oracle selective diagonal operator act on only the $%
n-$qubit spin system. Then by expanding (2.3) one can find that the oracle
selective diagonal operator $D_{S}$ really contains a variety of coupled
terms (e.g., $\{a_{k}^{s}I_{kz}\},$ $\{a_{k}^{s}a_{l}^{s}I_{kz}I_{lz}\},$ $%
etc.$) of both the Hilbert space of the $n-$qubit spin system and the
unstructured search space, indicating that there exists the interaction
between the two Hilbert spaces.

A slightly more complex instance than the above one may be helpful for
understanding more deeply the interaction between both the Hilbert spaces.
Suppose that an $n-$qutrit spin (or pseudospin) system that consists of $n$
spins (or pseudospins) with spin quantum number $I=1$ is used to solve the
unstructured search problem. It provides the fundamental quantum-computing
resource. A spin with $I=1$ has three basis states$\ $in its Hilbert space,
which may be set to the three eigenstates $|+1\rangle ,$ $|0\rangle ,$ and $%
|-1\rangle $ of the $z-$component spin operator $I_{z}\ $with eigenvalues $%
+1 $, $0$, $-1$, respectively, here the eigenvalue equation for the operator 
$I_{z}$ is given by $I_{z}|m\rangle =m|m\rangle \ $with $m=+1$, $0$, $-1$.
In quantum computation these three basis states $|+1\rangle ,$ $|0\rangle ,$
and $|-1\rangle $ may represent the three computational basis states $%
|0\rangle ,$ $|1\rangle ,$ and $|2\rangle $ of a single-qutrit system,
respectively. Then for the $n-$qutrit spin system it can turn out that the
oracle selective diagonal operator $D_{S}=|S\rangle \langle S|$ may be
expressed as%
\begin{equation}
D_{S}=\overset{n}{\underset{k=1}{\dbigotimes }}\left( \left( 1-\left\vert
a_{k}^{s}\right\vert \right) E_{k}+\frac{1}{2}a_{k}^{s}I_{kz}+\left( -1+%
\frac{3}{2}\left\vert a_{k}^{s}\right\vert \right) I_{kz}^{2}\right) 
\tag{2.6}
\end{equation}%
where the triple-value logical number $a_{k}^{s}=+1,$ $0,$ $-1;$ and the
unit operator $E_{k}$ and the spin operator $I_{kz}$ are the diagonal
operators of the $k-$th spin$-1$ (or pseudospin$-1$), i.e., $%
E_{k}=Diag(1,1,1),$ $I_{kz}=Diag(1,0,-1),$ and $I_{kz}^{2}=Diag(1,0,1).$
Correspondingly any candidate solution state $|S\rangle $ (or $|S)$) of the
search problem is given by%
\begin{equation}
|S\rangle =\overset{n}{\underset{k=1}{\dbigotimes }}\left( \left(
1-\left\vert a_{k}^{s}\right\vert \right) |T_{+1}^{k}\rangle +\frac{1}{2}%
a_{k}^{s}|T_{0}^{k}\rangle +\left( -1+\frac{3}{2}\left\vert
a_{k}^{s}\right\vert \right) |T_{-1}^{k}\rangle \right)  \tag{2.7}
\end{equation}%
where $|T_{+1}^{k}\rangle =|+1_{k}\rangle +|0_{k}\rangle +|-1_{k}\rangle ,$ $%
|T_{0}^{k}\rangle =|+1_{k}\rangle -|-1_{k}\rangle ,$ and $|T_{-1}^{k}\rangle
=|+1_{k}\rangle +|-1_{k}\rangle .$ Here one employs the triple-value logical
number vector $\left\{ a_{1}^{s},\text{ }a_{2}^{s},...,\text{ }%
a_{n}^{s}\right\} $ with the triple-value logical number $a_{k}^{s}=+1,$ $0,$
$-1$ $\left( 1\leq k\leq n\right) $ to express (or construct) the oracle
selective diagonal operator $D_{S}.$ Every candidate solution state $%
|S\rangle $ of the search problem is represented one-to-one by a
triple-value logical number vector $\left\{ a_{1}^{s},\text{ }a_{2}^{s},...,%
\text{ }a_{n}^{s}\right\} $ in an $n-$qutrit spin system. Furthermore,
readers may refer Refs. [6, 12] for a general construction of $D_{S}$ and $%
|S\rangle $ in a more complex quantum system. Now by using the triple-value
logical number $a_{k}^{s}=+1,$ $0,$ $-1$ the $k-$th component state $%
|s_{k}\rangle $ of the candidate solution state $|S\rangle =|s_{1}\rangle
|s_{2}\rangle ...|s_{n}\rangle $ may be written as $|s_{k}\rangle =\left(
1-\left\vert a_{k}^{s}\right\vert \right) |T_{+1}^{k}\rangle +\frac{1}{2}%
a_{k}^{s}|T_{0}^{k}\rangle +\left( -1+\frac{3}{2}\left\vert
a_{k}^{s}\right\vert \right) |T_{-1}^{k}\rangle =$ $|+1_{k}\rangle ,$ $%
|0_{k}\rangle ,$ and $|-1_{k}\rangle ,$ respectively. It is also easy to
find that the (index) number $S$ corresponding to the state $|S\rangle $ is
given by%
\begin{equation}
S=(1-a_{1}^{s})\times 3^{n-1}+(1-a_{2}^{s})\times
3^{n-2}+...+(1-a_{n}^{s})\times 3^{0}.  \tag{2.8}
\end{equation}%
Thus, the formulae (2.1), (2.6), (2.7), and (2.8) also set up the one-to-one
corresponding relations between $C_{S}\left( \theta \right) ,$ $D_{S},$ $%
|S\rangle ,$ and $S$ through the triple-value logical number vector $\left\{
a_{1}^{s},\text{ }a_{2}^{s},...,\text{ }a_{n}^{s}\right\} .$ Moreover, the
oracle selective diagonal operator of (2.6) and the candidate solution state
of (2.7) indeed reflect the symmetric structure of the tensor-product
Hilbert space of the $n-$qutrit spin system.

The triple-value logical number vector $\left\{ a_{1}^{s},\text{ }%
a_{2}^{s},...,\text{ }a_{n}^{s}\right\} $ still stands for any candidate
solution state $|S\rangle $ (or $|S)$) and it is able to represent the
unstructured search space $\left\{ |0),|1),...,|N-1)\right\} $ with $N=3^{n}$
of the search problem, because all these candidate solution states form an
orthonormal basis set of the unstructured search space. On the other hand,
in the oracle selective diagonal operator of (2.6) these $z-$component spin
operators $\left\{ I_{kz}\right\} $ and $\left\{ I_{kz}^{2}\right\} $ act on
only the $n-$qutrit spin system. By expanding (2.6) one can find that the
oracle selective diagonal operator consists of a variety of interaction
terms of both the Hilbert space of the $n-$qutrit spin system and the math
Hilbert space of the search problem. Thus, there exists the interaction
between the two Hilbert spaces.

As can be seen in (2.3) and (2.6), the oracle selective diagonal operator of
(2.6) associated with the $n-$qutrit spin system is quite different from
that one of (2.3) with the $n-$qubit spin system. The interacting terms
between the two Hilbert spaces in the oracle selective diagonal operator
with the $n-$qutrit spin system are much more complex than those in the
oracle selective diagonal operator with the $n-$qubit spin system. Both the $%
n-$qubit and $n-$qutrit spin systems have the tensor-product Hilbert spaces
with different symmetric structures, respectively. As shown in (2.3) and
(2.6), the tensor-product symmetric structure appearing in one oracle
selective diagonal operator is determined completely by the Hilbert space of
the quantum system associated with the oracle selective diagonal operator.
These indicate that \textit{the symmetric structure of tensor-product
Hilbert space of a quantum system shapes the interaction between the Hilbert
space of the quantum system and the math Hilbert space of the unstructured
search problem to be solved}.

Below an $n-$qubit spin (or pseudospin) system will be studied mainly, while
an $n-$qutrit spin (or pseudospin) system will not be further discussed
unless stated otherwise.

The search-space dynamical reduction was first constructed in Ref. [5a] in
the unit-number representation $\left\{ a_{1}^{s},\text{ }a_{2}^{s},...,%
\text{ }a_{n}^{s}\right\} $ in an $n-$qubit spin quantum system. The
interaction between the Hilbert space of the quantum system and the math
Hilbert space of the search problem is necessary to realize the search-space
dynamical reduction. As shown above, it is shaped by the symmetric structure
of the tensor-product Hilbert space of the quantum system. This is the
reason why the fundamental quantum-computing resource is necessary to
realize a search-space dynamical reduction. An efficient search-space
dynamical reduction is already realized in an $n-$qubit spin (or pseudospin)
system theoretically [2]. Here the unit-number representation $\left\{
a_{1}^{s},\text{ }a_{2}^{s},...,\text{ }a_{n}^{s}\right\} $ of the $n-$qubit
spin system is fundamentally important to realize the efficient search-space
dynamical reduction. It is used to express the oracle selective diagonal
operator $D_{S}$ and the unitary oracle selective diagonal operator $%
C_{S}\left( \theta \right) $ which are given by (2.3) and (2.1),
respectively. Then by manipulating unitarily the oracle selective diagonal
operator of (2.3) or the unitary oracle selective diagonal operator of
(2.1)\ one is able to realize an efficient search-space dynamical reduction.
Such a unitary manipulation is generally complex [2, 5a]. It consists of not
only the unitary operators (or transformations) but also their inverses.
Here the efficient search-space dynamical reduction is no longer described
in detail.

An efficient search-space dynamical reduction is the first step to realizing
an efficient unstructured quantum search process, i.e., the $HSSS$ quantum
search process. It eliminates the unstructured search space. Consequently it
eliminates the classical effect that is hidden behind the unstructured
search space. Therefore, it is the necessary condition for the $HSSS$
quantum search process as a whole to achieve a super-square or even an
exponential quantum-searching speedup. However, an efficient search-space
reduction alone is not sufficient to solve efficiently an unstructured
search problem. It can not yet form completely an efficient quantum search
process. Instead, its ultimate purpose is to generate the so-called
information-carrying unitary operators which further act as the basic
building blocks of the quantum-state-difference amplification (QUANSDAM)
process at the second step of the $HSSS$ quantum search process.

A search-space dynamical reduction is performed in the frame of unitary
quantum dynamics. But there may be two equivalent but different pictures [1,
2, 5] to describe it: the unitary state-transformation description and the
unitary quantum dynamical description. From the picture of the unitary state
transformation an efficient search-space dynamical reduction [2] realizes
that the information of one component state (or a few) of the solution state
of the search problem is efficiently transferred from the unstructured
search space to a polynomially-small subspace (or subset), when the
exponentially-large unstructured search space is eliminated. Equivalently,
from the picture of the unitary quantum dynamics, this efficient
search-space dynamical reduction generates an $\mathit{i}$nformation-$%
\mathit{c}$arrying ($IC$) unitary operator from a unitary sequence
consisting of the unitary oracle selective diagonal operators and the purely
quantum-mechanical ($QM$)\ unitary operators. This generated $IC$ unitary
operator carries the information of one component state (or a few) of the
solution state. It works in the two polynomially-small subspaces which
belong to the math Hilbert space of the search problem and the Hilbert space
of the quantum system, respectively. The generated $IC$ unitary operator
also may be further used to prepare a desired $IC$ unitary operator.

The simplest $IC$ unitary operator which may be efficiently prepared by the
efficient search-space dynamical reduction [2] may be briefly written as%
\begin{equation}
U_{\lambda }^{ic}(\theta _{m},a_{m}^{s})=\exp (-ia_{m}^{s}\theta
_{m}I_{m\lambda }),\text{ }1\leq m\leq n,\text{ }\lambda =x,y,z  \tag{2.9}
\end{equation}%
where the rotating angle $\theta _{m}$ is proportional to $1/2^{n}$ and
hence it is exponentially small. This $IC$ unitary operator is called the
basic $IC$ unitary operator. It is built out of the unitary oracle selective
diagonal operators of (2.1) and the $QM$ unitary operators. It was first
proposed in Ref. [5a]. It carries only the information ($a_{m}^{s}$) of the
component state $|s_{m}\rangle $ of the solution state. The operator $%
I_{m\lambda }$ in (2.9) may be a spin operator of the single $m-$th spin$%
-1/2 $ of the $n-$qubit spin system. Then in this case the $IC$ unitary
operator was prepared in Ref. [5], but the preparation could not be
efficient. Below consider only the case that the spin operator $I_{m\lambda
} $ of the $m-$th spin$-1/2$ stands for the pseudospin$-1/2$ operator of the 
$n-$qubit spin system, which may be written as $|0_{1}\rangle \langle
0_{1}|\tbigotimes ...\tbigotimes |0_{m-1}\rangle \langle 0_{m-1}|\tbigotimes
I_{m\lambda }\tbigotimes |0_{m+1}\rangle \langle 0_{m+1}|$ $\tbigotimes
...\tbigotimes |0_{n}\rangle \langle 0_{n}|.$ Then in this case the $IC$
unitary operator may be efficiently prepared in Ref. [2]. It works in a
two-dimensional subspace ($a_{m}^{s}=\pm 1$) of the math Hilbert space of
the search problem and also in a two-dimensional subspace $\{|0_{1}\rangle
...|0_{m-1}\rangle |s_{m}\rangle |0_{m+1}\rangle ...|0_{n}\rangle \}$ ($%
s_{m}=0,$ $1$) of the Hilbert space of the $n-$qubit spin system. Note that
when the $IC$ unitary operator is used, the $n-$qubit spin system should be
set to the initial state $|0\rangle $ in advance except the $m-$th spin$-1/2$%
.

As shown in (2.9), the $IC$ unitary operator depends upon the $m-$th spin
operator $I_{m\lambda }$ $\left( \lambda =x,y,z\right) .$ It can affect not
only the $m-$th spin$-1/2$ but also any coupled spin system which consists
of the $m-$th spin$-1/2$ and other spins that do not belong to the $n-$qubit
spin system. Moreover, by a suitable unitary transformation the $m-$th spin
operator $I_{m\lambda }$ may be changed to another spin operator $%
I_{m^{\prime }\lambda }$ that does not belong to the $n-$qubit spin system.
Then the generated unitary operator $U_{\lambda }^{ic}(\theta _{m^{\prime
}},a_{m}^{s})=\exp (-ia_{m}^{s}\theta _{m^{\prime }}I_{m^{\prime }\lambda })$
($m^{\prime }\notin \lbrack 1,n]$) also may act as an $IC$ unitary operator.
It may be used in a spin system that contains the $m^{\prime }-$th spin$-1/2$
but is different from the $n-$qubit spin system, here the $n-$qubit spin
system should be simply set to the initial state $|0\rangle $ in advance
when the $IC$ unitary operator is used.

The subspace of the math Hilbert space of the search problem for the $IC$
unitary operator $U_{\lambda }^{ic}(\theta _{m},a_{m}^{s})$ is
two-dimensional after the search-space dynamical reduction, while the
original unstructured search space of the search problem is exponentially
large ($2^{n}-$dimensional) before the search-space dynamical reduction.
This fact that the $IC$ unitary operator $U_{\lambda }^{ic}(\theta
_{m},a_{m}^{s})$ does not work in the whole unstructured search space but in
its two-dimensional subspace shows clearly that the search-space dynamical
reduction indeed eliminates the exponentially large unstructured search
space. Now the unstructured property of the search space does not have an
effect on the behavior of the $IC$ unitary operator $U_{\lambda
}^{ic}(\theta _{m},a_{m}^{s})$. In contrast, it has an essential effect on
the behavior of the oracle operation of a conventional quantum search
algorithm. This difference also shows that the $HSSS$ quantum search process
is essentially different from a conventional quantum search algorithm.

In the $HSSS$ quantum search process the basic $IC$ unitary operator $%
U_{\lambda }^{ic}(\theta _{m},$ $a_{m}^{s})$ may act as the basic building
block of a UNIDYSLOCK process or a QUANSDAM process. Here an exponential
QUANSDAM process is just the second step of the $HSSS$ quantum search
process. Its purpose is to extract the information $\left( a_{m}^{s}\right) $
of the component state $|s_{m}\rangle $ of the solution state in an
efficient way. A UNIDYSLOCK (or QUANSDAM) process may be constructed in a
quantum system such as a single-atom system. Its own $IC$ unitary operator
may be different from system to system. These different $IC$ unitary
operators could be prepared by starting from the basic $IC$ unitary operator
of (2.9).

In next sections one will see that the basic $IC$ unitary operator is the
necessary component for a QUANSDAM\ (or UNIDYSLOCK) process. Without it, in
a QUANSDAM process the quantum-state difference between a pair of
non-orthogonal quantum states is not able to build up and the $QSD$
amplification ability can not be promoted greatly by the $QM$ unitary
dynamical processes. One may say intuitively that the basic $IC$ unitary
operator provides the seed to grow the quantum-state difference in a
QUANSDAM process.

In the quantum-computing speedup theory the principle is general that an
essential quantum-computing speedup is achieved by the interaction between
the quantum physical laws (i.e., the unitary quantum dynamics and the
Hilbert-space symmetric structure and property) and the mathematical-logical
principle that a computational problem obeys. As described above, an
efficient search-space dynamical reduction is realized by the interaction
between the Hilbert space of a quantum system and the math Hilbert space of
an unstructured search problem in the frame of unitary quantum dynamics,
while the symmetric structure of the Hilbert space of the quantum system
shapes the interaction. Therefore, the principle indeed governs the
efficient search-space dynamical reduction. It is not only applied to the
first step of the $HSSS$ quantum search process, i.e., an efficient
search-space dynamical reduction, but also it is available over the whole $%
HSSS$ quantum search process. There is no exception for a UNIDYSLOCK process
and a QUANSDAM process.\newline
\newline
{\Large 3. The principle of a UNIDYSLOCK process and a QUANSDAM process}

Generally a unitary dynamical state-locking process (or briefly a UNIDYSLOCK
process) is defined as such a unitary process that transforms simultaneously
two or more orthogonal quantum states to their corresponding non-orthogonal
quantum states whose differences may be arbitrarily small [1, 7, 6]. At
first the author thought that a UNIDYSLOCK process could be realized by a
purely quantum-mechanical unitary dynamical process [7]. Unfortunately that
is not correct! After an intense research into it the author realized
eventually that a UNIDYSLOCK process does not make sense if one interprets
it from the point of view of the purely quantum-mechanical unitary dynamics.
Because a UNIDYSLOCK process is not only a quantum-mechanical unitary
dynamical process but also it obeys the mathematical-logical principle of
the search problem [5, 6, 1, 2], completely and exactly understanding a
UNIDYSLOCK process must consider both the aspect of the unitary quantum
dynamics and the aspect of the mathematical-logical principle and especially
the interaction between the quantum physical laws (i.e., the unitary quantum
dynamics and the Hilbert-space symmetric structure and property) and the
mathematical-logical principle. Below a detailed investigation is carried
out on the principle and mechanism of how a UNIDYSLOCK process and its
inverse process, i.e., a QUANSDAM process work in a quantum system on the
basis of the quantum-computing speedup theory [1].

Before a UNIDYSLOCK process is discussed in detail, one needs to set up a
quantitative measure for the quantum-state difference between a pair of
non-orthogonal quantum states, because the quantum-state-difference ($QSD$)
varying rate is the characteristic physical quantity of a UNIDYSLOCK process
and a QUANSDAM process. What one discusses in the paper is a pure quantum
system. In quantum mechanics for a pure quantum system the most naive method
to measure quantitatively the quantum-state difference between a pair of
quantum states is perhaps to use the scalar (or inner) product of the pair
of quantum states. For intuition in the paper the scalar product of a pair
of wave functions (i.e., quantum states) also is called the overlapping
integral of the pair of wave functions. Suppose that $\varphi _{1}(x)$ and $%
\varphi _{2}(x)$ are a pair of normalized wave functions in one-dimensional
coordinate space $(-\infty ,+\infty )$. Then in quantum mechanics the scalar
product of the pair of wave functions may be defined by [14, 17]%
\begin{equation}
\rho _{12}=\langle \varphi _{1}|\varphi _{2}\rangle =\int_{-\infty }^{\infty
}\varphi _{1}(x)^{\ast }\varphi _{2}(x)dx  \tag{3.1}
\end{equation}%
The absolute scalar product $\left\vert \rho _{12}\right\vert $ satisfies $%
0\leq \left\vert \rho _{12}\right\vert \leq 1.$ Its physical meaning is
explained as follows. If $\left\vert \rho _{12}\right\vert =1,$ then both
the wave functions $\varphi _{1}(x)$ and $\varphi _{2}(x)$ are the same one
up to a phase factor. If $\left\vert \rho _{12}\right\vert =0,$ then both
the wave functions are orthogonal to one another. If $0<\left\vert \rho
_{12}\right\vert <1,$ both the wave functions are neither equal to nor
orthogonal to one another and instead they are non-orthogonal to one
another. Thus, the scalar product (or the overlapping integral) $\rho _{12}$
may be used to measure quantitatively the quantum-state difference between a
pair of wave functions. As a typical example, the wave functions $\varphi
_{1}(x)$ and $\varphi _{2}(x)$ are taken as the two standard Gaussian
wave-packet states [15], respectively, each one of which has the
characteristic parameters $\{x_{l},$ $p_{l},$ $(\Delta x_{l})^{2},$ $T_{l}\}$
for $l=1$, $2$. Then it can turn out that the absolute overlapping integral $%
\left\vert \rho _{12}\right\vert $ may be written as%
\begin{equation*}
\left\vert \rho _{12}\right\vert =\left( \frac{4\Delta _{1}^{2}\Delta
_{2}^{2}}{\left( \Delta _{1}^{2}+\Delta _{2}^{2}\right) ^{2}+\beta _{12}^{2}}%
\right) ^{1/4}\times \exp \left( \frac{D^{\prime }}{\left( \Delta
_{1}^{2}+\Delta _{2}^{2}\right) ^{2}+\beta _{12}^{2}}\right)
\end{equation*}%
with%
\begin{equation*}
D^{\prime }=-\frac{1}{\hslash ^{2}}p_{12}^{2}\Delta _{1}^{2}\Delta
_{2}^{2}\left( \Delta _{1}^{2}+\Delta _{2}^{2}\right) -\frac{1}{4}\Delta
_{1}^{2}\left( x_{12}-\frac{2}{\hslash }p_{12}\beta _{2}\right) ^{2}-\frac{1%
}{4}\Delta _{2}^{2}\left( x_{12}-\frac{2}{\hslash }p_{12}\beta _{1}\right)
^{2}\newline
\end{equation*}%
where $\Delta _{l}^{2}=\left( \Delta x_{l}\right) ^{2},$ $\beta _{l}=\left( 
\frac{\hslash T_{l}}{2m}\right) ;$ $p_{12}=p_{1}-p_{2},$ $%
x_{12}=x_{1}-x_{2}, $ and $\beta _{12}=\beta _{1}-\beta _{2}.$

Below it first proves a particularly important result that a purely
quantum-mechanical unitary dynamical process cannot change two or more
orthogonal quantum states to their corresponding non-orthogonal states at
the same time. According to quantum mechanics [14, 17] it can turn out that
a pair of quantum states keep their scalar product unchanged under action of
an arbitrary unitary transformation. Suppose that a pair of initial quantum
states $|\Psi _{k}\rangle $ and $|\Psi _{l}\rangle $ which may be arbitrary
are acted on by an arbitrary unitary transformation at the same time:%
\begin{equation}
U:\left\{ 
\begin{array}{c}
|\Psi _{k}\rangle \rightarrow |\varphi _{k}\rangle \\ 
|\Psi _{l}\rangle \rightarrow |\varphi _{l}\rangle%
\end{array}%
\right.  \tag{3.2}
\end{equation}%
Then the scalar product of the final two states $|\varphi _{k}\rangle $ and $%
|\varphi _{l}\rangle $ satisfies the relation: 
\begin{equation}
\langle \varphi _{k}|\varphi _{l}\rangle =(\langle \Psi _{k}|U^{+})(U|\Psi
_{l}\rangle )=\langle \Psi _{k}|\Psi _{l}\rangle .  \tag{3.3}
\end{equation}%
This shows that the scalar product of the final two states is equal to that
one of the initial two states. Therefore, an arbitrary unitary
transformation cannot change the scalar product of a pair of quantum states.
This is a universal conclusion in quantum mechanics [14, 17]. It directly
leads to that a purely quantum-mechanical unitary dynamical process can not
transform simultaneously two or more orthogonal states to their
corresponding non-orthogonal states.

It is well-known in quantum mechanics that a quantum system obeys the
unitary quantum dynamics. (Throughout the paper quantum measurement is
merely used as a tool to obtain the final computational result at the end of
a quantum computational process. This is required by the quantum-computing
speedup theory [1]. Therefore, here quantum measurement is not considered.)
Then, as deduced from (3.2) and (3.3), in any unitary time evolution process
of a quantum system the quantum-state difference between a pair of quantum
states of the quantum system is never changed. This means that the
quantum-state difference is never changed during the unitary time evolution
process, although any one of the pair of quantum states may be a
superposition state, a quantum entanglement state, or a coherent state and
so on. This shows clearly that \textit{any quantum-state effects in a
quantum system can not cause varying of quantum-state difference during a
unitary quantum dynamical process, here the quantum-state effects may
include the quantum-state superposition, coherence interference,
entanglement and nonlocal effect (i.e., spooky action\ at a distance),
correlation and so on. }This is the first important property of the
quantum-state-difference varying.

There is no doubt that in quantum mechanics a UNIDYSLOCK process obeys the
unitary quantum dynamics. Then according to (3.2) and (3.3) it can not
change the scalar product of a pair of quantum states and hence it is not
able to transform simultaneously two or more orthogonal states to their
corresponding non-orthogonal states. However, according to its definition a
UNIDYSLOCK process is such a unitary quantum dynamical process that can
change two or more orthogonal states to their corresponding non-orthogonal
states at the same time. This means that a UNIDYSLOCK process does not make
sense in pure quantum mechanics.

As a quantum computational process (or sub-process) a UNIDYSLOCK process
obeys not only the unitary quantum dynamics but also the
mathematical-logical principle of the search problem [6, 7, 1]. It is
substantially different from a conventional quantum-mechanical unitary
dynamical process in that it must obey the mathematical-logical principle of
the search problem. Thus, both the aspects of the unitary quantum dynamics
and the mathematical-logical principle as well as the interaction between
the quantum physical laws (i.e., the unitary quantum dynamics and the
Hilbert-space symmetric structure and property) and the mathematical-logical
principle are key to understanding exactly and completely how a UNIDYSLOCK
process works in a quantum system. It must be from the point of view of the
mathematical-logical principle to comprehend that a UNIDYSLOCK process is
able to change the scalar product of a pair of quantum states. It must be
understood that the interaction can cause the quantum-state-difference
varying and may influence greatly the $QSD$ varying rate of a UNIDYSLOCK (or
QUANSDAM)\ process.

In general, a UNIDYSLOCK \ (or QUANSDAM) \ process \ consists of the
information-carrying unitary operators (i.e., the $IC$ unitary operators)
and the purely quantum-mechanical unitary operators (i.e., the $QM$ unitary
operators). Its core is the $IC$ unitary operators. Its characteristic
property that a UNIDYSLOCK process can change the scalar product of a pair
of quantum states (i.e., the quantum-state difference between the two
states) is uniquely original from the $IC$ unitary operators. As shown in
(3.2) and (3.3) above, this property has nothing to do with any $QM$ unitary
operator. A UNIDYSLOCK process could be best understood by performing its
inverse process, i.e., a QUANSDAM process in a quantum system. For
simplicity, here consider that the QUANSDAM process consists of only one
basic $IC$ unitary operator of (2.9) and it is applied to a single spin$-1/2$
(or pseudospin$-1/2$) system. The spin system has only two orthonormal basis
states $|0\rangle $ and $|1\rangle .$ Suppose that at the initial time the
spin system is in the state $|0\rangle $. Then it is acted on by the basic $%
IC$ unitary operator of (2.9). One therefore obtains the simple QUANSDAM
process in the system:%
\begin{equation}
U_{x}^{ic}(\theta _{m},a_{m}^{s})|0\rangle =\cos \frac{1}{2}\theta
_{m}|0\rangle -ia_{m}^{s}\sin \frac{1}{2}\theta _{m}|1\rangle .  \tag{3.4}
\end{equation}%
It seems that from the point of view of pure quantum mechanics there is not
a big difference between the QUANSDAM process and a usual unitary state
transformation with the (time or phase) parameter $\left( a_{m}^{s}\theta
_{m}\right) $. This is a trivial understanding for a QUANSDAM process. It is
too trivial to capture the essential aspect of a QUANSDAM process. However,
this simple QUANSDAM process may be comprehended more exactly and more
deeply from the point of view of the mathematical-logical principle. As
pointed out in the preceding section, the basic $IC$ unitary operator $%
U_{x}^{ic}(\theta _{m},a_{m}^{s})$ works in the two-dimensional subspace ($%
a_{m}^{s}=\pm 1$) of the math Hilbert space of the search problem. Then at
the end of the QUANSDAM process there must be two possible states to be
generated in the spin system, although only one of the two states really
appears in the system. These two possible states may be written as%
\begin{equation}
U_{x}^{ic}(\theta _{m},a_{m}^{s})|0\rangle =\left\{ 
\begin{array}{c}
|\psi _{+1}\rangle ,\text{ }if\text{ }a_{m}^{s}=+1 \\ 
|\psi _{-1}\rangle ,\text{ }if\text{ }a_{m}^{s}=-1%
\end{array}%
\right.  \tag{3.5}
\end{equation}%
where $|\psi _{\pm 1}\rangle =\cos \frac{1}{2}\theta _{m}|0\rangle \mp i\sin 
\frac{1}{2}\theta _{m}|1\rangle .$ It may be thought that both the final
states $|\psi _{+1}\rangle $ and $|\psi _{-1}\rangle $ are generated from
the initial two states which are the same and equal to the state $|0\rangle $
by the QUANSDAM process, respectively. Now one may examine the quantum-state
difference between the final two states $|\psi _{+1}\rangle $ and $|\psi
_{-1}\rangle $. It turns out that the scalar product of the final two states
is given by%
\begin{equation}
\langle \psi _{+1}|\psi _{-1}\rangle =\cos \theta _{m}.  \tag{3.6}
\end{equation}%
When $\theta _{m}=0,$ both the states $|\psi _{+1}\rangle $ and $|\psi
_{-1}\rangle $ are the same one and equal to the initial state $|0\rangle .$
Then the scalar product $\langle \psi _{+1}|\psi _{-1}\rangle $ is equal to
one. However, as shown in (3.6), when $\theta _{m}\neq 0,$ the scalar
product may not equal one. Particularly, when $\theta _{m}=\pi /2,$ it is
equal to zero and hence both the states $|\psi _{+1}\rangle $ and $|\psi
_{-1}\rangle $ are orthogonal to each other. These show clearly that the
QUANSDAM process is able to change the scalar product of the two states $%
|\psi _{+1}\rangle $ and $|\psi _{-1}\rangle \ $and so is the UNIDYSLOCK
process, i.e., the inverse of the QUANSDAM process. It is clear that under a
UNIDYSLOCK process the absolute scalar product $\left\vert \langle \psi
_{+1}|\psi _{-1}\rangle \right\vert $ is changed from zero to one.
Conversely, under a QUANSDAM process it is changed from one to zero. The
above investigation for the simple QUANSDAM process highlights the most
basic characteristic feature of a general QUANSDAM (or UNIDYSLOCK) process
that a QUANSDAM (or UNIDYSLOCK) process is able to change the quantum-state
difference between a pair of quantum states.

Here it needs to emphasize the unitarity of a UNIDYSLOCK process. Without
unitarity, a UNIDYSLOCK process will be trivial in quantum computation (as
explained below). It therefore rules out any irreversible physical processes
including quantum measurement and any irreversible mathematical-logical
operations as its building blocks. It is well known that two or more
orthogonal quantum states may be changed to a quantum state simultaneously
by a non-equilibrium irreversible process (i.e., a non-unitary dynamical
process). This is the only non-trivial physical process that can change the
quantum-state difference of a pair of quantum states except the UNIDYSLOCK
(or QUANSDAM) process. If now a UNIDYSLOCK process contained a
non-equilibrium irreversible sub-process, then certainly it would be able to
change many orthogonal states to their corresponding non-orthogonal states
at the same time. In this case the UNIDYSLOCK\ process is not significantly
different from a usual non-equilibrium irreversible process. Hence it
becomes trivial in quantum computation. Here the non-equilibrium
irreversible sub-process may be purely physical or it contains any
irreversible mathematical-logical operation. In either case it veils the
essential aspect of a UNIDYSLOCK\ process. It results in that one cannot
understand exactly the essential aspect of a UNIDYSLOCK process. A
non-equilibrium irreversible process could make a UNIDYSLOCK process more
quickly changing many orthogonal states to their corresponding
non-orthogonal states at the same time, but it also could make the inverse
of the UNIDYSLOCK process more hard amplifying the quantum-state difference.
Thus, it could degrade or even destroy the $QSD$ amplification ability of a
QUANSDAM process. Beside this problem the energy dissipation problem of
irreversibility (See, e.g., Ref. [20]) still needs to be considered for a
quantum computation to use a non-equilibrium irreversible process as its
building block.

A QUANSDAM (or UNIDYSLOCK) process generally contains many basic $IC$
unitary operators of (2.9) and $QM$ unitary operators. As shown in (3.2) and
(3.3), any $QM$\ unitary operator of the process can not change the scalar
product of a pair of quantum states. Then only the basic $IC$ unitary
operators of the process are able to change the scalar product of a pair of
quantum states (See, e.g., $\langle \psi _{+1}|\psi _{-1}\rangle $ in (3.6)$%
).$ That is, only the basic $IC$ unitary operators can cause the
quantum-state-difference varying. As pointed out in the preceding section,
the basic $IC$ unitary operator is built out of the unitary oracle selective
diagonal operators of (2.1) and the $QM$ unitary operators. Again here the $%
QM$ unitary operators can not change the quantum-state difference.
Therefore, the quantum-state-difference varying is uniquely original from
the unitary oracle selective diagonal operator.

As shown in the section 5, the unitary oracle selective diagonal operator
works in both the Hilbert space of the quantum system and the math Hilbert
space of the search problem. In the Hilbert space it is a single reversible
black-box functional operation, while in the math Hilbert space it may be
considered as a mathematical-parallel reversible black-box functional
operation as a whole. These reversible black-box functional operations (or
the component functional operations) in both the Hilbert spaces altogether
characterize completely the mathematical-logical principle of the
unstructured search problem, indicating that the unitary oracle selective
diagonal operator can characterize faithfully the mathematical-logical
principle of the search problem. One may say that the unitary oracle
selective diagonal operator is a reversible black-box functional operation
of the search problem. \textit{This reversible black-box functional
operation of the unstructured search problem really causes the change of the
quantum-state difference in a UNIDYSLOCK (or QUANSDAM)\ process.}\footnote{%
The reversible black-box functional operation also is called the
computational resource of a UNIDYSLOCK (or QUANSDAM) process [2].} This is
perhaps a most mysterious result for the Lecerf-Bennett reversible
computational theory [8]. This result is obtained on the basis of the
quantum-computing speedup theory [1].

As pointed out in the section 2, the basic $IC$ unitary operator of (2.9) is
generated by the search-space dynamical reduction. The latter is realized by
the interaction between the Hilbert space of the quantum system and the math
Hilbert space of the search problem in the frame of unitary quantum
dynamics. Thus, the unitary quantum dynamics, the Hilbert space of the
quantum system with its own symmetric structure, and the
mathematical-logical principle of the search problem may affect in principle
the basic $IC$ unitary operator. Consider that the basic $IC$ unitary
operator works in a two-dimensional reduction subspace of the math Hilbert
space and a two-dimensional subspace of the Hilbert space of the quantum
system and both the subspaces each have a minimum dimensional size. Then the
effect of the Hilbert-space symmetric structure could be secondary on the
basic $IC$ unitary operator. Therefore, the basic $IC$ unitary operator is
affected mainly by the unitary quantum dynamics, the Hilbert space of the
quantum system, and the mathematical-logical principle. Consider further
that the pure unitary quantum dynamics alone can not change the
quantum-state difference. Then these show that \textit{the
quantum-state-difference varying driven by the basic }$IC$\textit{\ unitary
operator is caused by the interaction between the unitary quantum dynamics
and the mathematical-logical principle and also the interaction between the
Hilbert space of the quantum system and the math Hilbert space of the search
problem.} This is the second important property of the
quantum-state-difference varying. This property also could be inferred
directly from the basic $IC$ unitary operator itself.

The property is very important that the effect of the symmetric structure of
the Hilbert space of the quantum system could be secondary on the
quantum-state-difference varying that is driven by the basic $IC$ unitary
operators of (2.9). It implies that a single quantum system such as a
single-atom system that does not have the symmetric structure of the Hilbert
space of a composite quantum system could be appropriate to realize an
exponential QUANSDAM process whose basic building block is the basic $IC$
unitary operator.

Below it is shown more clearly that the quantum-state-difference varying
driven by the basic $IC$\ unitary operator is still caused by the
interaction between the Hilbert space of the quantum system and the math
Hilbert space of the search problem, although the effect of the symmetric
structure of the Hilbert space of the quantum system could be secondary on
the quantum-state-difference varying.

As shown in (3.2) and (3.3), in quantum mechanics it is impossible that many
orthogonal states of a quantum system are simultaneously changed to their
corresponding non-orthogonal states by a unitary quantum dynamical process.
A\ UNIDYSLOCK process is a unitary quantum dynamical process. A mysterious
thing is that by the UNIDYSLOCK process many orthogonal quantum states may
be transformed simultaneously to their corresponding non-orthogonal states.
Here it should be pointed out that those orthogonal states that are
simultaneously changed to their corresponding non-orthogonal states by a
UNIDYSLOCK process are not all in the same Hilbert space of the quantum
system but in the joint Hilbert space which consists of the Hilbert space of
the quantum system and the math Hilbert space of the search problem. This
can be best illustrated with the simple QUANSDAM\ process of (3.5).

It is known from (3.5)\ that there are two possible quantum states (i.e.,
the states $|\psi _{+1}\rangle $ and $|\psi _{-1}\rangle $) at the end of
the QUANSDAM\ process of (3.4). One quantum state (either the state $|\psi
_{+1}\rangle $ or $|\psi _{-1}\rangle $) is in the two-state Hilbert space $%
\{|0\rangle ,$ $|1\rangle \}$ of the single spin$-1/2$ quantum system (i.e.,
the physical state space), while another is not in the Hilbert space $%
\{|0\rangle ,$ $|1\rangle \}$ but in the two-dimensional subspace of the
math Hilbert space of the search problem (i.e., the unphysical state space).
Both the states $|\psi _{+1}\rangle $ and $|\psi _{-1}\rangle $ can not
belong to the same Hilbert space $\{|0\rangle ,$ $|1\rangle \}$ at the same
time. If a QUANSDAM\ process consists of many steps, each step containing
one basic $IC$ unitary operator, then it is impossible that at the end of
one step the state $|\psi _{+1}\rangle $ appears in the Hilbert space $%
\{|0\rangle ,$ $|1\rangle \}$, while at the end of\ another step the state $%
|\psi _{-1}\rangle $ appears in the same subspace. The only exception is
that both the states $|\psi _{+1}\rangle $ and $|\psi _{-1}\rangle $ are the
same one. The reason for these is that the solution state of the search
problem is unique and deterministic and the information conservation law
must be obeyed (See also the next paragraph). Both the states $|\psi
_{+1}\rangle $ and $|\psi _{-1}\rangle $ should be considered as the members
of the two-dimensional subspace of the math Hilbert space, because one does
not know in advance which one of the two states does not appear in the
Hilbert space $\{|0\rangle ,$ $|1\rangle \}$ in the mathematical-logical
meaning of search. The physical behavior of the quantum state that appears
in the Hilbert space $\{|0\rangle ,$ $|1\rangle \}$ of the quantum system is
characterized completely by the quantum physical laws. This quantum state is
the joint point between both the Hilbert space of the quantum system and the
math Hilbert space of the search problem. It is really responsible for
extracting the information ($a_{m}^{s}$) of the $m-$th component state of
the solution state. On the other hand, the existence of the math Hilbert
space is necessary, so that the QUANSDAM process of (3.4) is able to change
the quantum-state difference between the pair of quantum states $|\psi
_{+1}\rangle $ and $|\psi _{-1}\rangle $. Thus, the math Hilbert space is
necessary to realize a general QUANSDAM (or UNIDYSLOCK) process.

Apparently the simple QUANSDAM process of (3.5) performs the conditional
unitary operation. That is, if $a_{m}^{s}=+1,$ it performs the unitary
operation $U_{x}^{ic}(\theta _{m},a_{m}^{s})|0\rangle =|\psi _{+1}\rangle $
and if $a_{m}^{s}=-1,$ it performs the unitary operation $U_{x}^{ic}(\theta
_{m},a_{m}^{s})|0\rangle =|\psi _{-1}\rangle .$ Actually it may be
considered that this simple QUANSDAM process performs a
mathematical-parallel black-box functional operation as a whole. Note that
the basic $IC$ unitary operator $U_{x}^{ic}(\theta _{m},a_{m}^{s})$ of
(2.9), which may be considered as a black-box functional operation, works in
both the two-dimensional subspace ($a_{m}^{s}=\pm 1$) of the math Hilbert
space and the Hilbert space of the single spin$-1/2$ quantum system. Then
this mathematical-parallel black-box functional operation owns two component
functional operations. One component functional operation that carries the
information of the component state of the real solution state is performed
unambiguously in the Hilbert space of the quantum system, while another that
carries the information of the component state of the candidate solution
states is performed unambiguously in the subspace of the math Hilbert space
but not in the Hilbert space of the quantum system. It may be thought that
these two component functional operations belong to the same subspace of the
math Hilbert space but one of which belongs to the quantum system and
another does not. It therefore may be considered that the QUANSDAM process
of (3.5) performs a mathematical-parallel black-box functional operation in
the subspace of the math Hilbert space. A mathematical-parallel functional
operation is essentially different from a conventional quantum parallel
operation [11, 10, 9]. It does not obey the (linear) superposition principle
in quantum physics. Each component functional operation of a
mathematical-parallel functional operation obeys the information
conservation law independently. The present analysis is available not only
for the simple QUANSDAM process of (3.5) but also for a general QUANSDAM\
process.

The above analysis shows that a QUANSDAM\ (or UNIDYSLOCK) process could
change the quantum-state difference between a pair of quantum states only
when the math Hilbert space of the search problem is involved. Therefore,
the change of the quantum-state difference between any two states, which are
in the Hilbert space of the quantum system and the math Hilbert space of the
search problem, respectively, during a QUANSDAM (or UNIDYSLOCK) process is
involved in the interaction between the Hilbert space of the quantum system
and the math Hilbert space of the search problem. This is part of the second
property of the quantum-state-difference varying (See above).

By combining the first and second properties of the quantum-state-difference
varying one can conclude that \textit{the quantum-state-difference varying
in a QUANSDAM\ (or UNIDYSLOCK) process is caused only by the interaction
between the quantum physical laws (i.e., the unitary quantum dynamics and
the Hilbert space of the quantum system) and the mathematical-logical
principle of the search problem.}

The above investigation into a UNIDYSLOCK (or QUANSDAM) process is limited
to the simple case that the reduction subspace of the math Hilbert space is
two-dimensional. Actually the dimension of the reduction subspace may be
more than two in a general QUANSDAM (or UNIDYSLOCK) process. Then in the
general case there are more than two candidate quantum states at the end of
the QUANSDAM (or UNIDYSLOCK) process. Generally, there are $%
%TCIMACRO{\U{2115} }%
%BeginExpansion
\mathbb{N}
%EndExpansion
$ candidate quantum states at the end of a QUANSDAM (or UNIDYSLOCK) process
if the reduction subspace is $%
%TCIMACRO{\U{2115} }%
%BeginExpansion
\mathbb{N}
%EndExpansion
-$dimensional. However, among these $%
%TCIMACRO{\U{2115} }%
%BeginExpansion
\mathbb{N}
%EndExpansion
$ states there is only one quantum state in the Hilbert space of the quantum
system and this quantum state carries the information of the solution state,
because there is only one solution state to the search problem. Then the
remaining $%
%TCIMACRO{\U{2115} }%
%BeginExpansion
\mathbb{N}
%EndExpansion
-1$ candidate states all are in the reduction subspace of the math Hilbert
space. Just like the simple QUANSDAM process of (3.5) this general QUANSDAM
process contains a mathematical-parallel black-box functional operation of
the reduction subspace of the math Hilbert space.

A typical example is given below. Suppose that an $n-$qutrit quantum system
is used to solve the unstructured search problem and that after the
search-space dynamical reduction the basic $IC$ unitary operator in form is
still given by $U_{x}^{ic}(\theta _{m},a_{m}^{s})$ of (2.9) with the
triple-value logical number $a_{m}^{s}=+1,$ $0,$ $-1$ (See the section 2).
This really means that the original unstructured search space of the search
problem is dynamically reduced to a three-dimensional reduction subspace of
the math Hilbert space of the search problem. Then in the present case the
simple QUANSDAM process of (3.5) should be modified to the form 
\begin{equation}
U_{x}^{ic}(\theta _{m},a_{m}^{s})|0\rangle =\left\{ 
\begin{array}{ccc}
|\psi _{+1}\rangle , & if & a_{m}^{s}=+1 \\ 
|\psi _{0}\rangle , & if & a_{m}^{s}=0 \\ 
|\psi _{-1}\rangle , & if & a_{m}^{s}=-1%
\end{array}%
\right.  \tag{3.7}
\end{equation}%
where the initial computational basis state $|0\rangle $ is equal to the
state $|+1\rangle $ (See the section 2). This shows that there are three
candidate quantum states at the end of the QUANSDAM process. Here only one
quantum state is in the Hilbert space of the $n-$qutrit quantum system,
while the rest two candidate states are in the three-dimensional reduction
subspace. This is different from the case in the $n-$qubit quantum system
above. It may be considered that the QUANSDAM process of (3.7) performs a
mathematical-parallel black-box functional operation in the
three-dimensional reduction subspace of the math Hilbert space.

It seems that the larger the reduction subspace of the math Hilbert space,
the weaker the interaction between the Hilbert space of the quantum system
and the math Hilbert space of the search problem and the smaller the $QSD$
amplification ability for a QUANSDAM process. This should be true for a
large reduction subspace of the math Hilbert space.

A QUANSDAM\ process like (3.4) can change the scalar product of a pair of
states, one of which is in the Hilbert space and another in the math Hilbert
space. However, it can prove below that a QUANSDAM\ (or UNIDYSLOCK) process
can not change the scalar product of any two quantum states in the Hilbert
space of a quantum system. In general a quantum state $|\psi _{k}\rangle $
of the quantum system could become a function of the unit number $%
b_{m}^{s}=\pm 1$ during a QUANSDAM (or UNIDYSLOCK) process. Then generally
one has $|\psi _{k}\rangle =|\psi _{k}\left( b_{m}^{s}\right) \rangle .$
Note that in the mathematical-logical meaning of search one does not know in
advance which one of the states $|\psi _{k}\left( +1\right) \rangle $ and $%
|\psi _{k}\left( -1\right) \rangle $ belongs to the quantum system. Thus,
one must consider both the cases $b_{m}^{s}=+1$ and $b_{m}^{s}=-1,$
respectively. Suppose that there are any two states $|\psi _{k}\left(
b_{m}^{s}\right) \rangle $ and $|\psi _{l}\left( b_{m}^{s}\right) \rangle $
of the quantum system with the unit number $b_{m}^{s}$ equal to some logical
value (either $+1$ or $-1$). When both the states are acted on by the basic $%
IC$ unitary operator $U_{\lambda }^{ic}(\theta _{m},a_{m}^{s})$ of (2.9) at
the same time, one obtains $U_{\lambda }^{ic}(\theta _{m},a_{m}^{s})|\psi
_{k}\left( b_{m}^{s}\right) \rangle =|\varphi _{k}\left(
a_{m}^{s},b_{m}^{s}\right) \rangle $ and $U_{\lambda }^{ic}(\theta
_{m},a_{m}^{s})|\psi _{l}\left( b_{m}^{s}\right) \rangle =|\varphi
_{l}\left( a_{m}^{s},b_{m}^{s}\right) \rangle .$ Now the scalar product of
the two states $|\varphi _{k}\left( a_{m}^{s},b_{m}^{s}\right) \rangle $ and 
$|\varphi _{l}\left( a_{m}^{s},b_{m}^{s}\right) \rangle $ is given by%
\begin{equation*}
\langle \varphi _{k}\left( a_{m}^{s},b_{m}^{s}\right) |\varphi _{l}\left(
a_{m}^{s},b_{m}^{s}\right) \rangle
\end{equation*}%
\begin{equation*}
=\left( \langle \psi _{k}\left( b_{m}^{s}\right) |U_{\lambda }^{ic}(\theta
_{m},a_{m}^{s})^{+}\right) U_{\lambda }^{ic}(\theta _{m},a_{m}^{s})|\psi
_{l}\left( b_{m}^{s}\right) \rangle
\end{equation*}%
\begin{equation*}
=\langle \psi _{k}\left( b_{m}^{s}\right) |\psi _{l}\left( b_{m}^{s}\right)
\rangle
\end{equation*}%
where the fixed logical value $a_{m}^{s}=+1$ or $-1.$ This formula shows
that the scalar product of the two states $|\varphi _{k}\left(
a_{m}^{s},b_{m}^{s}\right) \rangle $ and $|\varphi _{l}\left(
a_{m}^{s},b_{m}^{s}\right) \rangle $ is equal to the one of the two states $%
|\psi _{k}\left( b_{m}^{s}\right) \rangle $ and $|\psi _{l}\left(
b_{m}^{s}\right) \rangle $, indicating that the quantum-state difference
between the two states $|\psi _{k}\left( b_{m}^{s}\right) \rangle $ and $%
|\psi _{l}\left( b_{m}^{s}\right) \rangle $ of the quantum system can not be
changed by the basic $IC$ unitary operator $U_{\lambda }^{ic}(\theta
_{m},a_{m}^{s})$. It can further prove that the quantum-state difference
between any two states of the quantum system can not be changed by a general
QUANSDAM (or UNIDYSLOCK) process that consists of the basic $IC$ unitary
operators of (2.9) and the $QM$ unitary operators. This result leads to an
extension to the first property of the quantum-state-difference varying. As
shown above, for any fixed logical value $a_{m}^{s}=+1$ or $-1$ the basic $%
IC $ unitary operator $U_{\lambda }^{ic}(\theta _{m},a_{m}^{s})$ may be
really considered as a $QM$ unitary operator, when it is applied to a
quantum system. Then one can deduce from (3.2) and (3.3) that the
quantum-state difference between any two states of the quantum system never
changes under action of the basic $IC$ unitary operator on the quantum
system or more generally under action of a general QUANSDAM (or UNIDYSLOCK)
process. Therefore, during the QUANSDAM (or UNIDYSLOCK) process any
quantum-state effect of the quantum system can not cause the change of the
quantum-state difference.

That a UNIDYSLOCK (or QUANSDAM) process is able to change the quantum-state
difference between a pair of quantum states does not necessarily mean that
the UNIDYSLOCK (or QUANSDAM) process can change quickly (e.g., exponentially
fast) the quantum-state difference. How fast a UNIDYSLOCK (or QUANSDAM)\
process can change the quantum-state difference may be measured by the
quantum-state-difference ($QSD$) varying rate of the\ process. For a
QUANSDAM\ process it may be measured conveniently by the $QSD$\
amplification ability of the process. A\ general QUANSDAM process consists
of the basic $IC$ unitary operators and the $QM$\ unitary operators in
addition to the initial state (often the state is omitted without
confusion). Here only the basic $IC$ unitary operators can change the
quantum-state difference, while the $QM$ unitary operators can not. However,
without the $QM$ unitary operators the basic $IC$\ unitary operators alone
can not lead to a large $QSD$ amplification ability and as shown below, they
generate only a linear $QSD$ amplification ability. Then a super-linear or
even an exponential $QSD$ amplification ability must be involved in the $QM$
unitary operators in a QUANSDAM process. An exponential $QSD$ amplification
ability for a QUANSDAM process is really crucial to realize an exponential
quantum-searching speedup for the $HSSS$ quantum search process.

A\ general QUANSDAM process (omitting the initial state) may be written as%
\begin{equation}
QUANSDAM\left( K,\text{ }a_{m}^{s}\right)
=U_{K}V_{K}^{ic}(a_{m}^{s})U_{K-1}...U_{1}V_{1}^{ic}(a_{m}^{s})U_{0} 
\tag{3.8}
\end{equation}%
where $\{U_{k}\}$ are the $QM$\ unitary operators and $%
\{V_{k}^{ic}(a_{m}^{s})\}$ the $IC$\ unitary operators. Without losing
generality, the initial state of (3.8) may be simply set to the ground state
of the quantum system. The\ $IC$ unitary operator $V_{k}^{ic}(a_{m}^{s})$ in
(3.8) may take the basic $IC$ unitary operator of (2.9), the basic one in
(3.7), or more generally an $IC$ unitary propagator (See the section 4
below) which consists of the basic $IC$ unitary operators and the $QM$
unitary operators. Although the $QM$ unitary operators $\{U_{k}\}$ in (3.8)
act on the quantum system only, it may be considered that they are applied
to any state of the Hilbert space of the quantum system or any state of the
math Hilbert space in a mathematical-parallel form, which is dependent on
the logical number $a_{m}^{s}$ in the applied quantum state. The $QSD$
amplification ability of the QUANSDAM process of (3.8) could be dependent on
several factors. The first factor is the dimensional size of the reduction
subspace of the unstructured search space in which the basic $IC$ unitary
operator works. The second is the quantum system used to realize the
QUANSDAM\ process. More importantly the third is the mutual cooperation
between the $IC$ unitary operators $\{V_{k}^{ic}(a_{m}^{s})\}$ and the $QM$\
unitary operators $\{U_{k}\}$. Any $QM$ unitary operator like $U_{k}$ can
not change the quantum-state difference and the basic $IC$ unitary operator
can change only an exponentially-small value of quantum-state difference
(See (3.6) above), but the QUANSDAM process of (3.8) could greatly change
the quantum-state difference. This implies that the mutual cooperation
between the $IC$ unitary operators $\{V_{k}^{ic}(a_{m}^{s})\}$ and the $QM$\
unitary operators $\{U_{k}\}$ may be very important for the QUANSDAM\
process to achieve a super-square or even an exponential $QSD$ amplification
ability. It seems that when the reduction subspace is large, the mutual
cooperation becomes uneasy to realize. Therefore, the larger the reduction
subspace, the weaker the $QSD$ amplification ability of the QUANSDAM
process. When the reduction subspace is exponentially large, it should be
impossible to achieve an exponential $QSD$ amplification ability for the
QUANSDAM process no matter what these $QM$\ unitary operators $\{U_{k}\}$
are taken. These properties have nothing to do with any initial state of the
QUANSDAM process. Therefore, one must use a small reduction subspace to
realize the QUANSDAM process.

The purpose to investigate the QUANSDAM process of (3.8) for the given $IC$\
unitary operators $\{V_{k}^{ic}(a_{m}^{s})\}$ is to find an appropriate
quantum system and the appropriate $QM$ unitary operators $\{U_{k}\}$ of the
quantum system to realize the QUANSDAM process so that the QUANSDAM process
can achieve a super-square or even an exponential $QSD$ amplification
ability.

When $V_{k}^{ic}(a_{m}^{s})$ in (3.8) is taken as the basic $IC$ unitary
operator $U_{\lambda }^{ic}(\theta _{m},$ $a_{m}^{s})$ of (2.9), which works
in a two-dimensional reduction subspace, there are only two candidate final
states in the QUANSDAM process of (3.8) with the initial normalized state $%
|\Psi _{0}\rangle $: 
\begin{subequations}
\label{a}
\begin{equation}
U_{K}U_{\lambda }^{ic}(\theta _{m},a_{m}^{s})U_{K-1}...U_{1}U_{\lambda
}^{ic}(\theta _{m},a_{m}^{s})U_{0}|\Psi _{0}\rangle =\left\{ 
\begin{array}{c}
|\Psi _{+1}^{K}\rangle ,\text{ }if\text{ }a_{m}^{s}=+1 \\ 
|\Psi _{-1}^{K}\rangle ,\text{ }if\text{ }a_{m}^{s}=-1%
\end{array}%
\right.  \tag{3.9}
\end{equation}%
Here one may simply set the initial state $|\Psi _{0}\rangle $ to the ground
state of the quantum system. During the QUANSDAM process two intermediate
(different) states $|\Psi _{+1}^{k}\rangle =QUANSDAM\left( k,+1\right) |\Psi
_{0}\rangle $ and $|\Psi _{-1}^{k}\rangle =QUANSDAM\left( k,-1\right) $ $%
|\Psi _{0}\rangle $ with $1\leq k\leq K$ forever exclude one another in the
quantum system no matter what these $QM$ unitary operators $\{U_{k}\}$ take,
where $QUANSDAM\left( l,a_{m}^{s}\right) $ is given by (3.8)\ with the
settings $V_{k}^{ic}(a_{m}^{s})=U_{\lambda }^{ic}(\theta _{m},a_{m}^{s})$
and $K=l$. That is, only one of the two states is in the Hilbert space of
the quantum system, while another is in the math Hilbert space of the search
problem but is never in the quantum system. Therefore, in the sense of
quantum physics there does not exist any real quantum-state effect such as
the quantum-state superposition, coherence interference, entanglement and
nonlocal effect, or correlation between the two states $|\Psi
_{+1}^{k}\rangle $ and $|\Psi _{-1}^{k}\rangle $ over the whole QUANSDAM\
process. Here it must be emphasized that this result is universal for a
general QUANSDAM\ (or UNIDYSLOCK) process. Suppose that the reduction
subspace in which a general QUANSDAM (or UNIDYSLOCK) process works is $%
%TCIMACRO{\U{2115} }%
%BeginExpansion
\mathbb{N}
%EndExpansion
-$dimensional. Then there should be $%
%TCIMACRO{\U{2115} }%
%BeginExpansion
\mathbb{N}
%EndExpansion
$ candidate final states rather than two candidate final states in (3.9).
Moreover, only one of these $%
%TCIMACRO{\U{2115} }%
%BeginExpansion
\mathbb{N}
%EndExpansion
$ candidate states appears in the quantum system, while the remaining $%
%TCIMACRO{\U{2115} }%
%BeginExpansion
\mathbb{N}
%EndExpansion
-1$ candidate states do not. Clearly there does not exist any real
quantum-state effect between the candidate state that appears in the quantum
system and any one of the remaining $%
%TCIMACRO{\U{2115} }%
%BeginExpansion
\mathbb{N}
%EndExpansion
-1$ candidate states. Because these remaining $%
%TCIMACRO{\U{2115} }%
%BeginExpansion
\mathbb{N}
%EndExpansion
-1$ candidate states are not in the quantum system, there is not any real
quantum-state effect between any two of the remaining $%
%TCIMACRO{\U{2115} }%
%BeginExpansion
\mathbb{N}
%EndExpansion
-1$ candidate states. Therefore, one arrives at the result that there is not
any real quantum-state effect between any two of these $%
%TCIMACRO{\U{2115} }%
%BeginExpansion
\mathbb{N}
%EndExpansion
$ candidate states in the general QUANSDAM (or UNIDYSLOCK) process. This
result is important. By combining this result with the first property of
quantum-state-difference varying (See above the property and its extension)
one can conclude that \textit{no quantum-state effect of a quantum system
can cause the change of the quantum-state difference between the candidate
state that appears in the quantum system and any other candidate state of
the math Hilbert space during a QUANSDAM (or UNIDYSLOCK) process.}

How to extract the information ($a_{m}^{s}$) of the component state of the
solution state at the end of the QUANSDAM process of (3.9) by quantum
measurement? In general, the solution information ($a_{m}^{s}$) can be
obtained only by distinguishing one of the two candidate states $|\Psi
_{+1}^{K}\rangle $ and $|\Psi _{-1}^{K}\rangle $ from another at the end of
the QUANSDAM process. But here it must be pointed out that a single quantum
state (either $|\Psi _{+1}^{K}\rangle $ or $|\Psi _{-1}^{K}\rangle )$ of the
quantum system is not sufficient to obtain the solution information ($%
a_{m}^{s}$). Therefore, when both the states $|\Psi _{+1}^{K}\rangle $ and $%
|\Psi _{-1}^{K}\rangle $ are orthogonal to one another, the solution
information ($a_{m}^{s}$) can be obtained unambiguously by the quantum
measurement at the end of the QUANSDAM\ process. Here these two orthogonal
states could be first transferred to the eigenstates of the observable
Hermitian operator (i.e., a dynamical variable), respectively. Then by
quantum measuring eigenvalue of the observable operator one is able to
distinguish these two orthogonal states from one another. In this case the
quantum measurement is precise and may be considered to be deterministic. In
quantum mechanics it obeys the quantum-measurement postulate that one (or
another) of the eigenvalues of a dynamical variable is the only possible
value of a precise measurement of the dynamical variable [14, 17]. There is
the possibility that both the states $|\Psi _{+1}^{K}\rangle $ and $|\Psi
_{-1}^{K}\rangle $ are not orthogonal to one another. Then in this case in
technique it is still possible to use the quantum measurement such as the
POVM measurement [28] to distinguish these two non-orthogonal states from
one another. Such a quantum measurement is still probabilistic and obeys the
Born's rule [17, 14].

It may be thought that the unitary time evolution process of the QUANSDAM\
process of (3.9) owns two different branches, which correspond to the final
two candidate states $|\Psi _{+1}^{K}\rangle $ and $|\Psi _{-1}^{K}\rangle ,$
respectively. These two branches form a mathematical-parallel quantum
computation. They are given by ($1\leq k\leq K$) 
\end{subequations}
\begin{equation}
|\Psi _{+1}^{k}\rangle =U_{k}U_{\lambda }^{ic}(\theta
_{m},+1)U_{k-1}...U_{1}U_{\lambda }^{ic}(\theta _{m},+1)U_{0}|\Psi
_{0}\rangle ,  \tag{3.10a}
\end{equation}%
\begin{equation}
|\Psi _{-1}^{k}\rangle =U_{k}U_{\lambda }^{ic}(\theta
_{m},-1)U_{k-1}...U_{1}U_{\lambda }^{ic}(\theta _{m},-1)U_{0}|\Psi
_{0}\rangle .  \tag{3.10b}
\end{equation}%
Here only one of the two branches really describes the time evolution
process of the quantum system, while another describes the time evolution
process of the candidate state that belongs to the reduction subspace of the
math Hilbert space. Both are unitary. But the former is a real unitary
quantum dynamical process, while the latter is a mathematical unitary
dynamical process. Both the candidate states $|\Psi _{+1}^{k}\rangle $ and $%
|\Psi _{-1}^{k}\rangle $ obey different branches of the time evolution
process, respectively, leading to that they may be different in the course
of the QUANSDAM process even if their corresponding initial states are taken
as the same one. Consequently there is the quantum-state difference $\rho
_{12}\left( k\right) =\langle \Psi _{+1}^{k}|\Psi _{-1}^{k}\rangle $ between
the candidate states $|\Psi _{+1}^{k}\rangle $ and $|\Psi _{-1}^{k}\rangle \ 
$and more importantly the quantum-state difference is varying during the
QUANSDAM process, although the QUANSDAM process is unitary.

Owing to these $QM$ unitary operators $\{U_{k}\}$ the original Hilbert
subspace of the quantum system in which the basic $IC$ unitary operator of
(2.9) works may be changed during the QUANSDAM process of (3.9).
Correspondingly the original reduction subspace of the unstructured search
space, where the basic $IC$ unitary operator works, changes too. But the
dimensional size of the reduction subspace spanned by $\{|\Psi
_{+1}^{k}\rangle ,$ $|\Psi _{-1}^{k}\rangle \}$ of (3.10)\ during the
QUANSDAM\ process is always equal to the one of the original reduction
subspace except for the special case that $|\Psi _{+1}^{k}\rangle $ is the
same as $|\Psi _{-1}^{k}\rangle $.

If the reduction subspace in which a general QUANSDAM (or UNIDYSLOCK)
process works is $%
%TCIMACRO{\U{2115} }%
%BeginExpansion
\mathbb{N}
%EndExpansion
-$dimensional, then the unitary time evolution process of the QUANSDAM (or
UNIDYSLOCK)\ process owns $%
%TCIMACRO{\U{2115} }%
%BeginExpansion
\mathbb{N}
%EndExpansion
$ different branches. Only one of these $%
%TCIMACRO{\U{2115} }%
%BeginExpansion
\mathbb{N}
%EndExpansion
$ different branches is quantum-physical and it describes the time evolution
process of the candidate state that appears in the quantum system, while the
remaining $%
%TCIMACRO{\U{2115} }%
%BeginExpansion
\mathbb{N}
%EndExpansion
-1$ different branches each are mathematical and they describe the time
evolution processes of the remaining $%
%TCIMACRO{\U{2115} }%
%BeginExpansion
\mathbb{N}
%EndExpansion
-1$ candidate states of the math Hilbert space of the search problem,
respectively. Obviously, the reduction subspace may change during the
QUANSDAM (or UNIDYSLOCK) process. It is spanned by these $%
%TCIMACRO{\U{2115} }%
%BeginExpansion
\mathbb{N}
%EndExpansion
$ candidate states. Its dimensional size is always equal to $%
%TCIMACRO{\U{2115} }%
%BeginExpansion
\mathbb{N}
%EndExpansion
$, i.e., the one of the original reduction subspace except for the special
cases. Generally, these $%
%TCIMACRO{\U{2115} }%
%BeginExpansion
\mathbb{N}
%EndExpansion
$ candidate states can be distinguished from each other by the quantum
measurement (the usual one or the POVM one) at the end of the QUANSDAM
process, when they are orthogonal to each other or approximately orthogonal
to each other.

The $QSD$ amplification ability for the QUANSDAM process of (3.9) is just
the quantum-state-difference ($\rho _{12}\left( k\right) $) varying speed.
The latter is the characteristic physical quantity of a QUANSDAM\ (or
UNIDYSLOCK) process. Here $\rho _{12}\left( k\right) $ is dependent on the
number $k$ of the basic $IC$ unitary operators on the left-hand side of
(3.9). It is known from (3.10)\ that the initial $\rho _{12}\left( 0\right)
=1$ because $|\Psi _{+1}^{0}\rangle =|\Psi _{-1}^{0}\rangle =U_{0}|\Psi
_{0}\rangle .$ Since $\left\vert \rho _{12}\left( k\right) \right\vert
=\left\vert \langle \Psi _{+1}^{k}|\Psi _{-1}^{k}\rangle \right\vert \leq 1$
for $1\leq k\leq K,$ on average the varying rate of $\rho _{12}\left(
k\right) $ is $\left( \left\vert \rho _{12}\left( k\right) \right\vert
-\left\vert \rho _{12}\left( 0\right) \right\vert \right) /k$ per basic $IC$
unitary operator $U_{\lambda }^{ic}(\theta _{m},a_{m}^{s})$. If at the end
of the QUANSDAM\ process both the states $|\Psi _{+1}^{K}\rangle $ and $%
|\Psi _{-1}^{K}\rangle $ are orthogonal to one another, then $\rho
_{12}\left( K\right) =0.$ In this case on average the $QSD$ varying rate
over the whole QUANSDAM process is given by $-1/K\ $per basic $IC$ unitary
operator. Here one uses only the number of the basic $IC$ unitary operators
to calculate the average $QSD$ varying rate without taking into account the
number of the $QM$\ unitary operators $\{U_{k}\}$ in the QUANSDAM\ process.
The main reason for this is that any $QM$ unitary operator does not change
the quantum-state difference. Since the difference $\left\vert \rho
_{12}\left( k\right) \right\vert -\left\vert \rho _{12}\left( 0\right)
\right\vert $ is never more than one in magnitude, in a usual case (i.e., $%
\left\vert \left( \left\vert \rho _{12}\left( k\right) \right\vert
-\left\vert \rho _{12}\left( 0\right) \right\vert \right) \right\vert
\thickapprox 1$ with $k\thickapprox K$) the average $QSD$ varying rate for
the QUANSDAM process is mainly determined by the total number ($K$) of the
basic $IC$ unitary operators. As shown above, the average $QSD$\ varying
rate is inversely proportional to the total number $K$. Then this means that
the smaller the total number $K$, the larger the average $QSD$ amplification
ability of the QUANSDAM\ process.

From the point of view of quantum-computing complexity the complexity of a
QUANSDAM process is determined not only by the total number of the basic $IC$
unitary operators but also by the total number of the $QM$\ unitary
operators. However, in a usual case the influence of the $QM$ unitary
operators on the complexity could not be greater than that one of the basic $%
IC$ unitary operators. Then in such case the average $QSD$ amplification
ability (or the total number of the basic $IC$ unitary operators alone) may
be conveniently used to measure approximately the complexity of the QUANSDAM
process.

The $QSD$ amplification ability of a QUANSDAM\ process could be strongly
dependent upon the $QM$\ unitary operators. While a $QM$\ unitary operator
itself can not change the quantum-state difference, it is able to promote
(or speed up) the $IC$ unitary operator to change greatly the quantum-state
difference, that is, it can enhance the $IC$ unitary operator to amplify
greatly the $QSD$ varying speed in a QUANSDAM process. Undoubtedly the
average $QSD$ varying rate is convenient to measure the average $QSD$
amplification ability of a QUANSDAM process as a whole. However, it can not
measure exactly the effect of each $QM$ unitary operator on the $QSD$
amplification ability of the QUANSDAM process. It may be more exact to
employ the following $QSD$\ varying rate: 
\begin{equation}
\Delta \rho _{12}\left( k+1\right) =\left\vert \rho _{12}\left( k+1\right)
\right\vert -\left\vert \rho _{12}\left( k\right) \right\vert  \tag{3.11}
\end{equation}%
to measure the effect of each $QM$\ unitary operator $(U_{k})$ on the $QSD$
amplification ability of the QUANSDAM process. The physical meaning of $%
\Delta \rho _{12}\left( k+1\right) $ is that the $\left( k+1\right) -$th $IC$
unitary operator in the QUANSDAM\ process alone changes the quantum-state
difference when it acts on the quantum system, and the changing value is
given by $\Delta \rho _{12}\left( k+1\right) $. The reason why $\Delta \rho
_{12}\left( k+1\right) $ can be used to measure exactly the effect of the
single $QM$ unitary operator $U_{k}$ on the $QSD$ amplification ability is
that there is a difference for $\Delta \rho _{12}\left( k+1\right) $ between
there is and there is not the $QM$ unitary operator $U_{k}$ in the QUANSDAM
process. In technique it could be more convenient to use $\Delta \rho
_{12}^{2}\left( k+1\right) =\left\vert \rho _{12}\left( k+1\right)
\right\vert ^{2}-\left\vert \rho _{12}\left( k\right) \right\vert ^{2}$
rather than $\Delta \rho _{12}\left( k+1\right) .$ Beside the $QSD$\ varying
rate $\Delta \rho _{12}\left( k+1\right) $ one also may use the average $QSD$%
\ varying rate over the $L$ basic $IC$ unitary operators ($L<<K$) to measure
the average effect of the $L$ number of the $QM$ unitary operators on the $%
QSD$ amplification ability. Below it is described simply how a $QM$\ unitary
operator could be appropriate to enhance the $QSD$ amplification ability of
a QUANSDAM\ process.

First of all, consider the special case that all the $QM$\ unitary operators 
$\{U_{k}\}$ are the unit operators in the QUANSDAM process of (3.9). Then
from (3.9) one obtains%
\begin{equation}
\left( U_{\lambda }^{ic}(\theta _{m},a_{m}^{s})\right) ^{K}|\Psi _{0}\rangle
=\left\{ 
\begin{array}{c}
|\Psi _{+1}^{K}\rangle ,\text{ }if\text{ }a_{m}^{s}=+1 \\ 
|\Psi _{-1}^{K}\rangle ,\text{ }if\text{ }a_{m}^{s}=-1%
\end{array}%
\right.  \tag{3.12a}
\end{equation}%
This QUANSDAM process (without the initial state) consists of the $K$ basic $%
IC$ unitary operators of (2.9) alone. There is not any contribution made by
any $QM$\ unitary operator to its $QSD$ amplification ability. Therefore,
the QUANSDAM\ process of (3.12a) may be considered as the reference
QUANSDAM\ process. If a QUANSDAM process has the $QSD$ amplification ability
stronger than the one of (3.12a), then its $QM$ unitary operators make a
positive contribution to the $QSD$\ amplification ability. In contrast, if a
QUANSDAM process has the $QSD$ amplification ability weaker than the one of
(3.12a), then its $QM$ unitary operators are inappropriate for the QUANSDAM
process.

By using (2.9) the QUANSDAM process of (3.12a) (without the initial state)
is explicitly written as $\left( U_{\lambda }^{ic}(\theta
_{m},a_{m}^{s})\right) ^{K}=\exp (-ia_{m}^{s}K\theta _{m}I_{m\lambda }).$
Note that the angle $\theta _{m}\varpropto 1/2^{n}$ is exponentially small.
Then the number $K$ of the basic $IC$ unitary operators must be sufficiently
large so that the solution information $\left( a_{m}^{s}\right) $ can be
obtained by the QUANSDAM\ process. As a typical example, the QUANSDAM
process with $\lambda =x$ is applied to the initial state $|\Psi _{0}\rangle
=|0\rangle \ $and just like the simple QUANSDAM process of (3.4) one obtains 
\begin{equation}
\left( U_{x}^{ic}(\theta _{m},a_{m}^{s})\right) ^{K}|0\rangle =\cos \left( 
\frac{K}{2}\theta _{m}\right) |0\rangle -ia_{m}^{s}\sin \left( \frac{K}{2}%
\theta _{m}\right) |1\rangle .  \tag{3.12b}
\end{equation}%
This indicates that the two candidate states at the end of the QUANSDAM
process of (3.12a) are given by $|\Psi _{\pm 1}^{K}\rangle =\cos \left( 
\frac{K}{2}\theta _{m}\right) |0\rangle \mp i\sin \left( \frac{K}{2}\theta
_{m}\right) |1\rangle .$ Then the scalar product between the two states is
given by $\rho _{12}\left( K\right) =\langle \Psi _{+1}^{K}|\Psi
_{-1}^{K}\rangle =\cos \left( K\theta _{m}\right) .$ When $\cos \left(
K\theta _{m}\right) =0$ or when the number $K$ is given by $K=\frac{1}{2}\pi
/\left\vert \theta _{m}\right\vert ,$ the scalar product $\rho _{12}\left(
K\right) =0,$ indicating that both the candidate states $|\Psi
_{+1}^{K}\rangle $ and $|\Psi _{-1}^{K}\rangle $ are orthogonal to each
other. Now by the usual quantum measurement [17, 14] one is able to
determine the unit number $\left( a_{m}^{s}\right) $ unambiguously. Since
the angle $\theta _{m}$ is exponentially small, the number $K$ has to be
exponentially large. This shows that the QUANSDAM process of (3.12) that
consists of only the basic $IC$ unitary operators is an inefficient process
to amplify the quantum-state difference.

The $QSD$ varying rates for the QUANSDAM process of (3.12b) are calculated
below. The average $QSD$ varying rate over the whole process is given by $%
-1/K=-2\left\vert \theta _{m}\right\vert /\pi .$ It is exponentially small
in magnitude, indicating that the QUANSDAM process is inefficient. It is
easy to find that the scalar product $\rho _{12}\left( k\right) =\langle
\Psi _{+1}^{k}|\Psi _{-1}^{k}\rangle =\cos \left( k\theta _{m}\right) $ for $%
0\leq k\leq K.$ Then the average $QSD$\ varying rate of $\rho _{12}\left(
k\right) $ over the first $k$ basic $IC$ unitary operators is%
\begin{equation}
\left( \left\vert \rho _{12}\left( k\right) \right\vert -\left\vert \rho
_{12}\left( 0\right) \right\vert \right) /k=\left( \cos \left( k\theta
_{m}\right) -1\right) /k\thickapprox -\frac{1}{4}\pi \left\vert \theta
_{m}\right\vert k/K  \tag{3.13}
\end{equation}%
where $k\left\vert \theta _{m}\right\vert <<1.$ By using the formula (3.11)
it can turn out that the $QSD$ varying rate $\Delta \rho _{12}\left(
k+1\right) $ $\left( 0\leq k\leq K-1\right) $ is given by%
\begin{equation}
\Delta \rho _{12}\left( k+1\right) =\allowbreak -2\sin \left( \left( k+\frac{%
1}{2}\right) \theta \newline
_{m}\right) \sin \left( \frac{1}{2}\theta _{m}\right)  \tag{3.14}
\end{equation}%
where $\left( k+1\right) \left\vert \theta _{m}\right\vert \leq \pi /2.$
Note that $\theta _{m}$ is exponentially small. Then one has $\sin \left( 
\frac{1}{2}\left\vert \theta _{m}\right\vert \right) \thickapprox \frac{1}{2}%
\left\vert \theta _{m}\right\vert =\frac{1}{4}\pi /K.$ When $k\left\vert
\theta _{m}\right\vert <<1,$ one obtains from (3.14) that%
\begin{equation}
\Delta \rho _{12}\left( k+1\right) \thickapprox \allowbreak -\frac{1}{4}\pi
\left\vert \theta \newline
_{m}\right\vert \left( 2k+1\right) /K.  \tag{3.15}
\end{equation}%
Both $\left( \left\vert \rho _{12}\left( k\right) \right\vert -\left\vert
\rho _{12}\left( 0\right) \right\vert \right) /k<0$ ($1\leq k\leq K$) and $%
\Delta \rho _{12}\left( k+1\right) <0$ ($0\leq k\leq K-1$) are consistent
with the fact that the process of (3.12b) is a QUANSDAM process.

As shown in (3.15), when $k\left\vert \theta _{m}\right\vert <<1,$ the $QSD$
varying rate $\Delta \rho _{12}\left( k+1\right) $ is linearly dependent on
the value $k\left\vert \theta _{m}\right\vert $ or the number ($k$) of the
basic $IC$ unitary operators in the QUANSDAM\ process. It can prove that
this conclusion also holds for the QUANSDAM process of (3.12a) with an
arbitrary initial state $|\Psi _{0}\rangle $. As shown in (3.13), the
average $QSD$\ varying rate of $\rho _{12}\left( k\right) $ is linearly
dependent on the value $k\left\vert \theta _{m}\right\vert $ or the number $%
k $ too. Clearly, the $QSD$ varying rate $\Delta \rho _{12}\left( k+1\right) 
$ for the QUANSDAM\ process of (3.12) generally decreases linearly (or it
increases linearly in magnitude) as the value $k\left\vert \theta
_{m}\right\vert $ or the number $k$ of the basic $IC$ unitary operators.
Here the condition $k\left\vert \theta _{m}\right\vert <<1$ is quite
general, because one may set $K\left\vert \theta _{m}\right\vert =O\left(
1/p\left( n\right) \right) <<1$ and $0\leq k<K,$ here $p\left( n\right) $ is
a polynomial in the qubit number $n$ corresponding to the $2^{n}-$
dimensional unstructured search space. However, if $K\left\vert \theta
_{m}\right\vert =O\left( 1/p\left( n\right) \right) <<1,$ then $\rho
_{12}\left( K\right) =\cos \left( K\theta _{m}\right) \neq 0,$ indicating
that both the candidate states $|\Psi _{+1}^{K}\rangle $ and $|\Psi
_{-1}^{K}\rangle $ are not orthogonal to each other at the end of the
QUANSDAM\ process of (3.12). In this case in technique one is still able to
distinguish the two non-orthogonal states $|\Psi _{+1}^{K}\rangle $ and $%
|\Psi _{-1}^{K}\rangle $ from each other and obtain the solution information 
$\left( a_{m}^{s}\right) $ by the quantum measurement with unambiguous state
distinguishability [28], but the maximum successful probability is $P=1-$ $%
\left\vert \rho _{12}\left( K\right) \right\vert $.

There is not any $QM$ unitary operator in the QUANSDAM process of (3.12).
The characteristic feature for such QUANSDAM process is that the $QSD$\
varying rate $\Delta \rho _{12}\left( k+1\right) $ decreases linearly as the
value $k\left\vert \theta _{m}\right\vert $ or the number ($k$) of the basic 
$IC$ unitary operators in the QUANSDAM\ process. The QUANSDAM process may be
considered as the reference one for investigating the influence of a $QM$
unitary operator on the $QSD$\ amplification ability of a general QUANSDAM
process.

The QUANSDAM process of (3.12) may be called a linear QUANSDAM\ process due
to that its $QSD$ varying rate $\Delta \rho _{12}\left( k+1\right) $
decreases linearly as the number $(k)$ of the basic $IC$ unitary operators.
Just like the linear QUANSDAM process one may define other QUANSDAM
processes. A QUANSDAM\ process whose $\Delta \rho _{12}\left( k+1\right) $
decreases in a square (or quadratic) form as the number $k$ may be called a
square QUANSDAM process. Of course, here $\Delta \rho _{12}\left( k+1\right)
\varpropto \left\vert \theta \newline
_{m}\right\vert /K$ too. If the $QSD$ varying rate$\ \Delta \rho _{12}\left(
k+1\right) $ of a QUANSDAM process decreases in the form of a polynomial or
an exponential function as the number $k$, then the QUANSDAM process may be
called a polynomial or an exponential QUANSDAM\ process. As shown above,
there is not any $QM$ unitary operator in the linear QUANSDAM process of
(3.12). Then a super-linear QUANSDAM process exists only when it contains
the $QM$\ unitary operators. Note that a $QM$\ unitary operator itself can
not change the quantum-state difference. These imply that a super-linear
QUANSDAM process must be involved in the interaction between the $IC$
unitary operators and the $QM$\ unitary operators in the QUANSDAM process.

Here it must be pointed out that because of the parameter $\theta
_{m}\varpropto 1/2^{n}$ a square QUANSDAM process above does not mean that
the extraction of the solution information ($a_{m}^{s}$) can be realized in
a square speedup by the QUANSDAM process. Actually it can prove that a cubic
QUANSDAM process above is able to extract the solution information in a
square speedup. However, if one employs the $QSD$ varying rate $\Delta \rho
_{12}\left( k+1\right) /k$ or $\Delta \rho _{12}^{2}\left( k+1\right) /k$ to
measure a QUANSDAM process, then a square QUANSDAM process measured in this
way is able to obtain the solution information in a square speedup and hence
it is correctly correspondent to the amplitude amplification of a
conventional quantum search algorithm. In order to retain this
correspondence relation, whenever one compares a QUANSDAM process with the
amplitude amplification of a conventional quantum search algorithm in this
paper, it implies that the QUANSDAM process is measured by the $QSD$ varying
rate $\Delta \rho _{12}\left( k+1\right) /k$ (or $\Delta \rho
_{12}^{2}\left( k+1\right) /k)$. Here, if $k=0$, then $\Delta \rho
_{12}\left( k+1\right) /k$ is set to $\Delta \rho _{12}\left( k+1\right) .$

In principle an $IC$ unitary operator like $U_{\lambda }^{ic}(\theta
_{m},a_{m}^{s})$ of (2.9) acting on a quantum state with the solution
information ($a_{m}^{s}$) may become a QUANSDAM\ process or a UNIDYSLOCK
process. As known from (3.9), one has $|\Psi _{a_{m}^{s}}^{k}\rangle
=QUANSDAM\left( k,a_{m}^{s}\right) |\Psi _{0}\rangle $ with $0\leq k\leq K.$
Suppose now that $|\Phi _{a_{m}^{s}}^{k+1}\rangle =U_{\lambda }^{ic}(\theta
_{m},a_{m}^{s})|\Psi _{a_{m}^{s}}^{k}\rangle .$ Then $|\Psi
_{a_{m}^{s}}^{k+1}\rangle =U_{k+1}|\Phi _{a_{m}^{s}}^{k+1}\rangle .$ It can
be found that the quantum-state difference $\rho _{12}\left( k\right)
=\langle \Psi _{+1}^{k}|\Psi _{-1}^{k}\rangle =\langle \Phi _{+1}^{k}|\Phi
_{-1}^{k}\rangle ,$ indicating that the $QM$\ unitary operator $U_{k}$ does
not affect $\rho _{12}\left( k\right) .$ Though $U_{k}$ does not affect $%
\rho _{12}\left( k\right) ,$ it can affect $\rho _{12}\left( k+1\right) .$
If now $\left\vert \rho _{12}\left( k+1\right) \right\vert >\left\vert \rho
_{12}\left( k\right) \right\vert ,$ then $|\Phi _{a_{m}^{s}}^{k+1}\rangle
=U_{\lambda }^{ic}(\theta _{m},a_{m}^{s})U_{k}|\Phi _{a_{m}^{s}}^{k}\rangle $
is a UNIDYSLOCK\ process. If $\left\vert \rho _{12}\left( k+1\right)
\right\vert <\left\vert \rho _{12}\left( k\right) \right\vert ,$ then $|\Phi
_{a_{m}^{s}}^{k+1}\rangle =U_{\lambda }^{ic}(\theta
_{m},a_{m}^{s})U_{k}|\Phi _{a_{m}^{s}}^{k}\rangle $ is a QUANSDAM\ process.
Here the $QM$\ unitary operator $U_{k}$ plays a crucial role in that it can
increase the varying speed and adjust the varying direction of quantum-state
difference in a deterministic form for the next $IC$ unitary operator $%
U_{\lambda }^{ic}(\theta _{m},a_{m}^{s})$.

Consider the extended version of the reference QUANSDAM process of (3.12): 
\begin{equation}
|\tilde{\Psi}_{a_{m}^{s}}^{k}\rangle =\left( U_{\lambda }^{ic}(\theta
_{m},a_{m}^{s})\right) ^{k}|\tilde{\Psi}_{a_{m}^{s}}\rangle  \tag{3.16}
\end{equation}%
where the initial state $|\tilde{\Psi}_{a_{m}^{s}}\rangle $ is allowed to
carry the solution information ($a_{m}^{s}$). Unlike the QUANSDAM process of
(3.12), the process of (3.16) may be a QUANSDAM process or a\ UNIDYSLOCK
process. In either case it can prove that when $k\left\vert \theta
_{m}\right\vert <<1,$ for an arbitrary initial state $|\tilde{\Psi}%
_{a_{m}^{s}}\rangle $ the $QSD$ varying rate $\Delta \rho _{12}^{2}\left(
k+1\right) $ is linearly dependent on the value $k\left\vert \theta
_{m}\right\vert $ or the number $(k)$ of the basic $IC$ unitary operators in
(3.16). Thus, this process reproduces the characteristic feature of the
reference QUANSDAM process of (3.12).

A $QM$ unitary operator should make a positive contribution to the $QSD$
amplification ability for its own QUANSDAM\ process. Here the so-called
positive contribution means that, promoted by the $QM$ unitary operators,
the QUANSDAM process has a stronger $QSD$ amplification ability than the
linear one of the reference QUANSDAM\ process of (3.12). However, a linear
or super-linear QUANSDAM process could not be useful. A useful QUANSDAM\
process should have the $QSD$ varying rate $\Delta \rho _{12}\left(
k+1\right) /k\varpropto k^{2}$ or more in magnitude, so that it is able to
extract the solution information in a square speedup at least. Now a $QM$\
unitary operator (e.g., $U_{k}$) of a QUANSDAM process is said to be
appropriate if it is able to make the $QSD$ varying rate $\Delta \rho
_{12}\left( k+1\right) /k$ (or $\Delta \rho _{12}^{2}\left( k+1\right) /k$)
in magnitude exceeding $O\left( p_{2}\left( k\right) \right) ,$ where $%
p_{2}\left( k\right) $ is a quadratic polynomial in the number $k$. A
QUANSDAM process is obviously able to extract the solution information in a
super-square speedup if every $QM$ unitary operator of it is appropriate.

Suppose now that the $QM$\ unitary operators $\{U_{k}\}$ with $0\leq k<K$
are appropriate in the QUANSDAM process of (3.9). Then one may say that the
quantum state $|\Psi _{a_{m}^{s}}^{k}\rangle =U_{k}|\Phi
_{a_{m}^{s}}^{k}\rangle =QUANSDAM\left( k,a_{m}^{s}\right) |\Psi _{0}\rangle
\ $is appropriate. By applying the basic $IC$ unitary operator to the
appropriate quantum state $|\Psi _{a_{m}^{s}}^{k}\rangle $ one obtains $%
|\Phi _{a_{m}^{s}}^{k+1}\rangle =U_{\lambda }^{ic}(\theta
_{m},a_{m}^{s})|\Psi _{a_{m}^{s}}^{k}\rangle =U_{\lambda }^{ic}(\theta
_{m},a_{m}^{s})U_{k}|\Phi _{a_{m}^{s}}^{k}\rangle ,$ and both $\rho
_{12}\left( k+1\right) =\langle \Phi _{+1}^{k+1}|\Phi _{-1}^{k+1}\rangle $
and $\rho _{12}\left( k\right) =\langle \Phi _{+1}^{k}|\Phi _{-1}^{k}\rangle 
$ lead to that the $QSD$ varying rate $\Delta \rho _{12}^{2}\left(
k+1\right) /k$ can exceed $O\left( p_{2}\left( k\right) \right) $ in
magnitude. In contrast, if $U_{k}$ is not appropriate, then the state $|\Psi
_{a_{m}^{s}}^{k}\rangle =U_{k}|\Phi _{a_{m}^{s}}^{k}\rangle $ is not
appropriate and hence $\Delta \rho _{12}^{2}\left( k+1\right) /k$ can not
exceed $O\left( p_{2}\left( k\right) \right) $ in magnitude. If there is not
the $k-$th $QM$ unitary operator $U_{k},$ then one has $|\Phi
_{a_{m}^{s}}^{k+1}\rangle =U_{\lambda }^{ic}(\theta _{m},a_{m}^{s})|\Phi
_{a_{m}^{s}}^{k}\rangle =\left( U_{\lambda }^{ic}(\theta
_{m},a_{m}^{s})\right) ^{2}|\Psi _{a_{m}^{s}}^{k-1}\rangle .$ As shown in
(3.16), in this case the $QSD$ varying rate $\Delta \rho _{12}^{2}\left(
k+1\right) $ has the characteristic feature of the reference QUANSDAM
process of (3.12) or (3.16).

It is important how to find the appropriate $QM$\ unitary operators so that
one can construct a useful QUANSDAM process in a quantum system. In the next
section this research topic is further discussed.

There are two quite different types of QUANSDAM processes. One of which is
the amplitude-based QUANSDAM processes and another is the phase-based
QUANSDAM processes. The former is relatively simple and similar to the
amplitude amplification of a conventional quantum search algorithm and they
usually work in an $n-$qubit quantum system, while the latter is quite
complex and they work generally in a time- and space-dependent quantum
system. Below an amplitude-based QUANSDAM process is introduced simply. A
phase-based QUANSDAM process is closely related to the so-called
information-carrying unitary propagator. They are described in the next
section.

The fashion of an amplitude-based QUANSDAM\ process is that the same basic $%
IC$ unitary operator and the $QM$ unitary operator appear alternatively in
the QUANSDAM\ process. An amplitude-based QUANSDAM\ process is quite similar
to the amplitude amplification of a conventional quantum search algorithm.
Moreover, there is the corresponding relation between them. Actually, the
simple QUANSDAM process of (3.4) and the one of (3.12) may be considered as
the special amplitude-based QUANSDAM processes. The QUANSDAM\ process of
(3.9) is a general amplitude-based QUANSDAM\ process. Usually all the $QM$\
unitary operators $\{U_{k}\}$ in (3.9) are taken as the same one and doing
so could greatly simplify the exact calculation of the QUANSDAM\ process.
Though an amplitude-based QUANSDAM\ process could be calculated exactly, it
could not capture exactly the essential aspect of a general QUANSDAM\
process.

As an example, below analyze the amplitude-based QUANSDAM process of
(3.12b). It is known from (3.12b) that the state $|1\rangle $ carries the
solution information $\left( a_{m}^{s}\right) $. Here for convenience the
state $a_{m}^{s}|1\rangle $ in (3.12b) is called the information-carrying
state (or the $IC$ state briefly). In (3.12b) the $IC$ state $%
a_{m}^{s}|1\rangle $ has an amplitude of $-i\sin \frac{K}{2}\theta _{m}$.
The absolute amplitude value $\left\vert \sin \frac{K}{2}\theta
_{m}\right\vert $ may act as the measure of the quantum-state difference
between the final two candidate states $\{|\Psi _{\pm 1}^{K}\rangle \}$ of
the QUANSDAM\ process of (3.12b). When the number $K$ increases from $0$ to
its maximum value $K=\frac{1}{2}\pi /\left\vert \theta _{m}\right\vert $,
the amplitude value grows from the initial value $0$ to the final value $1/%
\sqrt{2},$ leading to that both the final states $\{|\Psi _{\pm
1}^{K}\rangle \}$ evolve from the identical initial states to the orthogonal
states.

For a general amplitude-based QUANSDAM\ process of (3.9) the situation may
not be so simple as the above one of (3.12b). Suppose that at the end of the
QUANSDAM process of (3.9) the state may be formally written as%
\begin{equation}
\Psi _{a_{m}^{s}}\left( \theta _{m},K\right) =\Psi _{a}\left( \theta
_{m},K\right) +a_{m}^{s}\Psi _{b}\left( \theta _{m},K\right)  \tag{3.17}
\end{equation}%
where $a_{m}^{s}\Psi _{b}\left( \theta _{m},K\right) $ is the $IC$ state.
The final state $\Psi _{a_{m}^{s}}\left( \theta _{m},K\right) $ is
normalized, but the $IC$ state may not. Then the amplitude of the $IC$ state
is just $\left\vert \left\vert \Psi _{b}\left( \theta _{m},K\right)
\right\vert \right\vert $ in magnitude. The normalization condition for the
final state $\Psi _{a_{m}^{s}}\left( \theta _{m},K\right) $ is given by $%
\left\vert \left\vert \Psi _{a_{m}^{s}}\left( \theta _{m},K\right)
\right\vert \right\vert ^{2}=1.$ By using (3.17) it is reduced to the form 
\begin{equation}
\left\vert \left\vert \Psi _{a}\left( \theta _{m},K\right) \right\vert
\right\vert ^{2}+\left\vert \left\vert \Psi _{b}\left( \theta _{m},K\right)
\right\vert \right\vert ^{2}=1  \tag{3.18a}
\end{equation}%
\begin{equation}
\Psi _{a}^{+}\left( \theta _{m},K\right) \Psi _{b}\left( \theta
_{m},K\right) +\Psi _{b}^{+}\left( \theta _{m},K\right) \Psi _{a}\left(
\theta _{m},K\right) =0  \tag{3.18b}
\end{equation}%
On the other hand, the orthogonal condition for the final two candidate
states $|\Psi _{+1}^{K}\rangle $ and $|\Psi _{-1}^{K}\rangle $ in (3.9) is
given by $\langle \Psi _{+1}^{K}|\Psi _{-1}^{K}\rangle =0.$ Now from (3.17)\
one obtains $|\Psi _{+1}^{K}\rangle =\Psi _{a}\left( \theta _{m},K\right)
+\Psi _{b}\left( \theta _{m},K\right) $ and $|\Psi _{-1}^{K}\rangle =\Psi
_{a}\left( \theta _{m},K\right) -\Psi _{b}\left( \theta _{m},K\right) .$
Thus, the orthogonal condition may be reduced to the form%
\begin{equation}
\langle \Psi _{+1}^{K}|\Psi _{-1}^{K}\rangle =1-2\left\vert \left\vert \Psi
_{b}\left( \theta _{m},K\right) \right\vert \right\vert ^{2}-2\Psi
_{a}^{+}\left( \theta _{m},K\right) \Psi _{b}\left( \theta _{m},K\right) =0 
\tag{3.19}
\end{equation}%
where the normalization condition (3.18) is already used. It is clearly
dependent on not only the absolute amplitude value $\left\vert \left\vert
\Psi _{b}\left( \theta _{m},K\right) \right\vert \right\vert $ but also\ the
scalar product $\Psi _{a}^{+}\left( \theta _{m},K\right) \Psi _{b}\left(
\theta _{m},K\right) .$ The orthogonal condition (3.19) is universal even
for a general QUANSDAM process. It could be helpful to determine the minimum
number $K$ of a general amplitude-based QUANSDAM\ process of (3.9).\newline
\newline
{\large 4. The phase-based QUANSDAM process and the information-carrying\
unitary propagators}

It was first proposed in Ref. [7] that a UNIDYSLOCK\ process could be
constructed in a single-atom quantum system which is a time- and
space-dependent quantum system. The basic theoretical method employed to
construct a UNIDYSLOCK\ process and its inverse in a single-atom system is
the unitary manipulation of a single atom in time and space. Since then, the
unitary manipulation of a single-atom system in time and space has been
insistently studied in theory in the past decade [7, 31, 19, 15]. The main
purpose for these works is to realize an exponential QUANSDAM (or
UNIDYSLOCK) process in a single-atom system and investigate the
quantum-computing speedup mechanism for the unitary quantum dynamics. The
work in this section continues along the same research direction and toward
the same goal. These works and the present work lay down the basis for the
coming exponential QUANSDAM process in a single-atom system.

Since they were first chosen in Ref. [7], Gaussian wave-packet states and
quadratic unitary propagators [7, 31, 19, 15] have been used as the basic
quantum states and the basic unitary propagators in the unitary manipulation
of a single atom in time and space, respectively. That a Gaussian
wave-packet state was used in the unitary manipulation in the early time [7]
is closely related to the construction of a spatially selective and
internal-state selective triggering pulse, with which the author attempted
to realize a reversible and unitary halting protocol in a single-atom system
(i.e., the so-called halting-qubit atom), and also related to the
realization of a UNIDYSLOCK process. The spatially selective and
internal-state selective triggering pulse [7, 31, 15] needs a narrow
wave-packet motional state of the single-atom system in one-dimensional
coordinate space. Here only the (internal-state-dependence) spatially
selective excitation of the triggering pulse requires the atomic wave-packet
motional state to have a narrow spatial wave-packet spread. It is therefore
important to choose a narrower atomic wave-packet motional state to realize
the spatially selective and internal-state selective triggering pulse and
study a UNIDYSLOCK (or QUANSDAM) process (See also Appendix A for the
relation between the spatially selective and internal-state selective
triggering pulse and the phase-based QUANSDAM\ process).

Initially the present author chose a Gaussian (shaped) wave-packet motional
state mainly based on the author's experience of experimental research on
the NMR application [32] of the Gaussian shaped pulses. It was shown in the
selective excitation experiments [32] that a Gaussian shaped pulse, which is
a radio-frequency electromagnetic wave pulse [23], has a narrower waveform
width ($\Delta t$) in the time domain (i.e., a shorter pulse width) than
other simple or complex (examined) shaped pulses under the same excitation
bandwidth ($\Delta \omega =\Delta E/\hslash $) in the frequency (or energy)
domain. Thus, it was thought by the author that a Gaussian (shaped)
wave-packet motional state of the single-atom system should have a narrower
spatial wave-packet spread ($\Delta x$) than other (shaped) wave-packet
motional states under the same conditions which include the same momentum
spread ($\Delta p$). That a Gaussian wave-packet motional state was chosen
as the basic quantum state to study a spatially selective and internal-state
selective triggering pulse and a UNIDYSLOCK\ process [7] is merely a correct
and important starting point. The theoretical basis and the experimental
basis for a Gaussian wave-packet state to be the basic quantum state and a
quadratic unitary propagator to be the basic unitary propagator mainly come
from the three aspects [15]. On the first aspect it has been shown in
quantum physics that a Gaussian wave-packet state can keep its Gaussian
shape unchanged under action of a quadratic unitary propagator (See, for
example, Refs. [21, 24]), while a quadratic unitary propagator can be
calculated exactly and conveniently by the Feynman path integral [18]. On
the second aspect in quantum physics the time evolution process of a
Gaussian wave-packet state can be calculated exactly and conveniently under
action of a quadratic unitary propagator. Here a Gaussian wave-packet state
may be a standard one or a non-standard one (See Ref. [15]). On the third
aspect in the coherent manipulation and control of a single-atom system
(mainly in experiment) a Gaussian wave-packet motional state has been
studied (See, for example, Refs. [34]); and a single atom in an external
harmonic potential field can be perfectly manipulated or controlled
experimentally (See Ref. [30b]). Gaussian wave-packet motional states [7,
31, 19, 15] continue to play a key and basic role in realizing the coming
exponential QUANSDAM process in a single-atom system [36].

A single atom is one of the simplest quantum systems that involve the
center-of-mass (COM) motion (or external motion) and the internal motion of
quantum system at the same time. It may have four degrees of freedom that
can be unitarily manipulated independently, one of which is the internal
motion degree of freedom and the others are the three independent COM motion
degrees of freedom in three-dimensional coordinate space. In particular, the
atomic COM motion is allowed to be manipulated unitarily in an arbitrary
manner. Moreover, both the atomic COM motion and internal motion can
interact with each other, when the external electromagnetic field is
appropriately applied to the single-atom system. Therefore, a complete
description for a single-atom system must consider both the atomic COM
motion and internal motion as well as the interaction between the two
motions in the study of the unitary manipulation of a single-atom system in
time and space. On the other hand, unlike a single atom motioning in
coordinate space each qubit of an $n-$qubit quantum system need not consider
explicitly its COM motion in quantum computation, even if the COM motion (or
external motion) could exist, since a quantum computation in an $n-$qubit
quantum system usually does not explicitly depend on the external motion of
each qubit of the quantum system. Therefore, in a usual case only the
internal motions of the qubits and the interaction between the internal
motions in the $n-$qubit quantum system need to be considered explicitly in
quantum computation.

Unlike a composite quantum system, a single quantum system such as a
single-atom system does not own the fundamental quantum-computing resource.
In the last section it is shown in theory that a single quantum system could
be appropriate to realize an exponential QUANSDAM (or UNIDYSLOCK) process,
when the basic building block of the QUANSDAM (or UNIDYSLOCK) process, i.e.,
the basic $IC$ unitary operator prepared by the search-space dynamical
reduction works in a smallest reduction subspace. Because the basic $IC$\
unitary operator works in a smallest reduction subspace, the effect of the
symmetric structure of the Hilbert space of the quantum system could become
secondary on the QUANSDAM (or UNIDYSLOCK) process which is performed in the
quantum system. Consequently one could choose an appropriate single quantum
system such as a single-atom system to realize an exponential QUANSDAM (or
UNIDYSLOCK) process. A single quantum system is simpler and easier to
manipulate and control unitarily.

As known in the section 2, the basic $IC$ unitary operator of (2.9) works in
the smallest reduction subspace of the math Hilbert space of the search
problem, i.e., a two-dimensional reduction subspace. Then in performance a
QUANSDAM (or UNIDYSLOCK) process whose basic building block is the basic $IC$
unitary operator could not be affected significantly by the symmetric
structure of the Hilbert space of the quantum system that performs the
QUANSDAM (or UNIDYSLOCK) process. Therefore, such QUANSDAM process even with
an exponential $QSD$ amplification ability could be realized in a
single-atom system. Below in principle it is described how an exponential
QUANSDAM process could be realized (mainly but no limited to) in a
single-atom system.

It is known from the basic $IC$ unitary operator of (2.9) that the $m-$th
spin (or qubit) of the $n-$qubit quantum system carries the information ($%
a_{m}^{s}$) of the $m-$th component state of the solution state. When a
QUANSDAM (or UNIDYSLOCK)\ process is constructed in a single-atom system, as
one key step of the construction the solution information ($a_{m}^{s}$)
needs first to be transferred in a unitary form from the $m-$th spin (or $m-$%
th qubit) of the $n-$qubit quantum system to the single-atom system and the
important is that the solution information is transferred to the COM motion
(degrees of freedom) of the single-atom system. The best way to realizing
this information transfer is perhaps that the solution information is first
transferred from the $m-$th spin to the atomic internal motion and then it
is further transferred from the atomic internal motion to the atomic COM
motion. When the solution information is transferred from the atomic
internal motion to the atomic COM motion, one needs to employ the
interaction between the internal motion and the COM motion of the
single-atom system to realize such information transfer. Such an interaction
tends to be internal-state selective (or dependent) in a\ single-atom
system. It can be easily realized in a single-atom system. It is well known
that the interaction between the internal and COM motions of a single atom
has been used extensively in the atomic laser light cooling [29], the atomic
quantum coherence interference [29a], the atomic quantum-computing
implementation [30], and the coherent manipulation and control of a
single-atom system [30b].

The interaction between the internal motion and the COM motion of a
single-atom system also plays an important role in realizing conveniently
the selective excitation of the COM energy eigenstates in the single-atom
system. When a single atom is in an external harmonic potential field, the
total atomic motion consists of the atomic internal motion and external COM\
harmonic motion. For simplicity here consider that the single-atom system is
in one-dimensional COM harmonic motion. On the one hand, according to the
standard one-dimensional harmonic-oscillator quantum theory [14] the atomic
COM harmonic motion is discrete and has equal-spacing energy levels. Such an
equal-spacing energy level structure tends to be an obstacle to realizing
conveniently the selective excitation of the COM\ energy states of the
single-atom system. On the other hand, the internal energy level structure
of a single atom is rich. The atomic internal energy levels are generally
discrete and not equal-spacing. For a single atom in an external harmonic
potential field, its total energy consists of the atomic internal energy and
external energy of the COM harmonic motion. The atomic energy level
structure therefore is quite different from the simple equal-spacing energy
level structure. Then the selective excitation of the atomic COM energy
states may be performed more conveniently with the help of the atomic
internal energy levels. Such a selective excitation is involved in both the
atomic internal motion and external COM harmonic motion. The interaction
between the atomic internal motion and external COM harmonic motion
therefore is necessary to realize the selective excitation. On the basis of
the interaction these well-established selective excitation techniques could
be used to realize it, which include the selective excitation techniques in
NMR spectroscopy [23] and the STIRAP technique [33] and so on.\ 

After the solution information ($a_{m}^{s}$) is transferred to the internal
motion of the single-atom system, the basic $IC$ unitary operator of (2.9)
originally working in the two-dimensional subspace of the $m-th$ qubit of
the $n-$qubit quantum system is changed to the one working in a
two-dimensional internal-state subspace of the Hilbert space of the atomic
internal motion. For convenience hereafter this new basic $IC$ unitary
operator of the single-atom system is still denoted as the same form as
(2.9), i.e., $U_{\lambda }^{ic}(\theta _{m},a_{m}^{s})=\exp
(-ia_{m}^{s}\theta _{m}I_{m\lambda }),$ where $I_{m\lambda }$ is a spin$-1/2$
(or pseudospin$-1/2$) operator of the atomic internal motion. Now by
starting from this basic $IC$\ unitary operator of the single-atom system
one may construct the desired QUANSDAM\ (or UNIDYSLOCK) process in the
single-atom system.

As pointed out in the preceding section, there are two different types of
QUANSDAM (or UNIDYSLOCK) processes. They are the amplitude-based and the
phase-based QUANSDAM (or UNIDYSLOCK) processes, respectively. A phase-based
QUANSDAM process may be quite different from an amplitude-based counterpart.
It works generally in a time- and space-dependent quantum system such as a
single-atom system, while an amplitude-based QUANSDAM process and the
amplitude amplification of a conventional quantum search algorithm [3, 13]
usually work in an $n-$qubit quantum system. As shown in the preceding
section, it is relatively simple to construct an amplitude-based QUANSDAM
process like (3.9). This is a bottom-to-top method to design a QUANSDAM
process. Suppose that in (3.9) the time evolution process for every $QM$\
unitary operator acting on the quantum system can be calculated exactly.
Then the amplitude-based QUANSDAM process of (3.9) may be calculated
exactly. However, the two final candidate states of (3.9) can be obtained
only after the whole QUANSDAM process of (3.9) is calculated exactly.
Therefore, it is not easy to find the two candidate orthogonal states at the
end of the amplitude-based QUANSDAM\ process with many $QM$ unitary
operators. In contrast, as shown below, the two candidate orthogonal states
at the end of a phase-based QUANSDAM\ process may be determined easily, but
it is an uneasy task how to construct efficiently the phase-based QUANSDAM\
process itself. This is a top-to-bottom method to design a QUANSDAM\ process.

Below it is described in detail how a phase-based QUANSDAM process works in
a single-atom system. First of all, from the point of view of quantum
mechanics a simple introduction is given to the momentum eigenfunctions and
their orthogonal relations for a free particle such as a single atom
motioning freely in one-dimensional coordinate space. It is well-known in
quantum mechanics [14] that a momentum eigenfunction of a free particle may
take the box normalization form or the Dirac $\delta -$function
normalization form. The former is a discrete form, while the latter is a
continuous form. Below the box normalization momentum eigenfunction is
mainly used. Now consider a momentum eigenfunction of the particle in
one-dimensional box with an arbitrary large length $L$ centered at the
origin, here the eigenfunction satisfies the periodic boundary condition.
This momentum eigenfunction in the box normalization form may be written as
[14]%
\begin{equation}
|\Psi _{k}\rangle =\frac{1}{\sqrt{L}}\exp (ip_{k}x/\hslash )  \tag{4.1}
\end{equation}%
where the discrete momentum eigenvalue $p_{k}=2\pi \hslash k/L$ with quantum
number $k=0,$ $\pm 1,$ $\pm 2,....$ It satisfies the orthonormal condition:%
\begin{equation}
\langle \Psi _{k}|\Psi _{l}\rangle =\delta _{kl}  \tag{4.2}
\end{equation}%
where the $\delta -$function $\delta _{kl}$ is given by $\delta _{kl}=0$ if $%
k\neq l$ and $\delta _{kl}=1\ $if $k=l$. The relation (4.2) shows that both
the momentum eigenfunctions $|\Psi _{k}\rangle $ and $|\Psi _{l}\rangle $
are orthogonal to each other when both the quantum numbers $k$ and $l$ are
different, i.e., $k\neq l.$ Moreover, when the box length $L\rightarrow
\infty ,$ any pair of momentum eigenfunctions are orthogonal to each other
no matter how small their momentum difference $p_{k}-p_{l}$ is as long as
the momentum difference does not equal zero.

Now suppose that an $IC$ unitary operator is taken as $\exp
(-ia_{m}^{s}p_{0}^{\prime }x/\hslash )$ for a single atom in the box with
one-dimensional coordinate space $\left( -L/2,L/2\right) $. This $IC$
unitary operator is called the $IC$ unitary momentum-displacement propagator 
\footnote{%
That $\exp \left( -ip_{0}x/\hslash \right) $ is called the unitary
momentum-displacement propagator is due to the unitary transformation $\exp
\left( ip_{0}x/\hslash \right) p\exp \left( -ip_{0}x/\hslash \right)
=p-p_{0}.$}. It is spatially dependent. It is quite different from the basic 
$IC$ unitary operator of (2.9) of the single-atom system. But it is
generally hard to prepare the $IC$ unitary momentum-displacement propagator
by using the basic $IC$ unitary operators and the $QM$ unitary operators. It
may be more direct and reasonable for the single-atom system to prepare the $%
IC$ unitary operator: 
\begin{equation}
U_{p}^{ic}(a_{m}^{s})=\exp (-ia_{m}^{s}p_{0}^{\prime }xS_{z}/\hslash ) 
\tag{4.3}
\end{equation}%
where $S_{z}$ is the $z-$component spin operator of the atomic internal
motion. This $IC$\ unitary operator may be called the
internal-motion-dependent $IC$ unitary momentum-displacement propagator.
Unlike the $IC$ unitary operator $\exp (-ia_{m}^{s}p_{0}^{\prime }x/\hslash
) $ it is able to reflect the interaction between the internal motion and
the COM motion of the single-atom system. As shown in Appendix A and also in
Refs. [15, 31], it could be generated approximately by using the basic $IC$
unitary operators and the suitable $QM$ unitary operators. However, so far
there is not a rigorous mathematical proof to show that the approximately
generated $IC$ unitary propagator $U_{p}^{ic}(a_{m}^{s})$ can lead to an
exponential QUANSDAM process.

Just like the basic $IC$\ unitary operator, the $IC$ unitary propagator $%
U_{p}^{ic}(a_{m}^{s})$ also carries the solution information ($a_{m}^{s}$).
Suppose that the single atom is in the product state $|\Psi _{k}\rangle
|0\rangle ,$ where $|0\rangle $ and $|\Psi _{k}\rangle $ are the atomic
internal ground state and COM momentum eigenfunction given by (4.1),
respectively. When $U_{p}^{ic}(a_{m}^{s})$ acts on the atomic product state $%
|\Psi _{k}\rangle |0\rangle $, one obtains the QUANSDAM process: 
\begin{equation}
U_{p}^{ic}(a_{m}^{s})|\Psi _{k}\rangle |0\rangle =\frac{1}{\sqrt{L}}\exp
(i\left( p_{k}-a_{m}^{s}p_{0}^{\prime }m_{z}\right) x/\hslash )|0\rangle 
\tag{4.4a}
\end{equation}%
where the eigen-equation $S_{z}|0\rangle =m_{z}|0\rangle $ with the
eigenvalue $m_{z}\neq 0$ is already used. There are two candidate momentum
wavefunctions $|\psi _{\pm 1}\rangle $ at the end of the QUANSDAM process.
These two wavefunctions may be explicitly obtained by rewriting (4.4a) as%
\begin{equation}
U_{p}^{ic}(a_{m}^{s})|\Psi _{k}\rangle |0\rangle =\left\{ 
\begin{array}{c}
|\psi _{+1}\rangle |0\rangle =\frac{1}{\sqrt{L}}\exp (i\left(
p_{k}-m_{z}p_{0}^{\prime }\right) x/\hslash )|0\rangle ,\text{ }if\text{ }%
a_{m}^{s}=+1 \\ 
|\psi _{-1}\rangle |0\rangle =\frac{1}{\sqrt{L}}\exp (i\left(
p_{k}+m_{z}p_{0}^{\prime }\right) x/\hslash )|0\rangle ,\text{ }if\text{ }%
a_{m}^{s}=-1%
\end{array}%
\right.  \tag{4.4b}
\end{equation}%
Here the important point is that only one of the two candidate states $%
\{|\psi _{+1}\rangle |0\rangle ,$ $|\psi _{-1}\rangle |0\rangle \}$ appears
in the single-atom system, while another does not. When the box length $%
L\rightarrow \infty ,$ both the wavefunctions $\{|\psi _{\pm 1}\rangle \}$
are the momentum eigenfunctions $\{\frac{1}{\sqrt{L}}\exp (i\left( p_{k}\mp
m_{z}p_{0}^{\prime }\right) x/\hslash )\},$ respectively. Thus, both the
momentum wavefunction $|\psi _{+1}\rangle $ with the momentum eigenvalue $%
\left( p_{k}-m_{z}p_{0}^{\prime }\right) $ and $|\psi _{-}\rangle $ with the
eigenvalue $\left( p_{k}+m_{z}p_{0}^{\prime }\right) $ are orthogonal to
each other no matter how small the momentum value $p_{0}^{\prime }$ is as
long as $p_{0}^{\prime }$ is not zero, indicating that the QUANSDAM process
of (4.4) changes the initial two identical COM momentum wavefunctions ($%
|\Psi _{k}\rangle $) to a pair of orthogonal wavefunctions $\{|\psi
_{+1}\rangle ,|\psi _{-1}\rangle \},$ respectively. For a finite box length $%
L,$ when $m_{z}p_{0}^{\prime }=p_{l}=2\pi \hslash l/L$ ($l\neq 0$ is a
quantum number), one knows from (4.1) and (4.4) that both the wavefunctions $%
\{|\psi _{+1}\rangle ,|\psi _{-1}\rangle \}$ are also the momentum
eigenfunctions. Then according to (4.2) they are also orthogonal to one
another, indicating that the QUANSDAM process also changes the initial two
identical states to a pair of orthogonal states, respectively. Here both the
final states $\{|\psi _{+1}\rangle |0\rangle ,|\psi _{-1}\rangle |0\rangle
\} $ have the same amplitude as the one of the initial state ($|\Psi
_{k}\rangle |0\rangle $) and keep their amplitudes unchanged during the
whole QUANSDAM process. What these states are changed by the QUANSDAM
process is only their spatially-dependent phases (or momentum eigenvalues).
Therefore, such a QUANSDAM process is called the phase-based QUANSDAM
process.

In principle, by quantum measurement one is able to distinguish the two
momentum eigenfunctions $|\psi _{+1}\rangle $ and $|\psi _{-1}\rangle $ in
(4.4b) from one another, because both the eigenfunctions are orthogonal to
one another even if the momentum value $p_{0}^{\prime }$ takes an
arbitrarily small nonzero value. Here there is no requirement that one have
to measure directly the momentum eigenvalue $p_{k}+m_{z}p_{0}^{\prime }$ (or 
$p_{k}-m_{z}p_{0}^{\prime })\ $so as to identify any one of the two
wavefunctions $|\psi _{+1}\rangle $ and $|\psi _{-1}\rangle $. Actually,
both the candidate momentum eigenstates $|\psi _{+1}\rangle |0\rangle $ and $%
|\psi _{-1}\rangle |0\rangle $ could be first transferred to the two
(candidate) internal energy eigenstates $|0\rangle |\varphi _{1}\rangle $
and $|0\rangle |\varphi _{2}\rangle $ of the single-atom system,
respectively, and then by quantum measuring the atomic internal energy one
is able to distinguish the two wavefunctions $|\psi _{+1}\rangle $ and $%
|\psi _{-1}\rangle $ from one another. Here $|\varphi _{1}\rangle $ and $%
|\varphi _{2}\rangle $ are the energy eigenstates of the atomic internal
motion. That the two wavefunctions can be distinguished from each other
unambiguously is only because the two wavefunctions are orthogonal to each
other. It has nothing to do with the quantum measurement of momentum
eigenvalue. Thus, rather than the quantum measurement, the phase-based
QUANSDAM process really leads to that the two wavefunctions can be
distinguished from each other unambiguously. Readers should be familiar with
the fact that when two non-orthogonal states are sufficiently close to one
another, it is hard to use any quantum measurement to distinguish them from
one another no matter which observable is measured.

Clearly, a phase-based \ QUANSDAM \ process is \ quite \ different from an
amplitude-based QUANSDAM process and the amplitude amplification of a
conventional quantum search algorithm. Like the amplitude amplification of a
conventional quantum search algorithm, an amplitude-based QUANSDAM\ process
could be affected greatly by the square speedup limit. This is a severe
shortcoming for an amplitude-based QUANSDAM\ process. It results in that it
is not easy to find an exponential amplitude-based QUANSDAM\ process. A
phase-based QUANSDAM process seems to avoid this square-speedup-limit
shortcoming. Therefore, it seems to capture the essential aspect of an
exponential QUANSDAM\ process. Apparently its potential to realize an
exponential QUANSDAM process and an exponential quantum-computing speedup
seems to be infinite.

The power of a phase-based QUANSDAM process to amplify the quantum-state
difference seems to be unlimited and even infinitely large in appearance.
For example, the phase-based QUANSDAM\ process of (4.4) consisting of only
one $IC$ unitary propagator $U_{p}^{ic}(a_{m}^{s})$ can change the same
initial states to a pair of orthogonal momentum eigenstates, respectively.
This, of course, concludes merely\ from the principle of a phase-based
QUANSDAM process in ideal conditions. In the phase-based QUANSDAM process
one does not consider how many basic $IC$ unitary operators are consumed to
prepare the ideal $IC$ unitary propagator $U_{p}^{ic}(a_{m}^{s})$ and how to
prepare the ideal initial momentum eigenfunction. One does not yet take into
account energy, space, and the quantum-computing resource, which could
probably be used by the QUANSDAM process. For example, in practice it is
impossible to prepare an ideal momentum eigenfunction which spreads over the
whole coordinate space $(-\infty ,+\infty )$. Actually the phase-based
QUANSDAM process of (4.4) has not yet touched the hard problem: how to
construct efficiently the phase-based QUANSDAM process itself.

The $IC$ unitary propagator $U_{p}^{ic}(a_{m}^{s})$ of (4.3) may be
considered as the special one of more general $IC$ unitary propagators in a
time- and space-dependent quantum system. The latter will be described in
detail below. In practice the simple QUANSDAM processes of (3.4) and (3.12)
are not useful and even the amplitude-based QUANSDAM process of (3.9) is
suspected to be useful. The phase-based QUANSDAM process of (4.4) seems to
be quite simple and extremely powerful, but in practice it is hard to
realize. These QUANSDAM processes need to be further developed or
generalized so that in practice they could be useful in a variety of quantum
systems. One of the possible strategies to generalize these QUANSDAM
processes is that the solution information ($a_{m}^{s}$) is loaded onto the
time evolution propagator of a quantum system and then the generated $IC$
unitary propagator acts as the basic building block to construct the desired
QUANSDAM process.

One simple method to load the solution information ($a_{m}^{s}$) onto the
unitary propagator of a quantum system with time-independent Hamiltonian $H$
is that (1) the information-carrying unitary propagator is first written
formally in the simple form%
\begin{equation}
U_{H}^{ic}(t_{m},a_{m}^{s})=\exp \left( -ia_{m}^{s}Ht_{m}/\hslash \right) 
\tag{4.5}
\end{equation}%
where $t_{m}$ is a real parameter and it is usually limited to be $%
t_{m}\thickapprox \theta _{m}$; (2) the $IC$ unitary propagator $%
U_{H}^{ic}(t_{m},a_{m}^{s})$ then is constructed by using the basic $IC$
unitary operators and the $QM$ unitary operators of the quantum system. Here
a quantum system with time-independent Hamiltonian is called a
time-independent quantum system. It can be found that the
internal-motion-dependent $IC$ unitary momentum-displacement propagator $%
U_{p}^{ic}(a_{m}^{s})$ of (4.3) is the special case of the $IC$ unitary
propagator $U_{H}^{ic}(t_{m},a_{m}^{s}),$ where $H=xS_{z}$ and $%
t_{m}=p_{0}^{\prime }.$ The basic $IC$ unitary operator $U_{\lambda
}^{ic}(\theta _{m},a_{m}^{s})$ of (2.9) is also the special case of the $IC$
unitary propagator $U_{H}^{ic}(t_{m},a_{m}^{s}),$ where $H=I_{m\lambda }$
and $t_{m}/\hslash =\theta _{m}.$ Therefore, the $IC$ unitary propagator $%
U_{H}^{ic}(t_{m},a_{m}^{s})$ is naturally a generalization of these $IC$
unitary operators $U_{p}^{ic}(a_{m}^{s})$ and $U_{\lambda }^{ic}(\theta
_{m},a_{m}^{s})$. There may be a variety of quantum-mechanical methods to
construct the $IC$ unitary propagator of (4.5). They may include the
eigenfunction expansion [17], the Green function method [14] and the Feynman
path integral [18], the Trotter-Suzuki method [16], and the multiple-quantum
operator algebra space [12, 22], just to name a few. But, for simplicity,
below it is described only how the energy eigenfunction expansion is used to
construct the $IC$ unitary propagator and to realize the
solution-information propagation process (See below).

Quantum mechanics provides many different methods to deal with the time
evolution process of a quantum system. Two basic methods among them that
have been used extensively are the eigenfunction expansion in the Hilbert
space of a quantum system [17] and the Green function method [14, 18]. The
eigenfunction expansion is considered as a well-established mathematical
(but not a physical) principle in quantum mechanics [17, 14]. It has been
used extensively to deal with the time evolution process of a quantum system
and especially it is a convenient method to treat the time evolution process
of a time-independent quantum system\ which may or may not space-dependent.
The energy eigenfunction expansion is the special case of the eigenfunction
expansion. Below it is used to realize the loading of the solution
information ($a_{m}^{s}$) onto the unitary propagator of a time-independent
quantum system, that is, it is used to construct the $IC$ unitary propagator
of (4.5). On the other hand, the Green function method is more convenient to
deal with the time evolution process of a time- and space-dependent quantum
system. Both the methods have important applications in the unitary
manipulation of a single atom in time and space (See Ref. [19] for the
eigenfunction expansion and Refs. [31, 15, 19] for the Green function method
and the Feynman path integral).

First of all, the eigenfunction expansion principle is simply introduced.
For simplicity, consider a time-independent and space-dependent quantum
system. Since its Hamiltonian $H$ is time-independent, its time evolution
propagator may be generally written as $U(t)=\exp \left( -iHt/\hslash
\right) .$ Suppose that $\{u_{k}\left( \mathbf{r}\right) \}$\ is the
complete orthonormal set of energy eigenfunctions of the Hilbert space of
the quantum system. Then the energy-eigenvalue equation is $Hu_{k}\left( 
\mathbf{r}\right) =E_{k}u_{k}\left( \mathbf{r}\right) ,$ where $E_{k}$ is
the energy eigenvalue associated with the energy eigenfunction $u_{k}\left( 
\mathbf{r}\right) $. According to the eigenfunction expansion principle [17,
14] an arbitrary quantum state $\Psi \left( \mathbf{r,}t_{0}\right) $ of the
quantum system can be expanded in terms of the energy eigenfunctions $%
\{u_{k}(\mathbf{r})\}:$%
\begin{equation}
\Psi \left( \mathbf{r,}t_{0}\right) =\sum_{k=0}^{\infty }A_{k}u_{k}(\mathbf{r%
})  \tag{4.6}
\end{equation}%
where $A_{k}=\langle u_{k}|\Psi \left( t_{0}\right) \rangle $ is an
expansion coefficient. Generally the number of expansion terms on the RH
side of (4.6) is equal to the number of the energy eigenfunctions in the
complete set $\{u_{k}\left( \mathbf{r}\right) \}$ or the dimensional size of
the Hilbert space of the quantum system. It is generally infinite for a
space-dependent quantum system such as a single atom motioning in
one-dimensional coordinate space. However, the number of expansion terms on
the RH side of (4.6) becomes finite if the number of the energy
eigenfunctions is finite in the complete set $\{u_{k}\left( \mathbf{r}%
\right) \}$. There are a lot of quantum systems for which the number of
expansion terms is finite. They tend to be space-independent. One typical
example is a multiple-spin$-1/2$ quantum system. An $n-$qubit quantum system
also may be considered as one of these quantum systems, which has a $2^{n}-$%
dimensional Hilbert space. Below this type of quantum systems are treated as
the special cases and will not be explicitly considered unless stated
otherwise.

The expansion series on the RH side of (4.6) is always convergent no matter
that its number of expansion terms is finite or infinite [17]. Now with the
help of the energy-eigenvalue equation $Hu_{k}\left( \mathbf{r}\right)
=E_{k}u_{k}\left( \mathbf{r}\right) $ it follows from (4.6) that the time
evolution process of the quantum system with arbitrary initial state $\Psi
\left( \mathbf{r,}t_{0}\right) $ may be expressed as [17, 14] 
\begin{equation}
\Psi \left( \mathbf{r,}t\right) =U(t)\Psi \left( \mathbf{r,}t_{0}\right)
=\sum_{k=0}^{\infty }A_{k}\exp \left( -iE_{k}t/\hslash \right) u_{k}(\mathbf{%
r}).  \tag{4.7}
\end{equation}%
Here$\ $without losing generality the initial time $t_{0}$ is set to $%
t_{0}=0 $ for the time-independent quantum system. This is the basic
mathematical formula for the energy eigenfunction expansion to calculate and
realize the time evolution process of a time-independent quantum system. Its
application to the unitary manipulation of a single atom with a Gaussian
wave-packet state in time and space may be seen in Ref. [19].

Now the basic equation of (4.7) is employed to load the solution information
($a_{m}^{s}$) onto the unitary propagator of a time-independent quantum
system. By applying the $IC$ unitary propagator $U_{H}^{ic}(t_{m},a_{m}^{s})$
of (4.5) to an arbitrary state $\Psi \left( \mathbf{r,}t_{0}\right) $ of the
quantum system which is given by the expansion (4.6) one obtains from the
basic equation of (4.7) the solution-information propagation process:%
\begin{equation}
U_{H}^{ic}(t_{m},a_{m}^{s})\Psi \left( \mathbf{r,}t_{0}\right)
=\sum_{k=0}^{\infty }A_{k}\exp \left( -ia_{m}^{s}E_{k}t_{m}/\hslash \right)
u_{k}(\mathbf{r}).  \tag{4.8}
\end{equation}%
Obviously, this solution-information propagation process is also a QUANSDAM
(or UNIDYSLOCK) process. Then a unitary sequence $USEQ\left( K\right) $ that
consists of the basic $IC$ unitary operators $\{U_{\lambda _{l}}^{ic}(\theta
_{m}^{l},a_{m}^{s})\}$ and the $QM$ unitary operators $\{V_{l}\}$ is
constructed,%
\begin{equation}
USEQ\left( K\right) =V_{K}U_{\lambda _{K}}^{ic}(\theta
_{m}^{K},a_{m}^{s})V_{K-1}...V_{1}U_{\lambda _{1}}^{ic}(\theta
_{m}^{1},a_{m}^{s})V_{0},  \tag{4.9}
\end{equation}%
such that when $USEQ\left( K\right) $ is applied to the same state $\Psi
\left( \mathbf{r,}t_{0}\right) ,$ one has%
\begin{equation}
USEQ\left( K\right) \Psi \left( \mathbf{r,}t_{0}\right) =\sum_{k=0}^{\infty
}A_{k}\exp \left( -ia_{m}^{s}E_{k}t_{m}/\hslash \right) u_{k}(\mathbf{r}). 
\tag{4.10}
\end{equation}%
Note that the state $\Psi \left( \mathbf{r,}t_{0}\right) $ is arbitrary. By
comparing (4.10) with (4.8) one finds that $U_{H}^{ic}(t_{m},a_{m}^{s})=USEQ%
\left( K\right) $. The equations (4.5)--(4.10) form the basis for the energy
eigenfunction expansion to construct the $IC$ unitary propagator of (4.5) in
a time-independent quantum system which may or may not be space-dependent.

As a special case, for an $n-$qubit quantum system the number of expansion
terms is finite on the RH side of (4.8). Then it is possible to construct a
unitary sequence $USEQ\left( K\right) $ of (4.9) with a finite number $K$
such that $U_{H}^{ic}(t_{m},a_{m}^{s})=USEQ\left( K\right) .$ For a
single-atom system which is a space-dependent quantum system the number of
expansion terms in (4.8) is generally infinite. Then it is generally hard to
generate a unitary sequence of (4.9) such that $%
U_{H}^{ic}(t_{m},a_{m}^{s})=USEQ\left( K\right) $ exactly, but it is still
possible to construct a unitary sequence $USEQ\left( K\right) $ with\ a
finite number $K$ such that $USEQ\left( K\right) $ is equal to $%
U_{H}^{ic}(t_{m},a_{m}^{s})$ approximately. Here the discussion is not
involved in the computational complexity of the unitary sequence $USEQ\left(
K\right) $. However, if the $IC$ unitary propagator $%
U_{H}^{ic}(t_{m},a_{m}^{s})$ is used as the basic building block of a
QUANSDAM\ process, then one must consider the computational complexity of
the unitary sequence $USEQ\left( K\right) $ that generates $%
U_{H}^{ic}(t_{m},a_{m}^{s})$ exactly or approximately. From the point of
view of the computational complexity it is required that the $IC$ unitary
propagator $U_{H}^{ic}(t_{m},a_{m}^{s})$ be generated by the unitary
sequence $USEQ\left( K\right) $ of (4.9) with a minimum number $K$ of the
basic $IC$ unitary operators and moreover, the number $K$ be polynomially
large at most.

Because the RH side of (4.8) is an infinite series, generally the $IC$
unitary propagator $U_{H}^{ic}(t_{m},a_{m}^{s})$ could only be constructed
approximately for a space-dependent quantum system. Note that the infinite
series of (4.8) also is a QUANSDAM process. As far as the QUANSDAM\ process
is concerned, it is more convenient and direct to realize approximately the
solution-information propagation\ process $U_{H}^{ic}(t_{m},a_{m}^{s})\Psi
\left( \mathbf{r,}t_{0}\right) $ of (4.8) than to construct approximately
the $IC$ unitary propagator $U_{H}^{ic}(t_{m},a_{m}^{s})$ separately.
According to the eigenfunction expansion principle [17] the infinite series
on the RH side of (4.6) is always convergent. This is the theoretical basis
for employing the energy eigenfunction expansion to construct approximately
the $IC$ unitary propagator $U_{H}^{ic}(t_{m},a_{m}^{s})$ and to realize
approximately the solution-information propagation process of (4.8). The
eigenfunction expansion principle itself is not involved in the convergent
speed of the infinite series of (4.6). But the latter is important to
determine whether the $IC$ unitary propagator $U_{H}^{ic}(t_{m},a_{m}^{s})$
can be efficiently constructed approximately and whether the
solution-information propagation\ process of (4.8) can be efficiently
realized approximately.

For simplicity, below consider a single-atom system with an external
harmonic potential field. Then the convergent speed of the infinite series
of (4.6)\ may be simply characterized through the truncation error:%
\begin{equation}
\varepsilon \left( M\right) =\left\vert \left\vert
\sum_{k=0}^{M-1}A_{k}u_{k}(\mathbf{r})-\Psi \left( \mathbf{r,}t_{0}\right)
\right\vert \right\vert =\sqrt{\sum_{k=M}^{\infty }\left\vert
A_{k}\right\vert ^{2}}.  \tag{4.11a}
\end{equation}%
Given any desired small value $\varepsilon >0$ the convergence of the
infinite series of (4.6) is said to be faster if the minimum number $M$ of
the expansion terms that satisfies $\varepsilon \left( M\right) <\varepsilon 
$ is smaller. Obviously, the convergent speed is directly dependent on the
expansion coefficients $\{A_{k}\}.$ It can prove that if $M$\ is
polynomially large and the truncation error $\varepsilon \left( M\right)
<\varepsilon $ can be neglected, then there is an efficient unitary sequence 
$USEQ\left( K\right) $ of (4.9) such that the solution-information
propagation process of (4.8) can be realized by the one of (4.10) up to a
global phase factor. The truncation error $\varepsilon \left( M\right) $ of
(4.11a) may act as a simple fast-convergent criterion for the infinite
series of (4.6). It may be better used for the special case that
contribution of the expansion terms with the lowest energy eigenstates to
the series of (4.6) is dominating. Though the fast-convergent criterion of
(4.11a) is simple and special, it could be used not only for the single-atom
system but also for other space-dependent quantum systems. For a more
general case the double-side truncation error may act as a better
fast-convergent criterion for the infinite series of (4.6). It is defined by%
\begin{equation}
\varepsilon \left( L,M\right) =\left\vert \left\vert
\sum_{k=L}^{L+M-1}A_{k}u_{k}(\mathbf{r})-\Psi \left( \mathbf{r,}t_{0}\right)
\right\vert \right\vert =\sqrt{\sum_{k=0}^{L-1}\left\vert A_{k}\right\vert
^{2}+\sum_{k=L+M}^{\infty }\left\vert A_{k}\right\vert ^{2}}.  \tag{4.11b}
\end{equation}%
When the energy quantum number $L=0$, this formula is reduced to (4.11a).
Thus, the fast-convergent criterion of (4.11a) is the special case of the
general one of (4.11b). It can turn out that if $M$\ is polynomially large
and for any given quantum number $L\geq 0$ the double-side truncation error $%
\varepsilon \left( L,M\right) <\varepsilon $ can be neglected, then there
exists an efficient unitary sequence $USEQ\left( K\right) $ of (4.9) such
that the solution-information propagation process of (4.8) can be realized
by the one of (4.10) up to a global phase factor.

If any quantum state $\Psi \left( \mathbf{r,}t_{0}\right) $ whose energy
eigenfunction expansion is given by the infinite series of (4.6) satisfies
the fast-convergent criterion $\varepsilon \left( L,M\right) <\varepsilon $,
where the quantum number $L\geq 0$ is some integer and $M$ is polynomially
large, then the state is called a fast-convergent quantum state. Because the
quantum number $L\geq 0$ in (4.11b) may be arbitrarily large and the
polynomially large number $M$ is not fixed, there are an infinite number of
linearly independent fast-convergent quantum states of the single-atom
system. All these fast-convergent quantum states form a fast-convergent
state set. Evidently the fast-convergent state set is a state subset of the
Hilbert space of the single-atom system. It contains an infinite number of
linearly independent fast-convergent quantum states.

One important point is that the fast-convergent criterion $\varepsilon
\left( L,M\right) <\varepsilon $ has nothing to do with any quantum-state
effect of the single-atom quantum system, because a wavefunction $\Psi
\left( \mathbf{r,}t_{0}\right) $ of the quantum system which is represented
by the expansion of (4.6) and satisfies the fast-convergent criterion may be
quantum or classical. Therefore, whether or not the wavefunction $\Psi
\left( \mathbf{r,}t_{0}\right) $ is a fast-convergent state has nothing to
do with any quantum-state effect of the quantum system. Moreover, it has
nothing to do with the energy, space, and even the fundamental
quantum-computing resource, which are associated with the unitary sequence $%
USEQ\left( K\right) $ of (4.9). It also has nothing to do with whether the
solution-information propagation process of (4.8) is a QUANSDAM process or a
UNIDYSLOCK\ process, when the wavefunction contains the solution information
($a_{m}^{s}$). Actually, it is determined completely by the shape of the
wavefunction for a given complete set of energy eigenfunctions $\{u_{k}(%
\mathbf{r})\}$.

The $IC$ unitary propagator $U_{H}^{ic}(t_{m},a_{m}^{s})$ may act as the
basic building block to construct a QUANSDAM\ process. By replacing the $IC$
unitary operator $V_{k}^{ic}(a_{m}^{s})$ with $%
U_{H}^{ic}(t_{m}^{k},a_{m}^{s})$ in the general QUANSDAM process of (3.8)
one obtains%
\begin{equation}
QUANSDAM\left( K,\text{ }a_{m}^{s}\right)
=U_{K}U_{H}^{ic}(t_{m}^{K},a_{m}^{s})U_{K-1}...U_{1}U_{H}^{ic}(t_{m}^{1},a_{m}^{s})U_{0}
\tag{4.12}
\end{equation}%
where the initial state which is omitted may be simply set to the ground
state of the quantum system and $U_{k}$ may be the $QM$ unitary operator or
propagator of the quantum system.

The total $QSD$ amplification ability of the QUANSDAM process of (4.12) is
determined by $(i)$ the unitary sequence $USEQ\left( K\right) $ of (4.9)\
that is used to realize the $IC$ unitary propagator $%
U_{H}^{ic}(t_{m}^{k},a_{m}^{s})$ and $(ii)$ the $QSD$ amplification ability
of each $IC$ unitary propagator $U_{H}^{ic}(t_{m}^{k},a_{m}^{s})$ that is
promoted by the $QM$ unitary propagators $\{U_{k}\}$.

Because the parameter $t_{m}^{k}$ ($t_{m}^{k}\thickapprox \theta _{m}$) in
the $IC$ unitary propagator $U_{H}^{ic}(t_{m}^{k},$ $a_{m}^{s})$ is
exponentially small, the $QSD$ amplification ability of the $IC$ unitary
propagator alone is weak and generally an $IC$ unitary propagator $%
U_{H}^{ic}(t_{m}^{k},$ $a_{m}^{s})$ alone is not able to realize an
exponential $QSD$ amplification. Then the interaction between the $IC$
unitary propagators $\{U_{H}^{ic}(t_{m}^{k},a_{m}^{s})\}$ and the $QM$
unitary propagators $\{U_{k}\}$ may be more important to realize an
exponential $QSD$ amplification of the QUANSDAM\ process of (4.12). This
interaction could be stronger than the one between the basic $IC$ unitary
operators and the $QM$ unitary operators in a QUANSDAM\ process such as
(3.9). However, from the point of view of the computational complexity the
realization of an exponential $QSD$ amplification makes sense only when the $%
IC$ unitary propagator $U_{H}^{ic}(t_{m}^{k},a_{m}^{s})$ can be generated by
the efficient unitary sequence $USEQ\left( K\right) $ of (4.9) or the
solution-information propagation process of (4.8) can be realized
efficiently.

As pointed out in the preceding section, an appropriate $QM$ unitary
operator in (3.9) is able to make the $QSD$ varying rate $\Delta \rho
_{12}\left( k+1\right) /k$ (or $\Delta \rho _{12}^{2}\left( k+1\right) /k$)
in magnitude exceeding $O\left( p_{2}\left( k\right) \right) ,$ where $%
p_{2}\left( k\right) $ is a quadratic polynomial in the number $k$ of the
basic $IC$ unitary operators in (3.9). But for the QUANSDAM\ process of
(4.12)\ the basic building block is directly the $IC$ unitary propagator $%
U_{H}^{ic}(t_{m}^{k},a_{m}^{s})$ rather than the basic $IC$ unitary
operator. For simplicity, here suppose that the $IC$ unitary propagator $%
U_{H}^{ic}(t_{m}^{k},a_{m}^{s})$ or the solution-information propagation
process of (4.8) can be realized efficiently. Then a $QM$\ unitary
propagator (e.g., $U_{k}$) in (4.12)\ is said to be appropriate if it can
make the $QSD$ varying rate $\Delta \rho _{12}\left( k+1\right) /k$ (or $%
\Delta \rho _{12}^{2}\left( k+1\right) /k$) exceeding $O\left( p_{2}\left(
k\right) \right) $ in magnitude, where $p_{2}\left( k\right) $ is a
quadratic polynomial in the number $k$ of the $IC$ unitary propagators $%
\{U_{H}^{ic}(t_{m}^{k},a_{m}^{s})\}$ in the process $QUANSDAM\left( k,\text{ 
}a_{m}^{s}\right) $ ($0\leq k<K$) of (4.12). If every $QM$ unitary
propagator in (4.12) is appropriate, then clearly the QUANSDAM process of
(4.12) can be used to extract the solution information in a super-square or
even an exponential speedup. Then these quantum states are clearly
appropriate, which are $|\Psi _{a_{m}^{s}}^{k}\rangle =QUANSDAM\left( k,%
\text{ }a_{m}^{s}\right) |\Psi _{0}\rangle $ ($0\leq k\leq K$) with the
initial state $|\Psi _{0}\rangle $ which usually may be the ground state of
the quantum system, if every $QM$ unitary propagator in (4.12) is
appropriate. It is obvious that each one of these appropriate states $%
\{|\Psi _{a_{m}^{s}}^{k}\rangle \}$ is a fast-convergent state.

How to find the appropriate $QM$ unitary propagators and the appropriate
states of the QUANSDAM process of (4.12) is a challenging task. These
appropriate $QM$ unitary propagators must ensure that on the one hand, the $%
QSD$ amplification ability of the QUANSDAM process is super-square or even
exponentially large and on the other hand, the solution-information
propagation process of (4.8) can be realized efficiently. Here the unitary
manipulation of a single atom in time and space [15, 31, 19, 7] that is
based on Gaussian wave-packet states and quadratic unitary propagators is a
systematic and powerful method to find these appropriate $QM$ unitary
propagators and these appropriate states for the QUANSDAM\ process in a
single-atom system. The related work will be reported in detail in future.%
\newline
\newline
{\large 5.\ The }${\large HSSS}${\large \ quantum search process}

Consider an unstructured search problem whose unstructured search space is $%
2^{n}-$dimensional and which may be solved in an $n-$qubit quantum system.
This search problem may be fully characterized by the Boolean function $f:$ $%
\{0,1,...,2^{n}-1\}\rightarrow \{0,1\}$ [13]. Now the unstructured search
space of the search problem may be given by $\{x\}=\{0,1,...,2^{n}-1\}$
(e.g., $x$ may be an index number). The unique solution $x_{0}$ to the
search problem, $x_{0}\in \{0,1,...,2^{n}-1\},$ may be defined simply by the
Boolean function $f\left( x\right) =1$ if $x=x_{0}$ and $f\left( x\right) =0$
if $x\neq x_{0}$. It also may be defined equivalently by the Boolean
functional operation:%
\begin{equation}
U_{f}:|x)|0)\rightarrow |x)|f\left( x\right) )=\left\{ 
\begin{array}{ccc}
|x_{0})|1) & if & x=x_{0} \\ 
|x)|0) & if & x\neq x_{0}%
\end{array}%
\right.  \tag{5.1}
\end{equation}%
It is clear that the solution space (or set) to the search problem is $%
\{|x_{0})|f\left( x_{0}\right) )\}$ or $\{|x_{0})|1)\}$ with $0\leq
x_{0}\leq 2^{n}-1,$ while the unstructured search space is $\{|x)\}$ with $%
0\leq x\leq 2^{n}-1.$ If one drops the functional state $|f\left(
x_{0}\right) )=|1)$ in the second register, then the solution space is just $%
\{|x_{0})\}$ with $0\leq x_{0}\leq 2^{n}-1.$ Here the solution $|x_{0})$ can
be any element of the solution space $\{|x_{0})\}$. Thus, the solution space
is just the unstructured search space. Without confusion, here $|x)$ denotes
the variable (number) state in the register and also the usual computational
basis vector in the math Hilbert space corresponding to the register, as can
be seen below. A similar denotation is also available for $|f\left( x\right)
).$

According to the Lecerf-Bennett reversible computational theory the Boolean
functional operation $U_{f}$ of (5.1) may be constructed in a reversible
form [8, 27]. Once it is constructed in this way, it may be applied to a
quantum system. This construction usually needs to use the intermediate
reversible computational steps and the auxiliary registers which are already
hidden in (5.1). Now the Boolean functional operation $U_{f}$ may be
performed in an $n-$qubit quantum system with the variable state space $%
\{|x\rangle \}$ and a functional quantum system with the functional state
space $\{|f\left( x\right) \rangle \}$ together. Here every intermediate
reversible computational step may be replaced with the corresponding unitary
operator (or quantum gate array [35]) that acts on the total quantum system
which may include the auxiliary quantum systems in addition to the $n-$qubit
quantum system and the functional quantum system; and in (5.1) the variable
state $|x)$ and the functional state $|f\left( x\right) )$ in the registers
are replaced with the usual computational basis state $|x\rangle $ of the $%
n- $qubit quantum system and the functional state $|f\left( x\right) \rangle 
$ of the functional quantum system, respectively.

Here one needs to pay attention to the difference between the solution space
(or set) $\{|x_{0})\},$ the unstructured search space $\{|x)\},$ and the
usual Hilbert space $\{|x\rangle \}$ of the quantum system. In classical
search the unstructured search space $\{|x)\}$ is usually just the solution
set $\{|x_{0})\}$. In a usual quantum search algorithm it may be considered
as the Hilbert space $\{|x\rangle \}$ of the $n-$qubit quantum system [3,
13] and the solution set is the usual computational basis state set of the
Hilbert space. Note that the Hilbert space is determined completely by a
basis state set. Then the unstructured search space, i.e., the Hilbert space
is still determined completely by the Boolean function $f\left( x\right) $
above [13]. In the $HSSS$ quantum search process the situation may be
different. The unstructured search space may be a linear complex vector
space $\{|x)\}$ (i.e., a math Hilbert space) and the solution set is the
usual computational basis vector set of the vector space. This linear
complex vector space has a similar mathematical definition as the usual
Hilbert space of a quantum system and it is unstructured, but it is
mathematical and does not have any physical meaning. Again the unstructured
search space, i.e., the math Hilbert space is still determined completely by
the Boolean function $f\left( x\right) $ above, because a linear vector
space is determined completely by a basis vector set. Therefore, there are
different\ considerations of the unstructured search space for the classical
search algorithm, the usual quantum search algorithm, and the $HSSS$ quantum
search process, respectively.

The mathematical Hilbert space (i.e., the math Hilbert space) is a
fundamental concept in the quantum-computing speedup theory [1]. It is
already introduced simply in the section 2 above. The concept of the math
Hilbert space does not exist independently in conventional quantum
computation. The reversible Boolean functional operation $U_{f}$ of (5.1)
also may be performed in the total math Hilbert space. It is known from
(5.1) that the Boolean functional operation works in the pivotal register
with the variable (number) state space $\{|x)\},$ the pivotal register with
the functional (number) state space $\{|f\left( x\right) )\},$ and the
auxiliary registers together. Now every register is replaced with a linear
complex vector space, i.e., a math Hilbert space. Then the Boolean
functional operation may be performed in the total math Hilbert space which
is the tensor product of the component math Hilbert spaces that correspond
to the register with $\{|x)\},$ the register with $\{|f\left( x\right) )\},$
and the auxiliary registers, respectively. The component math Hilbert spaces
each are large enough so that they can accommodate the variable vector space 
$\{|x)\},$ the functional vector space $\{|f\left( x\right) )\}$ and so on,
respectively.

Generally, in the quantum-computing speedup theory a math Hilbert space is
not equivalent to the corresponding Hilbert space of a quantum system. A
mathematical-parallel computation (or operation) is allowed to perform in a
math Hilbert space. In effect a mathematical-parallel operation is just a
usual classical-parallel operation. It is essentially different from a usual
quantum-parallel operation [11] as it may not obey the quantum superposition
principle.

Now one may perform the Boolean functional operation of (5.1) in the math
Hilbert space of the search problem or in the Hilbert space of the quantum
system. If the two Hilbert spaces are treated as the same one (See below
more clearly), then this is the routine treatment of a conventional quantum
search algorithm [3, 13]. If the two Hilbert spaces are treated separately,
then this is the treatment of the $HSSS$ quantum search process.

It is more general to use directly an oracle operation rather than a usual
functional operation as the basic building block of a quantum search
algorithm [3, 13]. Generally, in quantum computation an oracle operation
also may be thought of as a black-box functional operation. Moreover, it
could be constructed with the usual functional operations of a computational
problem to be solved. Below a black-box functional operation is constructed
by using the Boolean functional operation of (5.1). It may be further used
as the basic building block to construct a quantum search algorithm to solve
the unstructured search problem.

By using the Boolean functional operation $U_{f}$ of (5.1) one carries out a
sequence of the Boolean functional operations [5a, 1, 2]:%
\begin{equation*}
V_{0}:|x\rangle |0\rangle |0\rangle \rightarrow |x\rangle |0\rangle |1\rangle
\end{equation*}%
\begin{equation*}
U_{f}:|x\rangle |0\rangle |1\rangle \rightarrow |x\rangle |f\left( x\right)
\rangle |1\rangle
\end{equation*}%
\begin{equation*}
V\left( \theta \right) :|x\rangle |f\left( x\right) \rangle |1\rangle
\rightarrow \exp \left( -i\theta \delta \left( f\left( x\right) ,1\right)
\right) |x\rangle |f\left( x\right) \rangle |1\rangle
\end{equation*}%
\begin{equation*}
U_{f}:\exp \left( -i\theta \delta \left( f\left( x\right) ,1\right) \right)
|x\rangle |f\left( x\right) \rangle |1\rangle \rightarrow \exp \left(
-i\theta \delta \left( f\left( x\right) ,1\right) \right) |x\rangle
|0\rangle |1\rangle
\end{equation*}%
\begin{equation*}
V_{0}:\exp \left( -i\theta \delta \left( f\left( x\right) ,1\right) \right)
|x\rangle |0\rangle |1\rangle \rightarrow \exp \left( -i\theta \delta \left(
f\left( x\right) ,1\right) \right) |x\rangle |0\rangle |0\rangle
\end{equation*}%
where the $\delta -$function $\delta \left( f\left( x\right) ,1\right) =1$
if the Boolean function $f\left( x\right) =1$ and $\delta \left( f\left(
x\right) ,1\right) =0$ if $f\left( x\right) =0.$ The Boolean
function-operational sequence above may be written in a compact form%
\begin{equation}
BFSEQ=V_{0}U_{f}V\left( \theta \right) U_{f}V_{0}  \tag{5.2}
\end{equation}%
where $V_{0}=\exp \left( -i\pi /2\right) \exp \left( i\pi I_{x}\right)
=\left( 
\begin{array}{cc}
0 & 1 \\ 
1 & 0%
\end{array}%
\right) $ and $V\left( \theta \right) $ is the unitary diagonal matrix $%
V\left( \theta \right) =Diag\left( 1,1,1,\exp \left( -i\theta \right)
\right) .$ The Boolean function-operational sequence $BFSEQ$ is executed in
the main quantum system $(|x\rangle )$ and the two auxiliary qubits with the
initial state set to $|0\rangle |0\rangle .$ Here the single-qubit unitary
operation $V_{0}$ is performed only in the second auxiliary qubit and the
two-qubit phase-shift operation $V\left( \theta \right) $ is performed only
in the two auxiliary qubits. If now all the intermediate steps are hidden,
then the above function-operational process to perform the sequence $BFSEQ$
may be simply written as 
\begin{equation}
BFSEQ:|x\rangle |0\rangle |0\rangle \rightarrow \left\{ 
\begin{array}{ccc}
\exp \left( -i\theta \right) |x\rangle |0\rangle |0\rangle & if & f\left(
x\right) =1 \\ 
|x\rangle |0\rangle |0\rangle & if & f\left( x\right) =0%
\end{array}%
\right.  \tag{5.3}
\end{equation}%
The core of the Boolean function-operational sequence $BFSEQ$ is the Boolean
functional operation $U_{f}.$ The sequence $BFSEQ$ may act as the black-box
functional operation of a black-box quantum search algorithm to solve an
unstructured search problem. This black-box functional operation is defined
by the unitary transformation of (5.3) and is realized by the Boolean
function-operational sequence of (5.2). Now it is very clear that both the
unstructured search space $\{|x\rangle \}$ and the Hilbert space $%
\{|x\rangle \}$ of the $n-$qubit quantum system are the same one! This
black-box functional operation is characterized completely by the
mathematical-logical principle of the unstructured search problem or
equivalently by the Boolean function $f:$ $\{0,1,...,2^{n}-1\}\rightarrow
\{0,1\}$ for which there is a unique element $x_{0}$ such that $f\left(
x_{0}\right) =1$ $[13].$ In the unitary transformation of (5.3) the solution
state $|x\rangle =|x_{0}\rangle $ of the search problem which satisfies $%
f\left( x\right) =1$ if $x=x_{0}$ and $f\left( x\right) =0$ if $x\neq x_{0}$
can be an arbitrary computational basis state of the unstructured search
space $\{|x\rangle \}.$

The black-box functional operation of (5.3) may be further used as the basic
building block to construct a quantum search algorithm to solve the
unstructured search problem. If now one drops the auxiliary qubits ($%
|0\rangle |0\rangle $) from (5.3), then the black-box functional operation
is just the usual oracle operation $U_{o}\left( \theta \right) $ defined by
(2.2) of a conventional quantum search algorithm. Therefore, the usual
oracle operation $U_{o}\left( \theta \right) $ of (2.2) can represent
faithfully the black-box functional operation of (5.3).

It is known that the Boolean function $f\left( x\right) $ in (5.3) is
defined by $f\left( x\right) =1$ if and only if $x=x_{0}$ and $f\left(
x\right) =0$ if $x\neq x_{0}.$ If one hides the auxiliary qubits $\left(
|0\rangle |0\rangle \right) ,$ then the black-box functional operation of
(5.3) also may be expressed as%
\begin{equation}
BFSEQ:|x\rangle \rightarrow \left\{ 
\begin{array}{ccc}
\exp \left( -i\theta \right) |x\rangle & if & x=x_{0} \\ 
|x\rangle & if & x\neq x_{0}%
\end{array}%
\right.  \tag{5.4}
\end{equation}%
Later it will be shown that the black-box functional operation of (5.4) also
may be represented faithfully by the unitary oracle selective diagonal
operator $C_{S}\left( \theta \right) $ defined by (2.1) of the $HSSS$
quantum search process.

As pointed out in the quantum-computing speedup theory [1], the $HSSS$
quantum search process is essentially different from a conventional quantum
search algorithm [3, 13]. Below it is shown from the mathematical-logical
viewpoint that the unitary oracle selective diagonal operator $C_{S}\left(
\theta \right) $ of the $HSSS$ quantum search process is essentially
different from the usual oracle operation $U_{o}\left( \theta \right) $ of a
conventional quantum search algorithm. Here, as is well known, the unitary
oracle selective diagonal operator $C_{S}\left( \theta \right) $ and the
usual oracle operation $U_{o}\left( \theta \right) $ characterize
essentially the $HSSS$ quantum search process and the conventional quantum
search algorithm, respectively.

First of all, it must be pointed out that a unitary selective diagonal
operator $C_{k}\left( \theta \right) =\exp \left( -i\theta D_{k}\right) $
itself (See Refs. [ 5, 25b, 12] and also the section 2 above) does not own a
specific mathematical-logical meaning. From the mathematical-logical
viewpoint it has nothing to with the oracle operation $U_{o}\left( \theta
\right) $ of a conventional quantum search algorithm, because the latter has
an unambiguous mathematical-logical meaning. Its mathematical-logical
meaning comes from the computational problem to be solved, when it is used
to solve the problem. It could be used not only for solving an unstructured
search problem [5] but also for solving other hard problems (See, for
example, Ref. [25b]). If now the unitary selective diagonal operator $%
C_{k}\left( \theta \right) $ is employed to solve an unstructured search
problem, then it is given the mathematical-logical meaning of the search
problem, and here it represents the black-box functional operation of the
search problem. Suppose that $C_{k}\left( \theta \right) $ represents the
black-box functional operation of a conventional quantum search algorithm
[3, 13], then it is just the usual oracle operation $U_{o}\left( \theta
\right) .$ If here $C_{k}\left( \theta \right) $ represents the black-box
functional operation of the $HSSS$ quantum search process, then it is
renamed the unitary oracle selective diagonal operator $C_{S}\left( \theta
\right) .$ In this paper $C_{S}\left( \theta \right) $ represents only the
black-box functional operation of the $HSSS$ quantum search process.
Initially the author often explained misunderstandingly the diagonal matrix
of $C_{S}\left( \theta \right) $ as the matrix representation of the usual
oracle operation $U_{o}\left( \theta \right) $. However, it has been firmly
recognized by the author since its birth's day [5a] that the $HSSS$ quantum
search process is essentially different from a conventional quantum search
algorithm. Therefore, this misunderstanding explanation has never obstructed
the author's insistent effort to realize the search-space dynamical
reduction in the past one and half decades [5, 6, 1, 2]. The search scheme
to solve an unstructured search problem for a conventional quantum search
algorithm is to find directly the solution state $|S\rangle $ as a whole. In
contrast, the search scheme for the $HSSS$ quantum search process is first
to determine directly the information $\left( a_{k}^{s}\right) $ of each
component state $\left( |s_{k}\rangle \right) $ of the solution state $%
|S\rangle =|s_{1}\rangle |s_{2}\rangle ...|s_{n}\rangle $, then obtain the
unit-number vector $\{a_{1}^{s},a_{2}^{s},...,a_{n}^{s}\},$ and finally find
the solution state $|S\rangle $ of the unstructured search problem.
Therefore, what the $HSSS$ quantum search process treats is any single
component state of the solution state, while what the conventional quantum
search algorithm treats is the solution state as a whole.

In classical computation a parallel computation always can be expressed as a
consecutive series of computations. This principle is also correct in
quantum computation, that is, a quantum parallel functional operation [11,
10, 9] also can be expressed as a consecutive series of single (or
selective) functional operations. Here this principle is simply introduced
so as to help ones understand better the mathematical-logical difference
between the unitary oracle selective diagonal operator $C_{S}\left( \theta
\right) $ and the usual oracle operation $U_{o}\left( \theta \right) $. It
works in the frame of unitary quantum dynamics. It has nothing to do with
any quantum-state effects such as the quantum-state superposition, coherence
interference, entanglement and nonlocal effect, and correlation which are
the core components in the quantum parallel principle [11a]. The
quantum-computing speedup theory [1] considers that a quantum parallel
operation belongs to the unitary quantum dynamics that could be a driving
force to speed up a quantum computation, but it doesn't think that these
core components of the quantum parallel principle are responsible for an
essential quantum-computing speedup.

The quantum parallel operation $V_{f}$ of a function $f\left( x\right) $ in
an $n-$qubit quantum system may be generally expressed as [11a] 
\begin{equation}
V_{f}:\sum_{x\in \{0,1\}^{n}}a_{x}|x\rangle |y\rangle \rightarrow \sum_{x\in
\{0,1\}^{n}}a_{x}|x\rangle |y\tbigoplus f\left( x\right) \rangle  \tag{5.5}
\end{equation}%
where the usual computational basis state $|x\rangle
=|b_{1}^{x},b_{2}^{x},...,b_{n}^{x}\rangle $ with $b_{k}^{x}\in \{0,1\}$ $%
(1\leq k\leq n).$ The initial state on the left-hand side of (5.5) is an
arbitrary state of the quantum system. It is expanded in terms of the basis
states $\{|x\rangle |y\rangle \}$ of the Hilbert space (or its subspace) of
the quantum system according to the eigenfunction expansion principle [17,
14]. A selective (or single) reversible functional operation of the function 
$f\left( x\right) $ in the quantum system may be defined as%
\begin{equation}
V_{f\left( x_{1}\right) }:\sum_{x\in \{0,1\}^{n}}a_{x}|x\rangle |y\rangle
\rightarrow \sum_{x\in \{0,1\}^{n},\text{ }x\neq x_{1}}a_{x}|x\rangle
|y\rangle +a_{x_{1}}|x_{1}\rangle |y\tbigoplus f\left( x_{1}\right) \rangle .
\tag{5.6}
\end{equation}%
This single reversible functional operation $V_{f\left( x_{1}\right) }$
really performs selectively the reversible functional operation: 
\begin{equation}
V_{f\left( x_{1}\right) }:|x_{1}\rangle |y\rangle \rightarrow |x_{1}\rangle
|y\tbigoplus f\left( x_{1}\right) \rangle  \tag{5.7}
\end{equation}%
with the selected variable value $x_{1}$. For any other computational basis
state $|x\rangle \neq |x_{1}\rangle $ of the quantum system (or for any
other variable value $x\neq x_{1}$) the single reversible functional
operation makes no action. The selective functional operation $V_{f\left(
x_{1}\right) }$ is considered as a single reversible functional operation
rather than a parallel functional operation, although the initial state of
the functional operation may be a superposition state, because one action of
the functional operation on the initial state computes only one functional
value, i.e., $x_{1}\rightarrow f\left( x_{1}\right) $.

The construction of a selective reversible functional operation may be
simple or complex in a quantum system, but this doesn't matter for the
present purpose. For simplicity, here assume that any two selective
reversible functional operations do not have a cross-talk. For example, one
has%
\begin{equation*}
V_{f\left( x_{2}\right) }V_{f\left( x_{1}\right) }:\sum_{x\in
\{0,1\}^{n}}a_{x}|x\rangle |y\rangle \rightarrow \sum_{x\in \{0,1\}^{n},%
\text{ }x\neq x_{1},x_{2}}a_{x}|x\rangle |y\rangle
\end{equation*}%
\begin{equation}
+a_{x_{2}}|x_{2}\rangle |y\tbigoplus f\left( x_{2}\right) \rangle
+a_{x_{1}}|x_{1}\rangle |y\tbigoplus f\left( x_{1}\right) \rangle . 
\tag{5.8}
\end{equation}%
This formula shows that the second selective reversible functional operation 
$V_{f\left( x_{2}\right) }$ does not affect the function-operational result $%
|x_{1}\rangle |y\tbigoplus f\left( x_{1}\right) \rangle $ of the first
selective reversible functional operation $V_{f\left( x_{1}\right) }.$ Now
using this no-cross-talk assumption and noting that the initial state $%
\sum_{x\in \{0,1\}^{n}}a_{x}|x\rangle |y\rangle $ is arbitrary it can turn
out that the relation between the quantum parallel functional operation $%
V_{f}$ of (5.5) and the single reversible functional operations $%
\{V_{f\left( x\right) }\}$ of (5.6) may be simply written as%
\begin{equation}
V_{f}=\tprod\limits_{x\in \{0,1\}^{n}}V_{f\left( x\right) }.  \tag{5.9}
\end{equation}%
This formula shows that the quantum parallel functional operation $V_{f}$
may be decomposed as a consecutive sequence of many ($2^{n})$ single
reversible functional operations $\{V_{f\left( x\right) }\}$. For this
result a quantum parallel functional operation is similar to the classical
counterpart. The formula (5.9) is independent of any initial state in (5.5)
and it is purely unitary-dynamical. The superposition of functional states
on the right-hand side of (5.5) may be generated by applying the quantum
parallel operation $V_{f}$ to the initial state in (5.5) or by applying the
consecutive sequence of the single reversible functional operations on the
right-hand side of (5.9) to the same initial state. Obviously, the quantum
parallel functional operation $V_{f}$ is greatly economic in the generation
of the superposition of functional states in (5.5) when there are
exponentially many single reversible functional operations $\{V_{f\left(
x\right) }\}$ on the right-hand side of (5.9).

The relation (5.9) between the quantum parallel functional operation $V_{f}$
and the selective reversible functional operations $\{V_{f\left( x\right)
}\} $ is helpful for understanding correctly the mathematical-logical
difference between the unitary oracle selective diagonal operator $%
C_{S}\left( \theta \right) $ and the usual oracle operation $U_{o}\left(
\theta \right) .$ Suppose that the quantum parallel functional operation $%
V_{f}$ as a whole is irreducible. Then whether or not any selective
reversible functional operation $V_{f\left( x\right) }$ in (5.9) must be
irreducible too? Actually, the relation (5.9) does not require that each of
the selective reversible functional operations $\{V_{f\left( x\right) }\}$
be irreducible. Thus, any single reversible functional operation $V_{f\left(
x\right) }$ may or may not be irreducible, even if the quantum parallel
functional operation $V_{f}$ as a whole is irreducible. This property is
very helpful for understanding correctly why the unitary oracle selective
diagonal operator $C_{S}\left( \theta \right) $ is reducible, while the
usual oracle operation $U_{o}\left( \theta \right) $ is not. Later this
point will be discussed. It must be emphasized that here the so-called
`irreducible' (or `reducible') must be related to the mathematical-logical
meaning of a functional operation (or a computational problem).

With the help of the quantum parallel functional operation and the selective
reversible functional operation, below from the mathematical-logical
viewpoint it is shown that the unitary oracle selective diagonal operator $%
C_{S}\left( \theta \right) $ is essentially different from the usual oracle
operation $U_{o}\left( \theta \right) $. There is no doubt that when a
quantum algorithm solves a computational problem, it must obey the
mathematical-logical principle of the computational problem. However, it is
allowed in mathematics and computation that a same computational problem
could be solved by different algorithms. The $HSSS$ quantum search process
and a conventional quantum search algorithm are just two different quantum
search algorithms to solve a same unstructured search problem. If they solve
the same search problem, then they must obey the same mathematical-logical
principle of the search problem. For a conventional quantum search algorithm
the mathematical-logical principle of an unstructured search problem is
characterized essentially by the usual oracle operation $U_{o}\left( \theta
\right) $ defined by (2.2). It is shown above that the usual oracle
operation $U_{o}\left( \theta \right) $ is equivalent to the black-box
functional operation of (5.3) or (5.4) that is generated by the Boolean
functional operation $U_{f}$ of (5.1). For the $HSSS$ quantum search process
the mathematical-logical principle is characterized essentially by the
unitary oracle selective diagonal operator $C_{S}\left( \theta \right) $
defined by (2.1). Below it must prove that the unitary oracle selective
diagonal operator $C_{S}\left( \theta \right) $ is also equivalent to the
black-box functional operation of (5.3) or (5.4).

It is known that the usual oracle operation $U_{o}\left( \theta \right) $
has the mathematical- logical meaning of quantum parallel operation [3, 13].
In contrast, according to the expression $C_{S}\left( \theta \right) =\exp
\left( -i\theta D_{S}\right) $ ($D_{S}=|S\rangle \langle S|$ and $|S\rangle $
is the solution state) in (2.1) the unitary oracle selective diagonal
operator $C_{S}\left( \theta \right) $ may not have. Moreover, the usual
oracle operation $U_{o}\left( \theta \right) $ as a whole is irreducible,
while $C_{S}\left( \theta \right) $ is reducible. How can they each
characterize essentially the same mathematical-logical principle of the
search problem?

Because the black-box functional operation of (5.4) is a quantum parallel
operation which is applied to an $n-$qubit quantum system, just like the
quantum parallel functional operation $V_{f}$ of (5.9) it may be decomposed
as a consecutive sequence of many selective reversible functional
operations. According to (5.4) one may construct any selective black-box
functional operation $BFSEQ\left( y,\theta \right) $ by%
\begin{equation}
BFSEQ\left( y,\theta \right) :\sum_{x\in \{0,1\}^{n}}b_{x}|x\rangle
\rightarrow \sum_{x\in \{0,1\}^{n},\text{ }x\neq x_{0}}b_{x}|x\rangle
+b_{x_{0}}\exp \left( -i\theta \delta (y-x_{0})\right) |x_{0}\rangle 
\tag{5.10}
\end{equation}%
where the $\delta -$function $\delta (y-x_{0})=1$ if $y=x_{0}$ and $\delta
(y-x_{0})=0$ if $y\neq x_{0}$, and $|x_{0}\rangle $ is the solution state of
the search problem. The initial state on the left-hand side of (5.10) is
arbitrary in the quantum system. It is expanded in terms of the basis states 
$\{|x\rangle \}$ of the Hilbert space of the quantum system according to the
eigenfunction expansion principle. Here it is required that any two
selective black-box functional operations of (5.10) do not have a
cross-talk. A particularly important selective black-box functional
operation of (5.10) is $BFSEQ\left( x_{0},\theta \right) $ which is
selectively applied to the solution state $|x_{0}\rangle $. It is obtained
from (5.10) by setting $y=x_{0},$%
\begin{equation}
BFSEQ\left( x_{0},\theta \right) :\sum_{x\in \{0,1\}^{n}}b_{x}|x\rangle
\rightarrow \sum_{x\in \{0,1\}^{n},\text{ }x\neq x_{0}}b_{x}|x\rangle
+b_{x_{0}}\exp \left( -i\theta \right) |x_{0}\rangle .  \tag{5.11}
\end{equation}%
This selective functional operation alone performs the reversible functional
operation in the quantum system: 
\begin{equation}
BFSEQ\left( x_{0},\theta \right) :|x_{0}\rangle \rightarrow \exp \left(
-i\theta \right) |x_{0}\rangle  \tag{5.12}
\end{equation}%
with the solution state $|x_{0}\rangle $. Besides $BFSEQ\left( x_{0},\theta
\right) $ any other selective black-box functional operation with $y\neq
x_{0}$ also can be obtained from (5.10) and it is given by%
\begin{equation}
BFSEQ\left( y,\theta \right) :\sum_{x\in \{0,1\}^{n}}b_{x}|x\rangle
\rightarrow \sum_{x\in \{0,1\}^{n}}b_{x}|x\rangle ,\text{ for }y\neq x_{0}. 
\tag{5.13}
\end{equation}%
This selective reversible functional operation as a whole is a unit
operator. But its content could not be always identical to a unit operator.
On the basis of the no-cross-talk assumption it turns out that the quantum
parallel black-box functional operation of (5.4) may be written as a
consecutive sequence of the single black-box functional operations $%
\{BFSEQ\left( y,\theta \right) \}$ defined by (5.10):%
\begin{equation}
BFSEQ=\tprod\limits_{y\in \{0,1\}^{n}}BFSEQ\left( y,\theta \right) . 
\tag{5.14}
\end{equation}%
This formula is similar to (5.9). It is independent upon any quantum state
of the quantum system and hence it is purely unitary-dynamical. Actually,
with the help of (5.11) and (5.13) one can prove that the consecutive
sequence of (5.14) applying to an arbitrary initial state $\sum_{x\in
\{0,1\}^{n}}b_{x}|x\rangle $ of the quantum system generates just the same
state as the black-box functional operation of (5.4) applying to the same
initial state does,%
\begin{equation}
BFSEQ:\sum_{x\in \{0,1\}^{n}}b_{x}|x\rangle \rightarrow \sum_{x\in
\{0,1\}^{n},\text{ }x\neq x_{0}}b_{x}|x\rangle +b_{x_{0}}\exp \left(
-i\theta \right) |x_{0}\rangle .  \tag{5.15}
\end{equation}%
It is easy to find from (5.11) and (5.15) that both the quantum parallel
black-box functional operation ($BFSEQ$) of (5.4) and the selective
black-box functional operation $BFSEQ\left( x_{0},\theta \right) $ that is
selectively applied to the solution state $|x_{0}\rangle $ generate the same
unitary transformation of (5.4). This is the very reason why in the past the
author explained misunderstandingly the diagonal matrix of $C_{S}\left(
\theta \right) $ as the matrix representation of the usual oracle operation $%
U_{o}\left( \theta \right) .$ This point will be seen more clearly below.

According to (2.2) the usual oracle operation $U_{o}\left( \theta \right) $
applying to an arbitrary initial state of the quantum system generates%
\begin{equation}
U_{o}\left( \theta \right) :\sum_{x\in \{0,1\}^{n}}b_{x}|x\rangle
\rightarrow \sum_{x\in \{0,1\}^{n},\text{ }x\neq x_{0}}b_{x}|x\rangle
+b_{x_{0}}\exp \left( -i\theta \right) |x_{0}\rangle .  \tag{5.16}
\end{equation}%
Now from the expression $C_{S}\left( \theta \right) =\exp \left( -i\theta
D_{S}\right) =\exp \left( -i\theta |S\rangle \langle S|\right) $ of (2.1) ($%
|S\rangle $ is just the solution state $|x_{0}\rangle $ here) one knows that
the unitary oracle selective diagonal operator $C_{S}\left( \theta \right) $
selectively acts on only the solution state $|S\rangle $ without disturbing
any other states of the quantum system. When it acts on an arbitrary initial
state of the quantum system, one obtains%
\begin{equation}
C_{S}\left( \theta \right) :\sum_{x\in \{0,1\}^{n}}b_{x}|x\rangle
\rightarrow \sum_{x\in \{0,1\}^{n},\text{ }x\neq x_{0}}b_{x}|x\rangle
+b_{x_{0}}\exp \left( -i\theta \right) |x_{0}\rangle ,\text{ for }|S\rangle
=|x_{0}\rangle .  \tag{5.17}
\end{equation}%
It is easy to find from (5.11), (5.15), (5.16), and (5.17) that $BFSEQ\left(
x_{0},\theta \right) ,$ $BFSEQ,$ $U_{o}\left( \theta \right) $, and $%
C_{S}\left( \theta \right) $ all generate the same unitary transformation of
(5.4). Because $C_{S}\left( \theta \right) =\exp \left( -i\theta |S\rangle
\langle S|\right) $ is not involved in any state of the quantum system
except the solution state $|S\rangle =|x_{0}\rangle $, it should be
considered to be equivalent to the selective black-box functional operation $%
BFSEQ\left( x_{0},\theta \right) $ rather than the usual oracle operation $%
U_{o}\left( \theta \right) $ or the black-box functional operation $BFSEQ.$
As shown above, the usual oracle operation $U_{o}\left( \theta \right) $ is
equivalent to the black-box functional operation $BFSEQ$. They are the
quantum-parallel functional operations. On the other hand, the unitary
oracle selective diagonal operator $C_{S}\left( \theta \right) $ is
equivalent to the selective black-box functional operation $BFSEQ\left(
x_{0},\theta \right) $ because they are the single functional operations.
However, both $C_{S}\left( \theta \right) $ and $BFSEQ\left( x_{0},\theta
\right) $ are not equivalent to any one of $U_{o}\left( \theta \right) $ and 
$BFSEQ$, because the former is the selective functional operations, while
the latter is a quantum-parallel functional operation.

In appearance both $C_{S}\left( \theta \right) $ of (5.17) and $U_{o}\left(
\theta \right) $ of (5.16) generate the same unitary transformation. This
made the author explaining misunderstandingly $C_{S}\left( \theta \right) $
as the faithful representation of $U_{o}\left( \theta \right) $ in the past.
However, as shown above, they are not equivalent to one another at all.
Their mathematical-logical meanings are not the same one. The usual oracle
operation $U_{o}\left( \theta \right) $ is a quantum-parallel functional
operation and it already takes into account the mathematical-logical meaning
that the solution state ($|x_{0}\rangle $) can be an arbitrary usual
computational basis state of the unstructured search space which is just the
Hilbert space of the $n-$qubit quantum system in a conventional quantum
search algorithm. In contrast, the unitary oracle selective diagonal
operator $C_{S}\left( \theta \right) $ with $|S\rangle =|x_{0}\rangle $ is a
selective functional operation and it is not able to take into account the
mathematical-logical meaning that the solution state can be an arbitrary
usual computational basis state of the unstructured search space. Consider
the fact that the usual oracle operation $U_{o}\left( \theta \right) $ can
characterize essentially the mathematical-logical principle of the
unstructured search problem. Then it is clear that the unitary oracle
selective diagonal operator $C_{S}\left( \theta \right) $ with $|S\rangle
=|x_{0}\rangle $ alone is not able to characterize essentially the same
mathematical-logical principle of the search problem. Therefore, while the
usual oracle operation $U_{o}\left( \theta \right) $ can act as the basic
building block to construct a conventional quantum search algorithm, the
unitary oracle selective diagonal operator $C_{S}\left( \theta \right) $
with $|S\rangle =|x_{0}\rangle $ alone can not be used as the basic building
block to construct the $HSSS$ quantum search process.

In order that $C_{S}\left( \theta \right) $ can act as the basic building
block to construct the $HSSS$ quantum search process the
mathematical-logical meaning that the solution state $|x_{0}\rangle $ can be
an arbitrary usual computational basis state of the unstructured search
space must be taken into account in the $HSSS$ quantum search process. The $%
HSSS$ quantum search process considers that there are $N$ candidate solution
states for the $N-$dimensional unstructured search space. Correspondingly in
the unitary oracle selective diagonal operator $C_{S}\left( \theta \right)
=\exp \left( -i\theta |S\rangle \langle S|\right) $ the basis state $%
|S\rangle $ stands for any one of these $N$ candidate solution states
without limiting to the real solution state $|x_{0}\rangle $ only. This
really extends the mathematical-logical meaning for $C_{S}\left( \theta
\right) $ and in the meantime $C_{S}\left( \theta \right) $ is still a
selective black-box functional operation for any candidate solution state $%
|S\rangle $ (or $|S)$). Now one has $C_{S}\left( \theta \right) =BFSEQ\left(
x_{0},\theta \right) $ if $|S\rangle =|x_{0}\rangle $ and $C_{S}\left(
\theta \right) \neq BFSEQ\left( x_{0},\theta \right) $ if $|S\rangle \neq
|x_{0}\rangle .$ These (candidate) unitary oracle selective diagonal
operators altogether are considered as a single entity. This is crucial for
the search-space dynamical reduction in the $HSSS$ quantum search process.
With this extended mathematical-logical meaning the unitary oracle selective
diagonal operator $C_{S}\left( \theta \right) $ is able to represent
faithfully the black-box functional operation $BFSEQ$ of (5.4).

There is an essential difference between the unitary oracle selective
diagonal operator $C_{S}\left( \theta \right) $ with the real solution state 
$|S\rangle =|x_{0}\rangle $ and the one with a candidate solution state $%
|S\rangle \neq |x_{0}\rangle $. When the former acts on an initial state of
a quantum system, the generated state is still in the quantum system. In
contrast, when the latter is applied to the same initial state in the
unstructured search space, the generated state is merely a candidate state
and it is not in the quantum system (See the section 3 above).

The realization that $C_{S}\left( \theta \right) $ can represent faithfully
the black-box functional operation $BFSEQ$ can not be carried out alone in
the Hilbert space of a quantum system. The reason for this is that the
realization is involved in the mathematical-parallel computation (or
operation) that does not obey the superposition principle in quantum
mechanics. (Here in effect mathematical parallel computation is classical
parallel computation.) In principle the realization can be carried out alone
in the math Hilbert space of the search problem (See this section above and
also the section 3) as a mathematical-parallel computation is allowed in the
math Hilbert space, but it has nothing to do with a quantum system and
consequently it can not be implemented practically in quantum computation.
The unique possibility is that the realization is carried out in both the
Hilbert space of the quantum system and the math Hilbert space of the search
problem, as shown in the section 2 above.

The decomposition formula (5.14) of quantum-parallel operation provides the
answer for why the unitary oracle selective diagonal operator $C_{S}\left(
\theta \right) $ is reducible. The black-box functional operation $BFSEQ$ of
(5.4) is a quantum-parallel operation and it is irreducible as a whole. It
is known from (5.14) that $BFSEQ$ is a consecutive sequence of these $2^{n}$
selective black-box functional operations $\{BFSEQ\left( y,\theta \right) \}$
and $BFSEQ\left( x_{0},\theta \right) $ is merely one of these $2^{n}$
number of $\{BFSEQ\left( y,\theta \right) \}$. Then the formula (5.14) shows
that although the black-box functional operation $BFSEQ$ as a whole is
irreducible, the selective black-box functional operation $BFSEQ\left(
x_{0},\theta \right) $ is allowed to be reducible. Consider the fact that
the unitary oracle selective diagonal operator $C_{S}\left( \theta \right) $
with $|S\rangle =|x_{0}\rangle $ is equivalent to $BFSEQ\left( x_{0},\theta
\right) .$ Then one can conclude that $C_{S}\left( \theta \right) $ is
allowed to be reducible.

The quantum-computing speedup theory [1] then points out that whether or not
the unitary oracle selective diagonal operator $C_{S}\left( \theta \right) $
can be reduced is determined by the fundamental quantum-computing resource,
i.e., the symmetric structure of the Hilbert space of a composite quantum
system and moreover, if it can, then its reduction process is performed in
the frame of unitary quantum dynamics.

The practical implementation (or simulation) for $C_{S}\left( \theta \right) 
$ needs to employ two computing machines, one of which is the
quantum-physical computing machine (e.g., an $n-$qubit quantum system) and
another the quantum-computing math machine which owns the math Hilbert space
and is mathematical. The quantum-physical computing machine computes the
black-box functional operation $BFSEQ$ of (5.4) with the real solution state 
$|x_{0}\rangle $, while the quantum-computing math machine computes the
black-box functional operation $BFSEQ$ with any candidate solution state $%
|S\rangle \neq |x_{0}\rangle \ $and executes the mathematical-parallel
functional operation for $BFSEQ$ in the math Hilbert space. All these
computations of the black-box functional operation $BFSEQ$ with the sequence
(5.2) are efficient (in experiment or in the sense of theory) as long as the
reversible Boolean functional operation $U_{f}$ of (5.1) can be realized
efficiently. Therefore, $C_{S}\left( \theta \right) $ can be efficiently
implemented (or simulated) practically as long as $U_{f}$ can be realized
efficiently or more directly as long as the black-box functional operation $%
BFSEQ$ of (5.4) can be realized efficiently.

The main mathematical-logical differences between $U_{o}\left( \theta
\right) $ and $C_{S}\left( \theta \right) $ are summarized as follows. Both $%
U_{o}\left( \theta \right) $ and $C_{S}\left( \theta \right) $ can represent
faithfully the black-box functional operation $BFSEQ$ of (5.3) or (5.4), but
they use different Hilbert spaces. The usual oracle operation $U_{o}\left(
\theta \right) $ uses only the Hilbert space of the quantum system to
realize this faithful representation, while $C_{S}\left( \theta \right) $
needs to use explicitly both the Hilbert space of the quantum system and the
math Hilbert space of the search problem. Of course, it also may be thought
that in the usual oracle operation $U_{o}\left( \theta \right) $ both the
Hilbert spaces are coincident completely with one another, as shown in the
section 2 above. In the Hilbert space of the quantum system the usual oracle
operation $U_{o}\left( \theta \right) $ is a quantum parallel operation and
irreducible in the mathematical-logical meaning of unstructured search,
while $C_{S}\left( \theta \right) $ is a single black-box functional
operation and does not have the mathematical-logical meaning of quantum
parallel operation and moreover, it is allowed to be reducible in the
mathematical-logical meaning of unstructured search. Besides these
differences, in the math Hilbert space $C_{S}\left( \theta \right) $ may be
considered as a mathematical-parallel black-box functional operation as a
whole due to that the real solution state is unique and deterministic.

The difference of mathematical-logical meaning between $C_{S}\left( \theta
\right) $ and $U_{o}\left( \theta \right) $ is helpful for understanding
better the essential difference of the quantum-searching speedup mechanism
between the $HSSS$ quantum search process and a conventional quantum search
algorithm. Since the usual oracle operation $U_{o}\left( \theta \right) $ is
irreducible, there is the square speedup limit on a conventional quantum
search algorithm. Since $U_{o}\left( \theta \right) $ works only in the
Hilbert space of the quantum system, it may be considered as a $QM$ unitary
operator. This results in that it can not change the quantum-state
difference between any pair of quantum states of the quantum system. In
contrast, since the unitary oracle selective diagonal operator $C_{S}\left(
\theta \right) $ is allowed to be reducible, the search-space dynamical
reduction becomes feasible as long as there is the fundamental
quantum-computing resource. Since $C_{S}\left( \theta \right) $ works in
both the Hilbert space of the quantum system and the math Hilbert space of
the search problem, its behavior can not be completely described alone by
the quantum-mechanical principles. Instead, it is described completely by
both the quantum-mechanical principles and the mathematical-logical
principle of the search problem. This results in that though it is a unitary
operator, it is able to change the quantum-state difference between any pair
of quantum states that belong to the two Hilbert spaces, respectively. Due
to these two reasonings the $HSSS$ quantum search process with the basic
building block $C_{S}\left( \theta \right) $ may not be constrained by the
square speedup limit of a conventional quantum search algorithm.

With the basic building block $C_{S}\left( \theta \right) $ the $HSSS$
quantum search process sets up its own strategy to find the real solution
state among these $N$ $($here $N=2^{n})$ candidate solution states (or
equivalently distinguish the real solution state from the other $N-1$
candidate solution states). As well known before, the strategy consists of
the search-space dynamical reduction and the unitary dynamical state-locking
process and its inverse process. The theoretical basis behind the strategy
is the interaction between the quantum-physical laws (i.e., the unitary
quantum dynamics and the Hilbert-space symmetrical structure and property)
and the mathematical-logical principle of the search problem. The $HSSS$
quantum search process mainly consists of the two consecutive steps. The
first step is the efficient search-space dynamical reduction in the frame of
unitary quantum dynamics [2]. It eliminates the unstructured search space of
the search problem. This step becomes feasible because the unitary oracle
selective diagonal operator $C_{S}\left( \theta \right) $ may be reducible
and there is the fundamental quantum-computing resource in the $n-$qubit
quantum system used to solve the search problem. The second step is the
exponential quantum-state-difference amplification [36] which in principle
is already described in detail in the previous sections 3 and 4 in the paper.

Below explains the quantum-computing speedup mechanism for the search-space
dynamical reduction. It must be pointed out that without the search-space
dynamical reduction the computational complexity of an unstructured search
problem can not be essentially changed even\ if one employs the unitary
oracle selective diagonal operator $C_{S}\left( \theta \right) $ as the
basic building block to construct any quantum search algorithm to solve the
search problem. Therefore, it is still hard to extract the information of
the component states of the solution state for any quantum search algorithm
without the search-space dynamical reduction. Such quantum search algorithm
still must be subject to the square speedup limit of a conventional quantum
search algorithm. On the other hand, from the mathematical-logical viewpoint 
$C_{S}\left( \theta \right) $ is a mathematical-parallel black-box
functional operation in the unstructured search space (i.e., the math
Hilbert space) with exponentially large dimension $N=2^{n}$. It owns
exponentially many mathematical-parallel computational paths. The extraction
of the information of the component states of the solution state is a hard
task, since one needs first to compute these detailed computational paths
and then treat these computational results and all these are hard things.

When the exponentially large unstructured search space is cancelled, all
these exponentially many mathematical-parallel computational paths are
accordingly eliminated. That is, when the exponentially large unstructured
search space is reduced dynamically to a polynomially small subspace, the
black-box functional operation ($C_{S}\left( \theta \right) $) with
exponentially many computational paths is replaced with the one with only
polynomially many computational paths. Now it could become feasible to
compute a few or even a polynomially many computational paths. Therefore,
the search-space dynamical reduction greatly speeds up to extract the
information of the component states of the solution state. Moreover, due to
the search-space dynamical reduction the exponential
quantum-state-difference amplification [36] becomes feasible for the quantum
states that carry the information of the component state. These are the
contribution of the search-space dynamical reduction to the exponential
quantum-computing speedup in the $HSSS$ quantum search process.

Finally outlines the solution information flow over the $HSSS$ quantum
search process. The information flow is one-way. It starts from the
reversible Boolean functional operation that initially loads the solution
information of an unstructured search problem and may be considered as the
initial information source in the $HSSS$ quantum search process. The
solution information is first delivered to the black-box functional
operation (which also could be considered as the initial information
source). Then it is transferred in form to the unitary oracle selective
diagonal operator. Up to this stage the solution information is complete.
Then by the efficient search-space dynamical reduction in the frame of
unitary quantum dynamics the component of the solution information, i.e.,
the information of the component state of the solution state is output to
the information-carrying unitary operator. This step creates the information
source of the component state of the solution state. After the exponential
(unitary) QUANSDAM process, the information of the component state is
extracted by the quantum measurement.\newline
\newline
{\Large 6. Discussion}

A UNIDYSLOCK\ (or QUANSDAM) process is the characteristic quantum
computational process (or sub-process) of the quantum-computing speedup
theory. It obeys the unitary quantum dynamics, but it is able to change the
quantum-state difference between a pair of quantum states. This
characteristic property of a UNIDYSLOCK\ (or QUANSDAM) process is original
from the fundamental interaction between the quantum-physical laws (i.e.,
the unitary quantum dynamics and the Hilbert-space symmetric structure and
property) and the mathematical-logical principle that a computational
problem obeys. It is shown in the paper that any unitary dynamical process
in quantum mechanics can not change the quantum-state difference. It is also
shown that no quantum-state effect can cause the quantum-state-difference
varying during a UNIDYSLOCK\ (or QUANSDAM) process in a quantum system. This
conclusion is universal in the frame of unitary quantum dynamics for any
quantum computational process that employs a UNIDYSLOCK\ (or QUANSDAM)
process (via the unstructured quantum search process) to realize its
quantum-computing speedup. A UNIDYSLOCK\ (or QUANSDAM) process therefore is
able to distinguish the quantum-computing speedup theory from any
conventional quantum computational theory. The conventional quantum
computational theory based on the quantum parallel principle [11a] considers
that a quantum-computing speedup is essentially original from the
quantum-state effects of a quantum system. Here the quantum-state effects
include the quantum-state superposition, coherence interference,
entanglement and nonlocal effect, correlation and so on, and among these
quantum-state effects the quantum entanglement and nonlocal effect is
considered as the essential one. Then this really means that in the
conventional quantum computational theory there is not an essential
contribution for a UNIDYSLOCK (or QUANSDAM) process to the quantum-computing
speedup of any reversible (or unitary) quantum computational process.
Therefore, whether or not a UNIDYSLOCK\ (or QUANSDAM) process can make an
essential contribution to the quantum-computing speedup of a reversible or
unitary quantum computational process distinguishes strictly the
quantum-computing speedup theory from any conventional quantum computational
theory in the frame of unitary quantum dynamics.

Can the conventional quantum computational theory employ a UNIDYSLOCK (or
QUANSDAM) process to realize a quantum-computing speedup? There is no doubt
that a UNIDYSLOCK (or QUANSDAM) process makes no contribution to the
quantum-computing speedup of a conventional quantum computation if the
quantum computation does not contain any UNIDYSLOCK (or QUANSDAM) process.
However, even if a conventional quantum computation that is reversible or
unitary contained any UNIDYSLOCK (or QUANSDAM) process, the contribution of
the UNIDYSLOCK (or QUANSDAM) process to the quantum-computing speedup of the
conventional quantum computation would be secondary or negligible. The
reason for this is that if the contribution of the UNIDYSLOCK (or QUANSDAM)
process were dominating, then the one of the quantum-state effects to the
quantum-computing speedup would be secondary or negligible. There are only
two nontrivial physical processes that can change the quantum-state
difference between a pair of quantum states. One of which is a UNIDYSLOCK
(or QUANSDAM) process. It is unitary. Another is a non-equilibrium
irreversible process (which could contain the usual quantum measurement). It
is not unitary. Then the only possibility for a conventional quantum
computational process to employ a quantum-state-difference amplification to
realize its quantum-computing speedup is that the quantum-state-difference
amplification process\ must be irreversible! As shown in the section 3
above, without unitarity a quantum-state-difference amplification process,
e.g., a UNIDYSLOCK (or QUANSDAM) process could become trivial in quantum
computation. It also suffers from the energy dissipation problem of
irreversibility (See, e.g., Ref. [20]) in quantum computation. Even for the
case\ that the quantum-state-difference amplification process is a
non-equilibrium irreversible process one can not yet conclude that the
quantum-computing speedup is original from the quantum-state effects of the
quantum system, because in that case it is still possible that the
quantum-state-difference amplification could be original from the
irreversible dynamical process itself and it could have nothing to do with
any quantum-state effect of the quantum system. Physical world in nature is
fundamentally reversible. Therefore, although apparently the unitary quantum
dynamics often could not be obeyed strictly in nature, the quantum-computing
speedup theory [1] considers that it is not able to deny the fact that the
unitary quantum dynamics is a fundamental principle in nature. Thus, the
theory thought that \textit{even such a process as a non-equilibrium
irreversible process in nature that does not obey apparently the unitary
quantum dynamics is governed by the unitary quantum dynamics}.

Physicalization of a mathematical-logical functional operation is a
universal phenomenon in classical reversible computation [20] and in
conventional quantum computation [9, 10, 11]. It is not only related to
whether or not a mathematical-logical function is computable in a quantum
system but it also could affect the quantum-computing speedup mechanism. The
physicalization of a reversible functional operation is simply described as
follows. When it is realized in a physical system, a reversible functional
operation in mathematics becomes a pure reversible (or unitary) physical
process of the system. In the meantime, the function-operational space in
mathematics is mapped one-to-one onto the physical state space of the
system. Then the latter acts as the former in the realization of the
functional operation in the physical system. Thus, both the
function-operational space and the physical state space are the same one in
the realization. Both the spaces may have their own symmetric structures and
properties, respectively. But because they are the same one, the symmetric
structure and property of the physical state space may be washed out and
there is no consideration of the interaction between the two spaces. These
are the common basic characteristic features for the physicalization of a
reversible functional operation in the classical reversible computation and
the conventional quantum computation.

The physicalization of a reversible functional operation is necessary for
the conventional quantum computational theory based on the quantum parallel
principle [11a]. Though a huge number of the functional states\ can be
computed simultaneously by a quantum parallel functional operation, it is
hard to obtain the desired computational result from the final state of the
operation. The theory then employs a variety of quantum-state effects to
extract the desired computational result from the final state. Though these
quantum-state effects could be constrained by the mathematical symmetrical
structure and property of the computational problem to be solved, this
constraint (or its influence) is considered to be inessential or secondary
in the quantum-computing speedup mechanism based on the quantum parallel
principle [11a]. Therefore, it is generally considered in the theory that a
quantum-computing speedup is essentially original from these quantum-state
effects themselves. The latter mainly includes the quantum-state
superposition, coherence interference, entanglement and nonlocal effect, and
correlation. Furthermore, the quantum entanglement and nonlocal effect is
generally considered as the essential one among these quantum-state effects.
However, these quantum-state effects between any two or more different
(component) functional states of the final state of the quantum parallel
functional operation could exist only when these functional states are real
quantum states of the quantum system. Imaging that one component functional
state is a real quantum state, while another is an unphysical state. Then it
is impossible in quantum mechanics that these two functional states can
generate the quantum-state effects such as the quantum-state superposition,
coherence interference, entanglement and nonlocal effect, and correlation
and so on. Therefore, the quantum parallel operation of a\ function must be
physical. That is, the physicalization of a mathematical-logical functional
operation is necessary in the conventional quantum computation. This also
indicates clearly that there is not a separate math Hilbert space in the
conventional quantum computation. The physicalization of a reversible
functional operation could make the quantum-state effects of the functional
states dominating in the quantum-computing speedup mechanism of the
conventional quantum computational theory based on the quantum parallel
principle. But the symmetric structure of the Hilbert space of the quantum
system\ (i.e., the fundamental quantum-computing resource) may be washed out.

The quantum-computing \ speedup \ theory [1] does not think \ that these
quantum-state effects can be responsible for an essential quantum-computing
speedup. The difficulties and limitations of the quantum-computing speedup
mechanism based on the quantum parallel principle [11a] have been analyzed
and discussed in detail in Ref. [1]. Here these will not be further
discussed.

One important consequence for the physicalization of a reversible functional
operation in conventional quantum computation is that the quantum-state
difference between any pair of quantum states can not be changed by a
reversible functional operation, because the reversible functional operation
is treated as a pure $QM$\ unitary operator in conventional quantum
computation after it is physicalized in the quantum system. Another is that
in conventional quantum computation the fundamental quantum-computing
resource of a quantum system is not exploited to speed up a quantum
computation.

In contrast, in the quantum-computing speedup theory the reversible
functional operation of a\ computational problem is originally responsible
for the quantum-state-difference varying in a UNIDYSLOCK\ (or QUANSDAM)
process and the fundamental quantum-computing resource is responsible for an
essential quantum-computing speedup.

For a long time the physicalization of a reversible functional operation or
more generally the physicalization of computing has been mainstream in
quantum computational science. This can be shown by the Church-Turing
principle [11a]: "Every finitely realizable physical system can be perfectly
simulated by a universal model computing machine operating by finite means".
The physicalization of computing is the computing version of information
physicalization. While information physicalization seems to be quite
reasonable, the largest obstacle facing the computing physicalization comes
from the fact that computing must obey the mathematical-logical principle.
Quantum computing can not be described completely by the quantum-physical
laws alone and the mathematical-logical principle of a computational problem
to be solved can not be disregarded in the quantum-computing speedup
mechanism. This is consistent with the spirit of the quantum-computing
speedup theory [1].

The quantum-computing \ speedup theory considers that the fundamental
interaction between the quantum-physical laws (i.e., the unitary quantum \
dynamics and the Hilbert-space \ symmetric structure and property) \ and the
mathematical-logical principle that a computational problem obeys is the
origin of an essential quantum-computing speedup. Then in the theory both
the math Hilbert space of a computational problem and the Hilbert space of
the quantum system used to solve the computational problem must be treated
separately. One important reason why both the Hilbert spaces are treated
separately in the theory is that one needs to realize the search-space
dynamical reduction [5, 6, 1, 2]. In contrast, both the Hilbert spaces are
the same one in conventional quantum computation due to the physicalization
of computing. The math Hilbert space of a computational problem is a
fundamental concept in the quantum-computing speedup theory. The concept is
characteristic for the theory. It is not owned by the conventional quantum
computational theory. Due to the math Hilbert space the manner to realize a
reversible functional operation in the quantum-computing speedup theory is
essentially different from the one in conventional quantum computation. A
mathematical-parallel functional operation is the characteristic manner in
the quantum-computing speedup theory. In contrast, a quantum-parallel
functional operation is the characteristic manner in the conventional
quantum computational theory based on the quantum parallel principle. Due to
the interaction between the math Hilbert space and the Hilbert space of the
quantum system a reversible functional operation is able to change the
quantum-state difference between a pair of quantum states in the
quantum-computing speedup theory. This leads to that there is a UNIDYSLOCK
(or QUANSDAM) process which is characteristic for the theory. Of course, the
separate treatment for the two Hilbert spaces is not the purpose for the
theory. Instead, the theoretical purpose is to set up the interaction
between the two Hilbert spaces in the frame of unitary quantum dynamics so
that by the interaction the fundamental quantum-computing resource of the
quantum system can be harnessed to speed up (directly or indirectly) a
quantum computation. In contrast, there is not the interaction in
conventional quantum computation, because both the Hilbert spaces are the
same one due to the physicalization of computing. Thus, there is no
exploitation of the fundamental quantum-computing resource to speed up a
quantum computation in conventional quantum computation.

Here one must distinguish the fundamental quantum-computing resource from
the quantum-state effects in quantum-computing speedup mechanism. The former
is employed by the quantum-computing speedup theory [1] to speed up a
quantum computation, while the latter is considered by the conventional
quantum computational theory [11a] as the origin of a quantum-computing
speedup. It is well known in quantum mechanics\ that the Hilbert space of a
composite quantum system is formed of all the quantum states of the quantum
system. The symmetric structure and property of the Hilbert space therefore
is the global property of the quantum system. It is independent on any
concrete quantum state of the quantum system. In the quantum-computing
speedup theory the Hilbert-space symmetric structure is considered as the
fundamental quantum-computing resource and it is responsible for an
exponential quantum-computing speedup. It is also well-known in quantum
mechanics that two (appropriate) quantum states (or particles) of the
quantum system alone are sufficient to generate the quantum-state effects
such as the quantum-state superposition, coherence interference,
entanglement and nonlocal effect, and correlation and so on. These
quantum-state effects are generally considered as the partial (or partite)
property of the quantum system. The conventional quantum computational
theory considers that a quantum-computing speedup is original from these
quantum-state effects.

A reversible functional operation is the core of a quantum algorithm to
solve a computational problem in quantum computation. A computational
problem usually could be characterized completely by using the reversible
functional operations, when it is solved in quantum computation. In this
paper the author does not study how to realize a reversible functional
operation with a universal set of one- and two-qubit quantum logic gates (or
more complex gate sets) in a quantum system (See Ref. [35] and therein for
this research topic). Instead, what the author studies is that the
reversible functional operation of a computational problem is directly
considered as the starting point of research subject. Why the
quantum-computing speedup theory emphasizes the importance of a
computational problem? It is well known that what standard quantum mechanics
treats is atomic, molecular, nuclear system and so on. Correspondingly, what
the quantum-computing speedup theory treats is any concrete computational
problem. Each computational problem obeys its own mathematical-logical
principle and has its own mathematical symmetric structure.

A computational problem usually could be characterized completely by the
functional operations, when it is solved in computation. Therefore, these
functional operations usually could be used as the basic building blocks of
an algorithm to solve the computational problem in computational science,
here the algorithm may be classically irreversible, classically reversible,
or quantum. These functional operations that can characterize completely the
computational problem obey obviously the specific mathematical-logical
principle which could consist of a series of mathematical-logical rules or
conditions imposed by the computational problem. One may say that the
mathematical-logical principle is owned only by the computational problem.
It is well-known that a functional operation may be carried out in a Turing
machine in mathematical computation and it also may be performed in a
physical computing machine (i.e., a physical system) in physical
computation. However, the mathematical-logical principle of a computational
problem is essentially independent of any computing machine which may be
mathematical or physical, because a computational problem is independent of
any computing machine. For example, consider the problem that one wants to
determine whether an integer $p$ is prime or composite. As well known in
number theory, $(i)$ if the integer $p$ is prime, then it can be divided
only by one and itself; $(ii)$ if $p$ is not prime, then it must be a
product ($st$) of at least two integers $s$ and $t$ with $1<s,$ $t<p$. The
two properties $(i)$\ and $(ii)$ form the essential components of the
mathematical-logical principle of the present problem. Any one of these two
properties must be obeyed no matter which computing machine one uses to
solve the present problem. Of course, statement or expression for the
mathematical-logical principle of the computational problem could be
different from machine to machine, but its essence is the same! \newline
\newline
\newline
{\Large References}\newline
1. X. Miao, \textit{The universal quantum driving force to speed up a
quantum computation --- The unitary quantum dynamics},
http://arxiv.org/abs/quant-ph/ 1105.3573 (2011)\newline
2. X. Miao, \textit{Efficient dynamical reduction from the exponentially
large unstructured search space of a search problem to a polynomially small
subspace in an n-qubit spin system}, Unpublished work\newline
3. L. K. Grover, \textit{Quantum mechanics helps in searching for a needle
in a haystack}, Phys. Rev. Lett. 79, 325 (1997)\newline
4. (a) C. H. Bennett, E. Bernstein, G. Brassard, and U. Vazirani, \textit{%
Strengths \ and \ weaknesses \ of \ quantum\ \ computing}, \
http://arxiv.org/abs/quant-ph/ 9701001 (1997); (b) M. Boyer, G. Brassard, P.
H$\phi $yer, and A. Tapp, \textit{Tight bounds on quantum searching},
Fortschr. Phys. 46, 493 (1998); (c) C. Zalka, \textit{Grover}$^{\prime }s$%
\textit{\ quantum searching algorithm is optimal}, Phys. Rev. A 60, 2746
(1999); (d) L. K. Grover, \textit{How fast can a quantum computer search?}
http://arxiv.org/abs/quant-ph/9809029 (1998)\newline
5. (a)\ X. Miao, \textit{Universal construction for the unsorted quantum
search algorithms}, http://arxiv.org/abs/quant-ph/0101126 (2001); (b) 
\textit{Solving the quantum search problem in polynomial time on an NMR
quantum computer}, http:// arxiv.org/abs/quant-ph/0206102 (2002)\newline
6. X. Miao, \textit{Quantum search processes in the cyclic group state spaces%
}, http:// arxiv.org/abs/quant-ph/0507236 (2005)\newline
7. X. Miao, \textit{The basic principles to construct a generalized
state-locking pulse field and simulate efficiently the reversible and
unitary halting protocol of a universal quantum computer},
http://arxiv.org/abs/quant-ph/0607144 (2006)\newline
8. (a) Y. Lecerf, \textit{Machines de Turing r\'{e}versibles. R\'{e}cursive
insolubilit\'{e} en }$n$\textit{\ }$\in $\textit{\ }$N$\textit{\ de l' \'{e}%
quation }$u=\theta ^{n}u,$\textit{\ o\`{u} }$\theta $\textit{\ est un
isomorphisme de codes}, C. R. Acad. Sci., 257, 2597 (1963); (b) C. H.
Bennett, \textit{Logical reversibility of computation}, IBM J. Res. Develop.
17, 525 (1973)\newline
9. (a)\ P. Benioff, \textit{Quantum mechanical Hamiltonian models of Turing
machines}, J. Statist. Phys. 29, 515 (1982); (b)\ P. Benioff, \textit{%
Quantum mechanical Hamiltonian models of discrete processes that erase their
own histories: application to Turing machines}, Internat. J. Theor. Phys.
21, 177 (1982)\newline
10. R. P. Feynman, \textit{Quantum mechanical computers}, Found. Phys. 16,
507 (1986)\newline
11. (a)\ D. Deutsch, \textit{Quantum theory, the Church-Turing principle and
the universal quantum computer}, Proc. Roy. Soc. Lond. A 400, 96 (1985);
(b)\ D. Deutsch, \textit{Quantum computational networks}, Proc. Roy. Soc.
Lond. A 425, 73 (1989)\newline
12. X. Miao, \textit{Multiple-quantum operator algebra spaces and
description for the unitary time evolution of multilevel spin systems},
Molec. Phys. 98, 625 (2000)\newline
13. G. Brassard, P. H$\phi $yer, M. Mosca, and A. Tapp, \textit{Quantum
amplitude amplification and estimation},
http://arxiv.org/abs/quant-ph/0005055 (2000)\newline
14. L. I. Schiff, \textit{Quantum mechanics}, 3rd, McGraw-Hill book company,
New York, 1968\newline
15. X. Miao, \textit{Unitary manipulation of a single atom in time and space
--- The spatially-selective and internal-state-selective triggering pulses},
http://arxiv.org/ abs/quant-ph/1309.3758 (2013)\newline
16. (a)\ M. Suzuki, \textit{Fractal decomposition of exponential operators
with applications to many-body theories and Monte Carlo simulations}, Phys.
Lett. A 146, 319 (1990); (b) M. Suzuki, \textit{Decomposition formulas of
exponential operators and Lie exponentials with some applications to quantum
mechanics and statistical physics}, J. Math. Phys. 26, 601 (1985)\newline
17. J. von Neumann, \textit{Mathematical foundations of quantum mechanics},
(Translated by R. T. Beyer), Princeton University Press, 1955\newline
18. R. P. Feynman and A. R. Hibbs, \textit{Quantum mechanics and path
integrals}, McGraw-Hill, New York, 1965\newline
19. X. Miao, \textit{The STIRAP-based \ unitary \ decelerating \ and \
accelerating \ processes \ of a single free atom},
http://arxiv.org/abs/quant-ph/0707.0063 (2007)\newline
20. C. H. Bennett, \textit{The thermodynamics of computation --- a review},
Int. J. Theor. Phys. 21, 905 (1982)\newline
21. D. S. Saxon, \textit{Elementary quantum mechanics}, Holden Day, 1968%
\newline
22. X. Miao, \textit{Efficient multiple-quantum transition processes in an }$%
n-$\textit{qubit spin system}, http://arxiv.org/abs/quant-ph/0411046 (2004)%
\newline
23. R. Freeman, \textit{Spin Choreography}, Spektrum, Oxford, 1997\newline
24. E. J. Heller, \textit{Time-dependent approach to semiclassical dynamics}%
, J. Chem. Phys. 62, 1544 (1975)\newline
25. (a)\ E. Farhi, J. Goldstone, S. Gutmann, and M. Sipser, \textit{A limit
on the speed of quantum computation in determining parity},
http://arxiv.org/abs/quant-ph/ 9802045v2 (1998); (b) X. Miao, \textit{A
polynomial-time solution to the parity problem on an NMR quantum computer},
http://arxiv.org/abs/quant-ph/0108116 (2001)\newline
26. R. Beals, H. Buhrman, R. Cleve, M. Mosca, and R. De Wolf, \textit{%
Quantum lower bounds by polynomials}, Proc. 39th Annual Symposium on
Foundations of Computer Science, pp. 352 (1998); also see:
http://arxiv.org/abs/quant-ph/ 9802049 (1998)\newline
27. E. Fredkin and T. Toffoli, \textit{Conservative logic}, Internat. J.
Theor. Phys. 21, 219 (1982)\newline
28. (a)\ 1. I. D. Ivanovic, \textit{How to differentiate between
non-orthogonal states}, Phys. Lett. A 123, 257 (1987); (b) D. Dieks, \textit{%
Overlap and distinguishability of quantum states}, Phys. Lett. A 126, 303
(1988); (c) A. Peres, \textit{How to differentiate between non-orthogonal
states}, Phys. Lett. A 128, 19 (1988)\newline
29. (a) S. Chu, \textit{Nobel Lecture: The manipulation of neutral particles}%
, Rev. Mod. Phys. 70, 685 (1998); (b) C. N. Cohen-Tannoudji, \textit{Nobel
Lecture: Manipulating atoms with photons}, Rev. Mod. Phys. 70, 707 (1998);
(c)\ W. D. Phillips, \textit{Nobel Lecture: Laser cooling and trapping of
neutral atoms}, Rev. Mod. Phys. 70, 721 (1998)\newline
30. (a) J. I. Cirac and P. Zoller, \textit{Quantum computations with cold
trapped ions}, Phys. Rev. Lett. 74, 4091 (1995); (b) D. J. Wineland, C.
Monroe, W. M. Itano, D. Leibfried, B. E. King, and D. M. Meekhof, \textit{%
Experimental issues in coherent quantum-state manipulation of trapped atomic
ions}, J. Res. NIST, 103, 259 (1998)\newline
31. X. Miao, \textit{Unitarily manipulating in time and space a Gaussian
wave-packet motional state of a single atom in a quadratic potential field},
http: //arxiv.org/ abs/quant-ph/0708.2129 (2007)\newline
32. X. Miao and R. Freeman, \textit{Spin-echo modulation experiments with
soft Gaussian pulses}, J. Magn. Reson. A 119, 90 (1996); \textit{A spin-echo
technique for separation of multiplets in crowded spectra}, J. Magn. Reson.
A 116, 273 (1995)\newline
33. K. Bergmann, H. Theuer, and B. W. Shore, \textit{Coherent population
transfer among quantum states of atoms and molecules}, Rev. Mod. Phys. 70,
1003 (1998)\newline
34. (a)\ D. J. Heinzen and D. J. Wineland, \textit{Quantum-limited cooling
and detection of radio-frequency oscillations by laser-cooled ions}, Phys.
Rev. A 42, 2977 (1990); (b)\ J. I. Cirac, A. S. Parkins, R. Blatt, and P.
Zoller, \textit{"Dark" squeezed states of the motion of a trapped ion},
Phys. Rev. Lett. 70, 556 (1993); (c)\ C. Monroe, D. M. Meekhof, B. E. King,
and D. J. Wineland, \textit{A }$^{\prime \prime }$\textit{Schr\"{o}dinger cat%
}$^{\prime \prime }$\textit{\ superposition state of an atom,} Science 272,
1131 (1996)\newline
35. A. Barenco, C. H. Bennett, R. Cleve, D. DiVincenzo, N. Margolus, P.
Shor, T. Sleator, J. Smolin, and H. Weinfurter, \textit{Elementary gates for
quantum computation}, Phys. Rev. A 52, 3457 (1995)\newline
36. X. Miao, (in preparation)\newline
\newline
\newline
{\Large Appendix A: The information-carrying\ unitary propagators generated
approximately in a single-atom system}

The spatially-selective and internal-state-selective triggering pulse [15,
31] may be used to realize the $IC$ internal-state-dependent unitary
momentum-displacement propagator $U_{p}^{ic}(a_{m}^{s})$ of (4.3) in a
single-atom system. Here it is not considered whether this realization is
efficient or not. In Refs. [15, 31] the triggering pulse is constructed
approximately with the help of the Trotter-Suzuki method mainly [16] and by
using the interaction [29, 30] between the internal motion and the COM
motion of the single-atom system. Inspired by the construction of the
triggering pulse here a simple method is proposed to realize approximately
the $IC$ unitary propagator $U_{p}^{ic}(a_{m}^{s})$ of (4.3)$\ $in a
single-atom system. In this method a Baker-Campbell-Hausdorf (BCH)-type
operator identity is first obtained that is used to generate exactly the
unitary operator $\exp (-\tau ^{2}[A,B])$ from the unitary operators $\exp
(-iA\tau )$ and $\exp (-iB\tau )$ with the Hermitian operators $A$ and $B$.
This BCH-type operator identity may be explicitly expressed as%
\begin{equation}
\exp (-iA\tau )\exp (-iB\tau )\exp (+iA\tau )\exp (+iB\tau )=\exp (-\tau
^{2}[A,B])+O_{p}(\tau ^{3})  \tag{A1}
\end{equation}%
where the error operator $O_{p}(\tau ^{3})$ is exactly written as%
\begin{equation}
O_{p}(\tau ^{3})=\int_{0}^{\tau }d\lambda \exp (-i\lambda A)\exp (-i\lambda
B)Q_{R}(\lambda )\exp (-(\tau ^{2}-\lambda ^{2})[A,B])  \tag{A2}
\end{equation}%
with the operator $Q_{R}(\lambda )$ given by%
\begin{equation*}
Q_{R}(\lambda )=-2\lambda \exp (i\lambda A)\int_{0}^{\lambda }ds\{\exp
(isB)[[A,B],iB]\exp (-isB)
\end{equation*}%
\begin{equation*}
+\exp (-isA)[[A,B],iA]\exp (isA)\}\exp (i\lambda B)
\end{equation*}%
\begin{equation*}
+\int_{0}^{\lambda }ds\int_{0}^{s}ds^{\prime }\{\exp (is^{\prime
}B)[[A,B],iB]\exp (-is^{\prime }B)
\end{equation*}%
\begin{equation}
+\exp (is^{\prime }A)[[A,B],iA]\exp (-is^{\prime }A)\}\exp (i\lambda A)\exp
(i\lambda B).  \tag{A3}
\end{equation}%
This operator identity can be proven by using the operator algebra method
similar to those in Ref. [16b].

On the basis of the operator identity (A1) one may construct the $IC$
unitary propagator $\exp (-\tau ^{2}[A,B])$ with the suitable Hermitian
operators $A$ and $B$. Here one or two of the operators $A$ and $B$ must
carry the solution information $(a_{m}^{s})$ of the search problem. An
example is given below to show how to construct the $IC$ unitary propagator.

Case $(1):$ The single-atom system is a free atom. Suppose that in the
rotating frame the interaction between the atomic internal motion and COM
motion is simply given by $H_{I}=-KxI_{mx}$, where $I_{mx}$ is the $x-$%
component spin$-1/2$ operator of the atomic internal motion. The interaction
could be generated by the external electromagnetic field or static electric
(and/or magnetic) field \footnote{%
See, for example, B. W. Shore, \textit{The theory of coherent atomic
excitation}, Vol.1, Section 2.8 \& 3.3, Wiley, New York, 1990; J. D.
Jackson, \textit{Classical electrodynamics}, 2nd., Wiley \& Son, New York,
1975; L. Allen and J. H. Eberly, \textit{Optical resonance and two-level
atoms}, Dover, New York, 1987\newline
}. Here for simplicity it is not discussed in detail how to generate the
interaction. Then the Hamiltonian of the atom in the external field is given
by $H_{A}=\frac{1}{2m}p^{2}+H_{a}+H_{I},$ where in the rotating frame the
atomic internal Hamiltonian $H_{a}$ is zero under on-resonance condition.
Now in the operator identity (A1) one may set the operator $%
A=a_{m}^{s}\theta _{m}I_{my}/\tau $ and $B=H_{A}/\hslash =\left( \frac{1}{2m}%
p^{2}+H_{I}\right) /\hslash .$ Then it can prove that the commutator $[A,$ $%
B]=(ia_{m}^{s}\theta _{m}/\tau )(K/\hslash )I_{mz}x.$

Case $(2):$ The single-atom system is a single atom in an external harmonic
potential field. Here still suppose that in the rotating frame the
interaction between the atomic internal motion and COM motion is simply
given by $H_{I}=-KxI_{mx}.$ Then the total atomic Hamiltonian is given by $%
H_{A}=\frac{1}{2m}p^{2}+\frac{1}{2}m\omega ^{2}x^{2}+H_{a}+H_{I},$ where $%
H_{a}=0$ under on-resonance condition in the rotating frame. Now setting $%
A=a_{m}^{s}\theta _{m}I_{my}/\tau $ and $B=H_{A}/\hslash =\frac{1}{2m}p^{2}+%
\frac{1}{2}m\omega ^{2}x^{2}+H_{I}$ one still obtains the same commutator $%
[A,$ $B]=(ia_{m}^{s}\theta _{m}/\tau )(K/\hslash )I_{mz}x.$

In the above commutator $[A,$ $B]$ the spin operator $I_{m\lambda }$ ($%
\lambda =x,y,z$) acts on only the internal state of the single-atom system,
while the coordinate operator $x$ acts on only the COM motional state of the
system. From the commutator it follows that the unitary operator $\exp
(-\tau ^{2}[A,B])$ on the right-hand (RH) side of (A1) is given by 
\begin{equation}
\exp (-\tau ^{2}[A,B])=\exp \left( -ia_{m}^{s}(\theta _{m}\tau )(K/\hslash
)I_{mz}x\right) .  \tag{A4}
\end{equation}%
This unitary operator carries the solution information ($a_{m}^{s}).$ Thus,
it is really an information-carrying unitary operator, i.e., the
internal-state-dependent unitary momentum-displacement propagator $%
U_{p}^{ic}(a_{m}^{s})$ of (4.3). On the left-hand (LH) side of (A1) the
unitary operator $\exp (\pm iA\tau )$ is given by%
\begin{equation}
\exp (\pm iA\tau )=\exp (\pm ia_{m}^{s}\theta _{m}I_{my}).  \tag{A5}
\end{equation}%
By comparing it with (2.9) one sees that this unitary operator is in form
just the basic $IC$ unitary operator of (2.9). Since the operator $%
B=H_{A}/\hslash $ is the Hamiltonian of the single-atom system, it does not
carry the solution information ($a_{m}^{s}$). Then the unitary operator $%
\exp (\pm iB\tau )$ on the LH side of (A1) is the $QM$ unitary operator.
These show that the LH side of (A1) is the unitary sequence of the basic $IC$
unitary operators and the $QM$\ unitary operators, while the RH side of (A1)
is the $IC$ unitary propagator $\exp (-\tau ^{2}[A,B])$ in addition to the
error operator $O_{p}(\tau ^{3}).$ Therefore, the operator identity (A1)
could be used to construct the $IC$ unitary propagator $U_{p}^{ic}(a_{m}^{s})
$ of (4.3) by using the basic $IC$ unitary operators of (A5) and the $QM$\
unitary operators in the single-atom system. Of course, the pre-condition
for this is that the error operator $O_{p}(\tau ^{3})$ on the RH side of
(A1) can be neglected.

If one deletes the solution information ($a_{m}^{s}$) in the operator $A$
above, then the unitary propagator $\exp (-\tau ^{2}[A,B])$ generated by
(A1) is just the internal-state-selective triggering pulse [31, 15] and it
could be further developed as a spatially-selective and
internal-state-selective triggering pulse for a wave-packet motional state
such as a Gaussian wave-packet motional state of the single-atom system [15].

On the RH side of (A1) the $IC$ unitary operator $\exp (-\tau ^{2}[A,B])$ is
the main term, while the error operator $O_{p}(\tau ^{3})$ is usually a
secondary term. The latter is dependent on the parameter $\tau $ and the
operators $A$ and $B$. In a general case the operator identity (A1) could be
used to generate the lowest-order approximate unitary sequence for $\exp
(-\tau ^{2}[A,B]).$ Then by starting from the lowest-order approximation one
could employ the Trotter-Suzuki method [16a]\ to generate a higher-order
approximate unitary sequence for $\exp (-\tau ^{2}[A,B]).$

As a simple example, from the operator identity (A1) one may further obtain
the operator identity:%
\begin{equation*}
\left( \exp (-iA\tau /n)\exp (-iB\tau /n)\exp (+iA\tau /n)\exp (+iB\tau
/n)\right) ^{n^{2}}
\end{equation*}%
\begin{equation}
=\exp (-\tau ^{2}[A,B])+O_{p}(n^{2}\left( \tau /n\right) ^{3}).  \tag{A6}
\end{equation}%
Here in theory the exact expression for the error operator $%
O_{p}(n^{2}\left( \tau /n\right) ^{3})$ may be obtained from the operator
identity (A1). If in (A6) one takes $n\rightarrow \infty ,$ then one has%
\begin{equation*}
\lim_{n\rightarrow \infty }\left( \exp (-iA\tau /n)\exp (-iB\tau /n)\exp
(+iA\tau /n)\exp (+iB\tau /n)\right) ^{n^{2}}
\end{equation*}%
\begin{equation}
=\exp (-\tau ^{2}[A,B]).  \tag{A7}
\end{equation}%
Note that $\exp (-iA\tau /n)$ and $\exp (-iB\tau /n)$ are unitary. Then this
formula holds even when the Hermitian operators $A$ and $B$ are unbounded
(See, for example, T. F. Jordan, \textit{Linear operators for quantum
mechanics}, Chapt. VII, Dover, New York, 2006). It shows that when the
number $n$ is sufficiently large, the error operator $O_{p}(n^{2}\left( \tau
/n\right) ^{3})$ in (A6) can be neglected, and the $IC$ unitary propagator $%
\exp (-\tau ^{2}[A,B])$ can be generated approximately by the unitary
sequence on the LH side of (A6) which consists of the basic $IC$ unitary
operators of (A5) and the $QM$ unitary operators $\{\exp (\pm iB\tau /n)\}$.
Both the formulae (A6) and (A7) do not consider the computational complexity
for the approximate realization of the $IC$ unitary propagator $\exp (-\tau
^{2}[A,B]).$ Whether or not the $IC$ unitary propagator $\exp (-\tau
^{2}[A,B])$ can be realized efficiently by the unitary sequence on the LH
side of (A6) is largely dependent on the error operator $O_{p}(n^{2}\left(
\tau /n\right) ^{3}).$ Here there is not a rigorous mathematical proof to
show whether this realization is efficient or not.

Suppose now that $|0\rangle $ and $|\Psi (x,t_{0})\rangle $ are the internal
state and the COM motional state of the single-atom system, respectively,
where $|0\rangle $ is the eigenstate of the spin$-1/2$ operator $I_{mz}$ of
the atomic internal motion and the eigenvalue equation for the operator $%
I_{mz}$ is given by $I_{mz}|0\rangle =\frac{1}{2}|0\rangle .$ Now applying
both sides of (A6) to the initial product state $|\Psi (x,t_{0})\rangle
|0\rangle $ of the single-atom system one obtains the QUANSDAM process:%
\begin{equation*}
\left( \exp (-iA\tau /n)\exp (-iB\tau /n)\exp (+iA\tau /n)\exp (+iB\tau
/n)\right) ^{n^{2}}|\Psi (x,t_{0})\rangle |0\rangle
\end{equation*}%
\begin{equation}
=\exp (-\tau ^{2}[A,B])|\Psi (x,t_{0})\rangle |0\rangle +O_{p}(n^{2}\left(
\tau /n\right) ^{3})|\Psi (x,t_{0})\rangle |0\rangle .  \tag{A8}
\end{equation}%
Here using (A4) one finds that%
\begin{equation}
\exp (-\tau ^{2}[A,B])|\Psi (x,t_{0})\rangle |0\rangle =\exp \left(
-ia_{m}^{s}p_{0}x/\hslash \right) |\Psi (x,t_{0})\rangle |0\rangle  \tag{A9}
\end{equation}%
where $p_{0}=\frac{1}{2}K\tau \theta _{m}.$ The $IC$ unitary operator $\exp
\left( -ia_{m}^{s}p_{0}x/\hslash \right) $ in (A9) is an $IC$ unitary
momentum-displacement propagator. If now the initial motional state $|\Psi
(x,t_{0})\rangle $ is chosen as a momentum eigenfunction of the single-atom
system, then the unitary process (A9) is really the phase-based QUANSDAM
process of (4.4). In this case, if the error term on the RH side of (A8) can
be neglected, then the LH side of (A8) is really a phase-based QUANSDAM
process.

Since an ideal momentum eigenfunction is hard to prepare, one could employ
other motional state $|\Psi (x,t_{0})\rangle $ than a momentum eigenfunction
to realize the QUANSDAM process of (A8). For example, one could choose a
Gaussian wave-packet motional state to act as $|\Psi (x,t_{0})\rangle $ to
realize the QUANSDAM process. In this case the error term on the RH side of
(A8) may be calculated exactly in theory, although the calculation is not
easy at all when the number $n$ is large. In comparison with the QUANSDAM
process of (A7) the QUANSDAM process of (A8) in performance could be greatly
improved by a Gaussian wave-packet motional state [15]. So far there is not
yet a rigorous mathematical proof to show whether or not the QUANSDAM\
process of (A8) can achieve an exponential quantum-state-difference
amplification. \newline
\newline

\end{document}